\documentclass[aps,prd,eqsecnum,showpacs,amsmath]{revtex4}
\usepackage{graphicx}
\begin{document}

\thispagestyle{empty}

\title{Computing gravitational waves
from slightly nonspherical stellar collapse to a black hole: 
Odd-parity perturbation} 

\author{Tomohiro Harada $^{1}$\footnote{Electronic
 address:harada@gravity.phys.waseda.ac.jp},
Hideo Iguchi $^{2}$\footnote{Electronic
 address:iguchi@th.phys.titech.ac.jp},
and Masaru Shibata$^{3}$\footnote{Electronic
 address:shibata@provence.c.u-tokyo.ac.jp}}
\affiliation{$^{1}$Department of Physics, Waseda University, Shinjuku, Tokyo
169-8555, Japan\\
$^{2}$Department of Physics, Tokyo Institute of Technology, Meguro, Tokyo
152-8550, Japan\\
$^{3}$Department of Earth Science and Astronomy, Graduate School of Arts
and Sciences, University of Tokyo, Meguro, Tokyo 153-8902, Japan} 
\date{\today}

\begin{abstract}                
Nonspherical stellar collapse to a black hole is one of the most 
promising gravitational wave sources for gravitational wave detectors. 
We numerically study gravitational waves 
from a slightly nonspherical stellar collapse to a black hole 
in linearized Einstein theory.
We adopt a spherically collapsing star as
the zeroth-order solution and gravitational waves are computed 
using perturbation theory on the spherical background.  
In this paper, we focus on the perturbation of odd-parity modes. 
Using the polytropic equations of state with 
polytropic indices $n_p=1$ and 3, we qualitatively 
study gravitational waves emitted during the collapse of neutron stars 
and supermassive stars to black holes from a marginally stable 
equilibrium configuration.
Since the matter perturbation profiles can be chosen arbitrarily, 
we provide a few types for them. 
For $n_p=1$, the gravitational waveforms are mainly 
characterized by a black hole quasinormal mode ringing, 
irrespective of perturbation profiles given initially.
However, for $n_p=3$, the waveforms depend strongly on 
 the initial perturbation profiles.
In other words, the gravitational waveforms strongly depend 
on the stellar configuration and, in turn, on the ad hoc choice 
of the functional form of the perturbation in the case of 
supermassive stars.
\end{abstract}

\maketitle

\section{introduction}

Detection of gravitational waves is one of the greatest 
challenges in experimental and theoretical physics in this decade. 
Several kilometer-size laser interferometers, 
such as TAMA~\cite{TAMA2001}, 
the Laser Interferometric Gravitational Wave Observatory 
(LIGO)~\cite{abramovici1992}, and GEO~\cite{hough1992} 
are in operation now and VIRGO~\cite{bradaschia1990} will be 
in several years.  In addition to these ground-based detectors, 
the Laser Interferometer Space Antenna (LISA) with an arm 
length of $5\times 10^{6}~\mbox{km}$ 
has been proposed~\cite{thorne1995,schutz2001}, and is planned
to start taking observations in $2012$. 

Nonspherical stellar collapse is one of the most promising 
sources of gravitational waves for both 
ground-based detectors and space antennas.
Ground-based interferometric 
detectors have a good sensitivity in the 
frequency range between $\sim 10$ and 1 kHz. Thus, 
the stellar core collapse of a massive star to a 
neutron star~\cite{HD} or a black hole~\cite{SP} is one of the targets. 
According to \cite{SP},
at the formation of a massive black hole,
quasinormal modes of a black hole are excited and
gravitational waves of high amplitude associated with
the quasinormal modes are emitted. 
(See~\cite{ks1992} for a review of black hole quasinormal modes.)
The frequency of
gravitational waves associated with the 
fundamental quadrupole quasinormal modes of 
rotating black holes is \cite{Leaver} 
\begin{equation}
f \sim (0.03-0.08)M^{-1} 
\simeq (300-800)
\biggl({M \over 20M_{\odot}}\biggr)^{-1} {\rm Hz},\label{eq1.1}
\end{equation}
where the frequency is higher for more rapidly rotating black holes.
Equation (\ref{eq1.1}) indicates
that formation of black holes of mass 
$\agt 20M_{\odot}$ may be a promising source for 
laser-interferometric detectors. 

The frequency band of space antennas is between 
$\sim 10^{-4}$ and $\sim 0.1$ Hz~\cite{thorne1995}. 
This suggests that the formation of supermassive black holes 
may be one of the most promising sources. Although the actual 
scenarios by which supermassive black holes form are 
still uncertain, viable stellar 
dynamical and hydrodynamical routes
leading to the formation of supermassive black holes 
have been proposed~\cite{rees1984,rees1998,rees2001}. 
In typical hydrodynamical scenarios, 
a supermassive gas cloud is built up from multiple collisions of 
stars or small gas clouds in stellar clusters to form a supermassive star. 
Supermassive stars ultimately collapse to 
black holes following quasistationary 
cooling and contraction to the onset of radial instability~\cite{ZN,st1983}. 
Such dynamical formation of supermassive 
black holes may be a strong gravitational wave source for 
LISA~\cite{st1979,bs1999,ssu2001,sbss2002,ss2002}.

The most hopeful approach for the 
computation of gravitational waves from stellar collapse to a black hole 
is to numerically solve the fully nonlinear coupled equations of
Einstein and the general relativistic hydrodynamic equations. 
There has been much progress in this field in the last 
few years~\cite{MS}. 
However, it is not technically easy to compute gravitational waves
with high precision in numerical relativity, since the amplitude of 
gravitational waves associated with the stellar collapse is not
very large and as a result, they could be contaminated 
by numerical noises and/or gauge modes. 
To cross-check the numerical results and also to compute 
precise gravitational waveforms,  
it is desirable to have another method.

As an alternative approach, linear perturbation theory 
has been developed~\cite{cpm1978,sm1987,gm2000}. In this approach, 
we decompose the fully nonlinear metric and matter field of 
slightly nonspherical profiles into 
a spherically symmetric dynamical field 
and linearized nonspherical perturbations. Because of 
progress in numerical techniques and computational resources, 
Einstein's equations in spherically symmetric 
spacetime can be now accurately computed at low computational cost. 
Furthermore, due to the spherical symmetry of the background spacetime, 
the perturbations can be expanded in spherical harmonics.
Thus all the equations for gravitational waves reduce to 
two simple (1+1) wave equations of even and odd parities,
which can be numerically solved with high precision. 
Although this method is applicable only to 
slightly nonspherical problems, the gravitational waveforms of
small amplitude associated with stellar collapse can be
computed with high accuracy. The result is also 
useful for calibration of fully nonlinear numerical results. 

The history of progress in linear perturbation theory
of a spherically symmetric spacetime 
is as follows. The first work in this field was done by 
Regge and Wheeler~\cite{rw1957} and Zerilli~\cite{Z1970}. 
They derived the linear perturbation 
equations of odd parity~\cite{rw1957} and of even parity~\cite{Z1970}
in Schwarzschild background spacetime.
Subsequently, a gauge-invariant formalism of linear perturbations was
developed by Moncrief~\cite{mon1974}. Extending his work, 
Cunningham, Price, and Moncrief~\cite{cpm1978} 
derived perturbation equations on the Oppenheimer-Snyder solution
for a collapsing uniform dust ball~\cite{os1939}, and computed 
gravitational waves emitted during the gravitational collapse
of a dust ball to a black hole. 
Gerlach and Sengupta~\cite{gs1979} subsequently developed 
a gauge-invariant formulation of the 
linear perturbation on general spherically symmetric spacetimes.
Using the Gerlach-Sengupta formalism, Seidel and co-workers~\cite{sm1987}  
computed gravitational radiation from stellar core collapses, focusing
mainly on the waveforms associated with the formation of neutron stars. 
They numerically solved the spherically symmetric general relativistic
hydrodynamic equations using the May-White scheme~\cite{mw1966,vanriper1979}. 
Harada {\it et al.}~\cite{hcnn1997} studied scalar gravitational radiation
from a collapsing homogeneous dust ball 
in scalar-tensor theories of gravity, 
using a similar method to that of Cunningham, Price, and 
Moncrief~\cite{cpm1978}.
Iguchi, Nakao, and Harada~\cite{inh1998,hin2002}
studied nonspherical perturbations of a collapsing
inhomogeneous dust ball, which is described
by the Lema\^{\i}tre-Tolman-Bondi solution~\cite{ltb1933}. Recently, 
Gundlach and Mart\'{\i}n-Garc\'{\i}a~\cite{gm2000} have developed
a covariant gauge-invariant formulation of nonspherical
perturbations on spherically symmetric spacetimes 
with a perfect fluid, and derived coordinate-independent 
matching conditions for perturbations at the stellar surface.

The dynamics of spherically symmetric spacetimes 
have been often studied using a method developed by
May and White~\cite{mw1966}.
In this scheme, spherically symmetric spacetimes are described 
in terms of the so-called Misner and Sharp coordinate system~\cite{ms1964}, 
in which a spacelike comoving slicing and 
an orthogonal time coordinate are adopted. 
This formulation is robust for the simulation of 
oscillating spherical stars and stellar core collapse to 
neutron stars. However, it is not robust enough to 
carry out simulations for 
black hole formation, because the computation often crashes 
before all of the matter is swallowed into a black hole due to
inappropriate choice of the slicing 
condition~\footnote{For example, 
for the marginally bound collapse of an inhomogeneous dust ball,
which is described by the Lema\^{\i}tre-Tolman-Bondi 
solution~\cite{ltb1933},
the central singularity appears at $t=(6\pi \rho_{c,i})^{-1/2}$,
while the surface is swallowed by an event horizon
at $t=(6\pi \bar{\rho}_{i})^{-1/2}-4M/3$, where
$t$ is the orthogonal time coordinate and 
$\rho_{c,i}$ and $\bar{\rho}_{i}$ are the initial central density 
and average density of the dust cloud at $t=0$, respectively.
It is obvious that the former can be earlier than the latter
for the initial data with certain central concentration.}.  

To compute black hole formation, a null formulation proposed by 
Hernandez and Misner is well suited~\cite{hm1966}. 
In this formulation, 
spacetime is foliated by an outgoing null coordinate and thus 
the whole region outside the black hole horizon can be covered. 
The singularity avoidance of the null foliation is assured until 
the foliation reaches an event horizon if the cosmic censorship holds. 
Using this formulation, Miller and Motta~\cite{mm1989} 
performed the numerical simulation
of collapse to black hole formation. 
Baumgarte, Shapiro, and Teukolsky~\cite{bst1995} used
this formulation
to study neutrino emission in the delayed collapse of hot neutron stars
to black holes. 
This formulation was also applied to the study of 
cosmic censorship~\cite{harada1998,hin2002}
and the formation of primordial black holes~\cite{nj1999}. 
Linke {\it et al.}~\cite{lfjmp2001} also computed the spherical collapse 
of supermassive stars using an outgoing null coordinate
but different radial coordinate, i.e., Bondi metric to
study the neutrino emissivity during the collapse. 

In this paper, we present a new implementation 
in linearized Einstein theory, in which 
spherical background spacetimes are computed with the 
Hernandez-Misner scheme, while
nonspherical linear perturbations are treated using
the single-null coordinate system. 
With the Hernandez-Misner scheme, 
it is possible to follow spherical stellar collapse
to a black hole until almost all the matter has collapsed below 
the event horizon.
The null coordinate system is well-suited for 
computation of gravitational waves emitted near the event horizon, which
we want to study here. 
As a related work to the present treatment,
Siebel {\it et al.}~\cite{siebel2002}
presented simulations of gravitational collapse of neutron stars 
to black holes and the computation of quasinormal ringing
in the spherically symmetric Einstein-fluid-Klein-Gordon system
using an outgoing null coordinate without linear approximation.

This paper is organized as follows. 
In Sec.~II we briefly review the evolution equations for 
spherically symmetric spacetimes 
in terms of the Hernandez-Misner null formulation. 
In Sec.~III we describe the evolution equations 
for odd-parity gauge-invariant perturbations 
in the single-null coordinate system 
and then derive the explicit matching conditions at the stellar surface.
In Sec.~IV we explain the method for computation of gravitational waves 
in our gauge-invariant formalism. 
In Sec.~V we describe numerical techniques 
adopted in the current implementation. 
In Secs.~VI and VII, we present the numerical results of test simulations, 
and gravitational waveforms from the
collapse of a supermassive star and neutron star to black holes, 
respectively. Section VIII is devoted to a summary. 
We adopt geometrical units in which $G=c=1$, 
where $G$ and $c$ denote the gravitational constant and 
speed of light, respectively.

\section{background spacetime}

\subsection{$2 + 2$ split of spherically symmetric spacetimes}

We decompose a spherically symmetric spacetime ${\cal M}$
into a product as ${\cal M}={\cal M}^2\times {\cal S}^2$. Namely 
a metric is written as
\begin{equation}
g_{\mu\nu}\equiv \mbox{diag}(g_{AB},R^{2}\gamma_{ab}),
\end{equation}
where $g_{AB}$, $R$, and $\gamma_{ab}$ are the (1+1) Lorentz metric, 
scalar function on ${\cal M}^2$, and 
unit curvature metric on ${\cal S}^2$.
The greek indices $\mu, \nu, \ldots$,
capital latin indices $A, B, \ldots$, and
small Latin indices $a, b, \ldots$ 
denote the spacetime components, ${\cal M}^2$ components and
${\cal S}^2$ components, respectively. 
The covariant derivatives on ${\cal M}$, ${\cal M}^{2}$,
and ${\cal S}^{2}$ are denoted 
as $_{;\mu}$, $_{|A}$ and $_{:a}$, i.e., we define them from the conditions 
$g_{\mu\nu;\lambda}=0$, $g_{AB|C}=0$, and $\gamma_{ab:c}=0$. 

The stress-energy tensor for general spherically symmetric 
spacetimes is given by 
\begin{equation}
t_{\mu\nu}=\mbox{diag}(t_{AB}, (t^{a}_{a}/2)R^{2}\gamma_{ab}). 
\end{equation}
The totally antisymmetric covariant unit tensors 
$\epsilon_{AB}$ on ${\cal M}^{2}$ and $\epsilon_{ab}$ on 
${\cal S}^{2}$ are defined as 
\begin{eqnarray}
\epsilon_{AC}\epsilon^{CB}=-g_{A}^{~B}, \quad
\epsilon_{ac}\epsilon^{cb}=\gamma_{a}^{~b}.
\end{eqnarray}

\subsection{Hernandez-Misner formulation of general relativistic
hydrodynamics}

We choose the Hernandez-Misner coordinate system of the form 
\begin{equation}
ds^{2}=-e^{2\psi}du^{2}-2e^{\psi+\lambda/2}dud x +R^{2}
(d\theta^{2}+\sin^{2}\theta d\phi^{2}),
\end{equation}
where $ x $ is a comoving coordinate, and $\psi$, $\lambda$, and $R$ 
are functions of $u$ and $x$. 

We assume that the stars are composed 
of a perfect fluid, for which the energy-momentum 
tensor is written as 
\begin{equation} 
t_{\mu\nu}=(\epsilon+p)u_{\mu}u_{\nu}+pg_{\mu\nu},
\end{equation}
where $\epsilon$, $p$, and $u^{\mu}$ are the energy density,
pressure and four velocity of the fluid.

We define the following new variables 
\begin{eqnarray}
U& \equiv & e^{-\psi}R_{,u}, \\
\Gamma& \equiv & e^{-\lambda/2}R_{,x} -U=
\sqrt{1-\frac{2m}{R}+U^{2}}, 
\label{eq:Gamma+U}
\end{eqnarray}
where $m$ is the Misner-Sharp quasilocal mass~\cite{ms1964}.
Then, the field equations are written in the form~\cite{hm1966}
\begin{eqnarray}
U_{,u}&=&-\frac{e^{\psi}}{1-c_{s}^{2}}
\left[\frac{\Gamma e^{-\lambda/2}}{\epsilon+p}p_{,x}
+\frac{m+4\pi R^{3}p}{R^{2}}\right] 
\nonumber \\
& &-\frac{e^{\psi}c_{s}^{2}}{1-c_{s}^{2}}
\left(e^{-\lambda/2}U_{,x}+\frac{2U\Gamma}{R}\right), \\
R_{,u}&=&e^{\psi}U, \\
m_{,u}&=&-e^{\psi}4\pi R^{2}p U, \\
 e^{-\lambda/2} &=&\frac{\Gamma+U}{R_{,x}}, \\
n&=&\frac{e^{-\lambda/2}}{4\pi R^{2}}f, \\
\left(\frac{\epsilon}{ n }\right)_{,u}
&=&-p\left(\frac{1}{ n }\right)_{,u}, 
\label{eq:firstlaw}\\
m_{,x}&=&4\pi R^{2}e^{\lambda/2}(\epsilon\Gamma-p U), \\
\psi_{,x}&=&
\frac{1}{\Gamma}U_{,x}+e^{\lambda/2}\frac{m+4\pi p R^{3}}
{\Gamma R^{2}}.
\end{eqnarray}
$n$ is the baryon rest-mass density,
$f=f(x)$ is an arbitrary function associated with 
the rescaling of the radial coordinate,
and $c_{s}$ is the sound speed which is defined by 
\begin{equation}
c_{s}^{2}\equiv \left(\frac{\partial p}{\partial \epsilon}\right)
_{s=\mbox{const}},
\end{equation}
where $s$ denotes the entropy. 

The regularity condition at $x=0$ gives the boundary conditions as 
\begin{eqnarray}
R&=&0, \\
U&=&0, \\
\Gamma&=&1, \\
m&=&0. 
\end{eqnarray}
The boundary condition at the stellar surface $x=x_{\rm s}$
is given by 
\begin{equation}
p=0.
\end{equation}

In the original form of the Hernandez-Misner formalism~\cite{hm1966},
$f$ is chosen to be unity. In this case, 
$x$ coincides with the conserved mass $\mu$ 
contained in the interior to a shell. 
Another candidate for $f$ is
\begin{equation}
f(x)=\frac{4\pi n(u_{0},x) x^{2}}{(\Gamma+U)(u_{0},x)}. \label{eq221}
\end{equation}
In this case, $x$ coincides with the circumferential
radius of the shell on the null surface $u=u_{0}$. 
In this paper we adopt Eq.~(\ref{eq221}) for $f$, since 
it has a nicer feature for the integration of nonspherical 
perturbations which we will describe in the next section. 

We assume that the exterior of the star is a vacuum. 
Then, the zeroth-order solution is the Schwarzschild spacetime as 
\begin{equation}
ds^{2}=-\left(1-\frac{2M}{R}\right)dT^{2}
+\left(1-\frac{2M}{R}\right)^{-1}dR^{2} 
+R^{2}(d\theta^{2}+\sin^{2}\theta d\phi^{2}),
\end{equation}
where $M$ is the gravitational mass of the system. 
To compute gravitational waves in this background, 
it is convenient to introduce null coordinates 
$\bar{u}$ and $\bar{v}$ defined as
\begin{eqnarray}
\bar{u}&=&T-R_{*}, \\
\bar{v}&=&T+R_{*}, 
\end{eqnarray}
where 
\begin{equation}
R_{*}\equiv R+2M\ln\left(\frac{R}{2M}-1\right).
\end{equation}
Then, the line element is of the form 
\begin{equation}
ds^{2}=-\left(1-\frac{2M}{R}\right)d\bar{u}d\bar{v}
+R^{2}(d\theta^{2}+\sin^{2}\theta d\phi^{2}). 
\end{equation}
In the following, we refer to the standard outgoing null coordinate 
$\bar{u}$ as an observer time coordinate
according to~\cite{bst1995}. 

The ratio of the proper time interval $d\tau_{\rm s}$ 
of an observer at the stellar surface to
the observer time interval $d\bar{u}$ is given by
\begin{equation}
\frac{d\tau_{\rm s}}{d\bar{u}}=(\Gamma+U)(u,x_{\rm s}).
\end{equation}  
The lapse function $\alpha$ at $x$ is defined by the ratio 
of the proper time interval $d\tau$ of each fluid element to 
the observer time interval $d\bar{u}$ as 
\begin{equation}
\alpha(u,x)\equiv \frac{d\tau}{d\bar{u}}=\frac{d\tau}{d\tau_{\rm s}}
\frac{d\tau_{\rm s}}{d\bar{u}}=
\frac{e^{\psi(u,x)}}{e^{\psi(u,x_{\rm s})}}(\Gamma+U)(u,x_{\rm s}).
\end{equation}
We note that the lapse function $\alpha$ is directly 
related to the observed redshift $z$ as $\alpha=1/(1+z)$. 

In solving the dynamics of a spherical star, 
the boundary condition for $\psi$ is arbitrary.
In the present computation, 
we choose $e^{\psi}(u,x_{\rm s})=1$ 
for the boundary condition of $\psi$. Then, 
we can identify the null coordinate $u$ with 
the proper time $\tau_{\rm s}$ of the comoving observer at 
the stellar surface.

\section{Odd-parity perturbations}

\subsection{Gauge-invariant perturbations}

Perturbed metric and matter fields of odd parity are denoted by 
\begin{eqnarray}
\Delta g_{\mu\nu}&=&\left(
\begin{array}{cc}
0 & h_{A}S_{a} \\
h_{A}S_{a} & h (S_{a:b}+S_{b:a}) 
\end{array}
\right), \\
\Delta t_{\mu\nu}&=&\left(
\begin{array}{cc}
0 & \Delta t_{A}S_{a} \\
\Delta t_{A}S_{a} & \Delta t (S_{a:b}+S_{b:a})
\end{array}
\right), 
\end{eqnarray}
where $Y$, $S_{a}\equiv \epsilon_{a}^{~~b}Y_{:b}$ and $S_{a:b}+S_{b:a}$
are the scalar, vector and tensor harmonics, respectively. Here, 
the suffices $l$ and $m$ are omitted for simplicity.
The scalar harmonic function $Y$ satisfies
\begin{equation}
\gamma^{ab}Y_{:ab}=-l(l+1)Y.
\end{equation}

The gauge-invariant perturbations are defined as
\begin{eqnarray}
k_{A}&\equiv & h_{A}-h_{|A}+2h v_{A}, \\
L_{A}&\equiv &\Delta t_{A}-Q h_{A}, \\
L & \equiv & \Delta t -Q h,
\end{eqnarray}
where 
\begin{equation}
v_{A}\equiv \frac{R_{|A}}{R}.
\end{equation}
$L$ is identically zero for $l=1$, and 
no perturbation of odd parity appears for $l=0$.

For a perfect fluid, 
the matter perturbations of odd parity are only specified 
by the four-velocity perturbations as 
\begin{equation}
\Delta u_{\mu}=(0,\beta S_{a}),
\end{equation}
where $\beta$ is a function of $x$ and $u$ and 
completely determines the matter perturbations.
The concrete form is determined by solving the field equations 
[cf. Eq. (\ref{eq:j})]. 
In terms of $\beta$, the gauge-invariant perturbations are written as 
\begin{eqnarray}
L_{A}&=&\beta (\epsilon+p)u_{A}, 
\end{eqnarray}
and $L=0$.

\subsection{Field equations: interior}

The covariant field equations 
for nonspherical perturbations in the stellar interior 
were derived by Gerlach and Sengupta~\cite{gs1979} for 
general matter fields and have been recently reformulated
by Gundlach and Mart\'{\i}n-Garc\'{\i}a~\cite{gm2000} for 
a perfect fluid.  
We follow~\cite{gm2000} to derive the basic equations for the perturbations.

The perturbation equation for the matter field, 
$\Delta (t^{\mu}_{~\nu;\mu})=0$, is integrated to give
\begin{equation}
\beta=-\frac{e^{-\lambda/2}}{R^{2}(\epsilon+p)}j,
\label{eq:j}
\end{equation}
where $j=j(x)$ is an arbitrary function of $x$.
Integration of $j$ by $dx$ yields a conserved quantity $J$ as 
\begin{equation}
J=\int j dx, 
\end{equation}
where $\oint J S_{\phi}\sin\theta d\theta d\phi$ corresponds to 
the $z$ component of angular momentum for $l=1$. 

At $R=0$, $\beta$ should satisfy the regularity condition as
\begin{equation}
\beta=R^{l+1}\bar{\beta},
\end{equation}
where $\bar{\beta}$ is a regular function at $R=0$.
The regularity of $\beta$ also leads to the following condition 
\begin{equation}
e^{-\lambda/2}j=R^{l+3}\bar{j},
\end{equation}
where $\bar{j}$ is a regular function at $R=0$.
In terms of the function $j$, the gauge-invariant matter perturbations 
are described as
\begin{eqnarray}
L_{u}&=&\frac{e^{\psi-\lambda/2}}{R^{2}}j ,\\
L_{x}&=&\frac{1}{R^{2}}j. 
\end{eqnarray}
Thus, in the following, we specify $j$ instead of $\beta$. 

The metric perturbations of odd parity are characterized by 
a master variable as 
\begin{equation}
\Pi \equiv e^{-\psi-\lambda/2}\left[\left(
\frac{k_{x}}{R^{2}}\right)_{,u}-
\left(\frac{k_{u}}{R^{2}}\right)_{,x}
\right],
\label{eq:defpi}
\end{equation}
where $\Pi$ is related to a variable $\tilde{\pi}_{1}$ introduced 
in~\cite{cpm1978} as $\tilde{\pi}_{1}=l(l+1)R^{2}\Pi $.
$\Pi$ should satisfy the regularity condition at $R=0$ as 
\begin{equation}
\Pi=R^{l-2}\bar{\Pi},
\label{eq:barPi}
\end{equation}
where $\bar{\Pi}$ is a regular function at $R=0$. 
$\bar{\Pi}$ satisfies the following wave equation for $l\ge 2$
\begin{eqnarray}
& &-2e^{-\psi-\lambda/2}\bar{\Pi}_{,x u}+e^{-\psi-\lambda/2}(e^{\psi-\lambda/2}\bar{\Pi}_{,x})_{,x} \nonumber \\ 
& &\quad +\frac{2(l+1)}{R}\{-(\Gamma+U)e^{-\psi}\bar{\Pi}_{,u}
+\Gamma e^{-\lambda/2}\bar{\Pi}_{,x}\}
\nonumber \\
& &\quad -(l+2)\left[4\pi (\epsilon-p)
+(l-2)\frac{2m}{R^{3}}\right]\bar{\Pi} \nonumber \\
&=&16\pi\left[Re^{-\psi-\lambda/2}
(e^{\psi}\bar{j})_{,x}
+\{(l+1)\Gamma+2U\}\bar{j}\right].
\label{eq:waveeqforbarpi}
\end{eqnarray}
The relation between $k_{A}$ and $\Pi$ is given for $l\ge 1$ by
\begin{eqnarray}
(l-1)(l+2)k_{u}&=&
16\pi R^{2}L_{u}+(R^{4}\Pi)_{,u}
 -e^{\psi-\lambda/2}(R^{4}\Pi)_{,x}, 
\label{eq319}\\
(l-1)(l+2)k_{x}&=&16\pi R^{2}L_{x}-(R^{4}\Pi)_{,x}.\label{eq320}
\end{eqnarray}
For $l\ge 2$, the gauge-invariant 
metric perturbations are obtained 
from Eqs.~(\ref{eq:waveeqforbarpi}), (\ref{eq319}) and (\ref{eq320}). 
However, for $l=1$, Eqs.~(\ref{eq319}) and (\ref{eq320}) give 
\begin{equation}
R^{4}\Pi=16\pi J, \label{eq321}
\end{equation}
where we assume that the perturbation is regular at $R=0$. 
Equation (\ref{eq321}) implies that 
there is no gravitational-wave mode for $l=1$.

\subsection{Field equations: exterior}

We have two metric perturbations $k_{T}$ and $k_{R}$ in the exterior.
Defining the master variable $\Pi$ as
\begin{equation}
\Pi\equiv \left(\frac{k_{R}}{R^{2}}\right)_{,T}
-\left(\frac{k_{T}}{R^{2}}\right)_{,R},
\end{equation}
we find the following wave equations for $l\ge 2$ 
\begin{equation}
-\Phi_{,TT}+\Phi_{,R_{*}R_{*}}-V(R)\Phi=0,
\label{eq:waveextr}
\end{equation}
where
\begin{eqnarray}
\Phi&\equiv& R^{3}\Pi, \\
V(R)&\equiv& \left(1-\frac{2M}{R}\right)\left(\frac{l(l+1)}{R^{2}}
-\frac{6M}{R^{3}}\right).
\end{eqnarray}
Here, we note that $\Phi$ is related to the variable 
$\tilde{\psi}$ defined in~\cite{cpm1978} as 
$\tilde{\psi}=l(l+1)\Phi$. 
Using the double-null coordinates $(\bar{u},\bar{v})$,
we obtain
\begin{equation}
4 \Phi_{,\bar{u}\bar{v}}+V(R)\Phi=0.
\label{eq:waveexuv}
\end{equation}

Equation~(\ref{eq:waveextr}) has a static solution with an appropriate
fall-off at infinity as 
\begin{equation}
\Phi_{\rm static}=\frac{q}{l(l+1)}\left(\frac{2M}{R}\right)^{l}
F\left(l-1,l+3,2l+2;\frac{2M}{R}\right),
\label{eq:exteriorstatic}
\end{equation}
where $F(a,b,c;z)$ denotes the hypergeometric function and 
$q$, which has the dimension of length, 
corresponds to the multipole moment of the system.
Because $[l(l+1)]^{-1}$ is factorized out in the above equation, 
the definition of the moment is the same 
as that defined in~\cite{cpm1978}.
This static solution is used for providing initial conditions 
of metric perturbations (see Sec. V).

The relation between $k_{A}$ and $\Pi$ is given for $l\ge 1$ by
\begin{eqnarray}
(l-1)(l+2)k_{T}&=&-\left(1-\frac{2M}{R}\right)(R^{4}\Pi)_{,R}
\nonumber \\
&=&(R^{4}\Pi)_{,\bar{u}}-(R^{4}\Pi)_{,\bar{v}}, \label{eq328} \\
(l-1)(l+2)k_{R}&=&-\left(1-\frac{2M}{R}\right)^{-1}(R^{4}\Pi)_{,T}
\nonumber \\
               &=&-\left(1-\frac{2M}{R}\right)^{-1}
[(R^{4}\Pi)_{,\bar{u}}+(R^{4}\Pi)_{,\bar{v}}]. \label{eq329}
\end{eqnarray}
For $l=1$, we find the solutions of Eqs.~(\ref{eq328}) and 
(\ref{eq329}) as 
\begin{equation}
R^{4}\Pi=\mbox{const}.\label{eq330}
\end{equation}
The integration constant in Eq. (\ref{eq330}) is related to the total 
angular momentum of the fluid perturbation for $l=1$. The master variable 
$\Pi$ is time independent in the exterior for $l=1$ as 
shown in Eq. (\ref{eq330}). 
This constancy implies the conservation of the angular momentum
of linear perturbation in the spherically symmetric background.

\begin{figure}[htbp]
\includegraphics[scale=1]{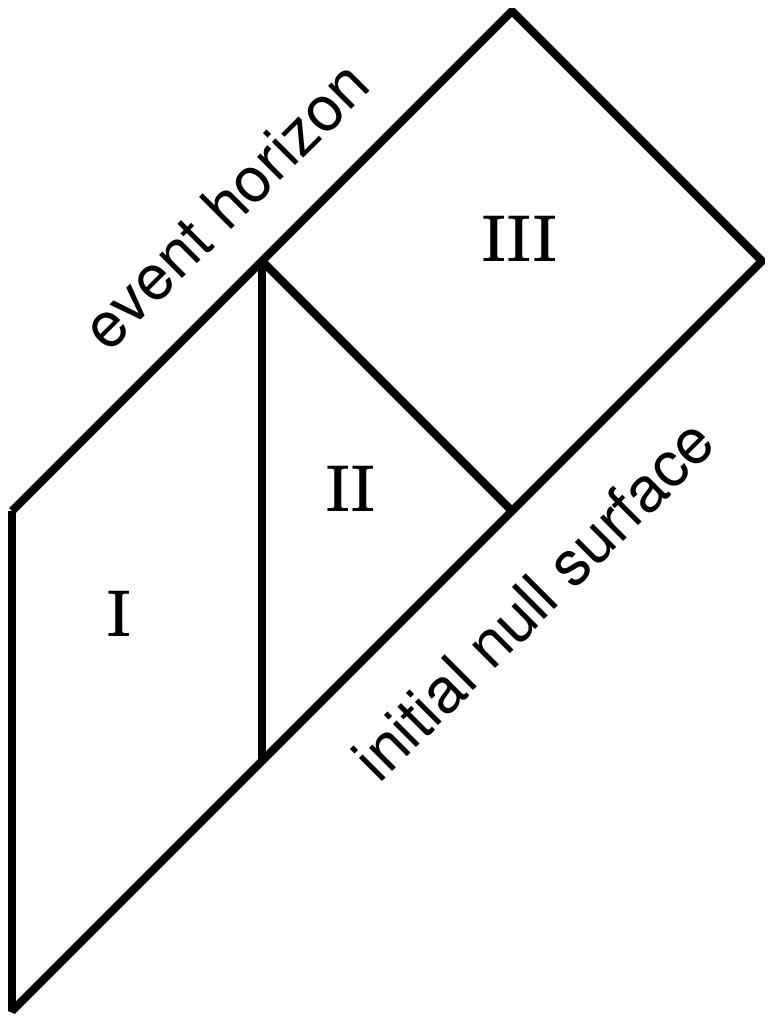}
\caption{\label{fg:penrose} Spacetime structure
of spherically symmetric black hole formation is depicted. 
Region I denotes 
the stellar interior, region II denotes the intermediate exterior region 
from which we can emit an ingoing light ray which 
hits the stellar surface before the horizon formation, 
and region III denotes the far exterior region. 
The boundary between regions I and II is the stellar surface.}
\end{figure}

To compute nonspherical metric perturbations, 
we divide the background spacetime into regions I--III 
(see Fig.~\ref{fg:penrose}).
Region I is defined as the interior of the star.
Region II is an intermediate
exterior region from which one can emit an ingoing null ray 
that encounters the stellar surface before the stellar 
surface is swallowed into the event horizon. 
Region III is the exterior region outside region II. 
Region II is introduced 
to help the matching procedure at the 
stellar surface in numerical computation. 
Hereafter, we will refer to the ingoing null surface
which divides the exterior regions into 
regions II and III as the junction null surface.
Such an elaborate procedure is needed to calculate 
the late-time gravitational radiation extracted at the
point far from the star and to assure that the matching 
condition is satisfied at the stellar surface
simultaneously.  

For the convenience of computation, we introduce new null 
coordinates $\tilde{u}$ and $\tilde{v}$ in
region II. We identify these null coordinates 
$\tilde{u}$ and $\tilde{v}$
with the values of the proper time $\tau_{\rm s}$ of an observer
comoving at the stellar surface $x=x_{\rm s}$ at its intersection of 
an outgoing ray ($\tilde{u}={\rm const}$) and an ingoing ray 
($\tilde{v}={\rm const}$), respectively. Namely, 
the stellar surface is given by $\tilde{u}=\tilde{v}$. 

When we define functions $A$ and $B$ as 
\begin{eqnarray}
A(\tilde{u})&\equiv &\frac{d\bar{u}}{d\tilde{u}}=
\frac{1}{\Gamma+U}(\tilde{u},x_{\rm s}), \\
B(\tilde{v})&\equiv &\frac{d\bar{v}}{d\tilde{v}}
=\frac{1}{\Gamma-U}(\tilde{v},x_{\rm s}),
\end{eqnarray}
we can rewrite the wave equation for $\Phi$ as
\begin{equation}
4\Phi_{,\tilde{u}\tilde{v}}+A(\tilde{u})
B(\tilde{v})V(R)\Phi=0.
\label{eq:evolvePhiII}
\end{equation}
We integrate Eq.~(\ref{eq:evolvePhiII}) in region II.
The event horizon is given by a finite value of $\tilde{u}$.
It is found that the effective potential term $\tilde V \equiv A(\tilde{u})
B(\tilde{v})V(R)$ is regular on the event horizon
in this coordinate system, which helps numerical integration 
of the wave equation. 

\subsection{Matching}
The matching condition 
at the stellar surface $x=x_{\rm s}$ 
for the odd-parity perturbations 
is obtained from the continuity condition for $\Pi$, $n^{A}L_{A}$,  
and $n^{A}\Pi_{|A}-16\pi R^{-2} u^{A}L_{A}$ for $l\ge 1$, and
for $n^{A}k_{A}$ and $u^{A}k_{A}$ for $l\ge 2$, 
where $n_{A}\equiv -\epsilon_{AB}u^{B}$~\cite{gm2000}.
The explicit equations are 
\begin{eqnarray}
&& \Pi_{{\rm in}}=\Pi_{{\rm ex}}, \label{eq:matchingPi}  \\
&&-e^{-\psi}\Pi_{{\rm in},u}+e^{-\lambda/2}\Pi_{{\rm in},x}-16\pi R^{-4}
e^{-\lambda/2}j
\nonumber \\
&=& -\frac{\Pi_{{\rm ex},\bar{u}}}{\Gamma+U}
+\frac{\Pi_{{\rm ex},\bar{v}}}{\Gamma-U}
=-\Pi_{{\rm ex},\tilde{u}}+\Pi_{{\rm ex},\tilde{v}}. 
\end{eqnarray}
Using $u^{A}\Pi_{{\rm in}|A}=u^{A}\Pi_{{\rm out}|A}$,
we can derive an alternative form of the matching condition as 
\begin{eqnarray}
& &-2e^{-\psi}\Pi_{{\rm in},u}+e^{-\lambda/2}\Pi_{{\rm in},x}
-16\pi R^{-4}e^{-\lambda/2}j
\nonumber \\
&=& -2 \frac{\Pi_{{\rm ex},\bar{u}}}{\Gamma+U}=-2 \Pi_{{\rm ex},\tilde{u}}
\label{eq:matchingPiU}. 
\end{eqnarray}

For $l=1$, the matching conditions lead to 
\begin{eqnarray}
\Pi_{{\rm in}}&=&16\pi \frac{J(x)}{R^{4}},\\
\Pi_{{\rm ex}}&=&16\pi \frac{J(x_{\rm s})}{R^{4}}.
\end{eqnarray}
Thus, the gauge-invariant variable $\Pi$ 
is completely determined by the initial distribution 
of perturbed angular momentum in the star.

\section{Gravitational waves}

To compute gravitational waves in the wave zone, 
it is convenient to adopt the radiation gauge. 
In this gauge, the following tetrad components denote 
the $+$ and $\times$ modes of gravitational waves; 
\begin{eqnarray}h_{+}&\equiv& \frac{1}{2}(h_{\hat{\theta}
\hat{\theta}}-h_{\hat{\phi}\hat{\phi}}), \\
h_{\times}&\equiv &h_{\hat{\theta}\hat{\phi}}.
\end{eqnarray}  
Hereafter, we adopt the following choice for
the orthonormal bases:
\begin{eqnarray}
& &Y_{l0}~~~{\rm for~}m=0~~{\rm and}~~
\nonumber \\
& &Y_{l,m_{\pm}}\equiv {1 \over \sqrt{2}}(Y_{l,m}\pm Y_{l,-m})~~~
{\rm for~}m\not=0. \label{eq404}
\end{eqnarray}

The metric perturbations $h_{+}$ and 
$h_{\times}$ in the radiation gauge are written in terms of $\Phi$ as
\begin{eqnarray}
h_{+}&=&-\frac{1}{(l-1)(l+2)}
\frac{1}{R}(\Phi+\mbox{const}){\cal A}_{+}(\theta,\phi)
+O(R^{-2}),\\
h_{\times}&=&-\frac{1}{(l-1)(l+2)}
\frac{1}{R}
(\Phi+\mbox{const}){\cal A}_{\times}(\theta,\phi)
+O(R^{-2}),
\end{eqnarray}
where the angular dependence can be explicitly calculated as
\begin{eqnarray}
{\cal A}_{+}(\theta,\phi)&\equiv &S_{\theta:\theta}-\frac{1}{\sin^{2}\theta}S_{\phi:\phi} \nonumber \\
&=&2 \left(\frac{1}{\sin\theta}Y_{,\theta\phi}
-\frac{1}{\sin\theta\tan\theta}Y_{,\phi}\right)
, \\
{\cal A}_{\times}(\theta,\phi)&\equiv & \frac{S_{\theta:\phi}+S_{\phi:\theta}}{\sin\theta} \nonumber \\
&=& \frac{2}{\tan\theta}Y_{,\theta}
-2m^{2}\frac{1}{\sin^{2}\theta}Y+l(l+1)Y.
\end{eqnarray}
Here, $Y$ denotes one of the bases shown in Eq.~(\ref{eq404}). 
The luminosity of gravitational waves $P_{l,n}$ for each mode 
in terms of the master variable
is given by (see~\cite{cpm1978,inh1998} for $m=0$
and also~\cite{ll1975})
\begin{eqnarray}
\frac{d P_{l,n}}{d\Omega}&=&\frac{1}{16\pi(l-1)^{2}(l+2)^{2}}
\Phi_{,\bar{u}}^{2}({\cal A}_{+}^{2}+{\cal A}_{\times}^{2})
(\theta,\phi), \\
P_{l,n}&=&\frac{1}{16\pi}\frac{l(l+1)}{(l-1)(l+2)}
\Phi_{,\bar{u}}^{2},
\end{eqnarray}
where the subscript $n$ denotes $0$ and $m_{\pm}$. 

\section{Numerical method}

\subsection{Numerical integration}

The spherically symmetric stellar collapse is 
computed using the single-null 
comoving coordinates. Our method is essentially the same 
as that used by Baumgarte, Shapiro, and Teukolsky~\cite{bst1995}.
An artificial viscosity term is incorporated
to deal with shock waves.
The details of numerical method which we adopt are found in~\cite{bst1995}.

To solve the perturbation equations 
for the interior of the star (region I),
it is convenient to decompose 
Eq.~(\ref{eq:waveeqforbarpi}) into 
the first-order differential equations as 
\begin{eqnarray}
& &-2e^{-\psi-\lambda/2}Q_{,u}+e^{-\psi-\lambda/2}(e^{\psi-\lambda/2}Q)_{,x} 
+\frac{2(l+1)}{R}\{-(\Gamma+U)e^{-\psi}P
+\Gamma e^{-\lambda/2}Q\} \nonumber \\
& & \hskip 1cm \quad -(l+2)\left[4\pi(\epsilon-p) 
+(l-2)\frac{2m}{R^{3}}\right]\bar{\Pi} \nonumber \\
& & \quad= 16\pi\left[Re^{-\psi-\lambda/2}
(e^{\psi}\bar{j})_{,x}
+\{(l+1)\Gamma+2U\}\bar{j}\right], \label{eq:evolveQ}\\
& & -2e^{-\psi-\lambda/2}P_{,x}+e^{-\psi-\lambda/2}(e^{\psi-\lambda/2}Q)_{,x} 
+\frac{2(l+1)}{R}\{-(\Gamma+U)e^{-\psi}P
+\Gamma e^{-\lambda/2}Q\} \nonumber \\
& & \hskip 1cm \quad -(l+2)\left[4\pi(\epsilon-p)
+(l-2)\frac{2m}{R^{3}}\right]\bar{\Pi} \nonumber \\
& &\quad = 16\pi\left[Re^{-\psi-\lambda/2}
(e^{\psi}\bar{j})_{,x}
+\{(l+1)\Gamma+2U\}\bar{j}\right], 
\label{eq:evolveP} \\
& &\bar{\Pi}_{,u}=P, \\
& &\bar{\Pi}_{,x}=Q.
\label{eq:Q}
\end{eqnarray}
We note that 
the regularity at the center requires
\begin{equation} 
P=e^{\psi-\lambda/2}Q. 
\label{eq:centerPQ}
\end{equation}

Equation (\ref{eq:evolveQ}) constitutes a hyperbolic-type 
partially differential equation to which 
we apply the first-order up-wind scheme to stably evolve the function $Q$.
Other equations constitute ordinarily differential equations, 
and thus the integration is 
carried out with the second-order Runge-Kutta method. 
The Courant-Friedrich-Lewy (CFL) condition 
for the stability of integration limits 
the $n$-th time step $\Delta u^{n}$ as 
\begin{equation}
\Delta u^{n}= 2 C\min_{i} e^{-\psi^{n}_{i}+(\lambda^{n}_{i}/2)}\Delta x_{i},
\end{equation}
where $C (\le 1)$ is the Courant number. 

In the exterior of the star (regions II and III), 
the double-null coordinates are adopted.
In region II, 
Eq.~(\ref{eq:evolvePhiII}) is decomposed into
the first-order differential equations as 
\begin{eqnarray}
\tilde{Z}_{,\tilde{v}}&=&-\frac{1}{4}A(\tilde{u})B(\tilde{v})V(R)\Phi, 
\label{eq:tildeZV}\\
\tilde{W}_{,\tilde{u}}&=&-\frac{1}{4}A(\tilde{u})B(\tilde{v})V(R)\Phi, \\
\Phi_{,\tilde{u}}&=&\tilde{Z},\\
\Phi_{,\tilde{v}}&=&\tilde{W}, 
\label{eq:PhitildeV}
\end{eqnarray}
and in region III, 
Eq.~(\ref{eq:waveexuv}) is decomposed as 
\begin{eqnarray}
Z_{,\bar{v}}&=&-\frac{1}{4}V(R)\Phi, 
\label{eq:ZV}\\
W_{,\bar{u}}&=&-\frac{1}{4}V(R)\Phi, \\
\Phi_{,\bar{u}}&=&Z,\\
\Phi_{,\bar{v}}&=&W.
\end{eqnarray}
For integration of these equations, 
we use the finite differencing scheme proposed by 
Hamad\'e and Stewart~\cite{hs1996}.

The coordinates $(\tilde{u},\tilde{v})$ depend on the
spacetime trajectory of the stellar surface. Thus, they
are determined after the spherically symmetric 
stellar dynamics is solved. 
For this reason, we divide a numerical simulation 
into two steps. In the first step, we carry out numerical computation 
for the zeroth-order background solution taking into account the 
CFL condition for the first-order nonspherical perturbations, 
and in the second step, we evolve the nonspherical perturbations. 

For accurate numerical integration, the distribution of 
grid points in region I plays a quite important role.
We adopt the equally spaced grid 
in terms of the initial circumferential radius.
In region II, the distribution of grid points 
is automatically determined in computing 
zeroth-order solution. 
In region III, the equally spaced grid 
in terms of $\bar{u}$ and $\bar{v}$ works well.

\subsection{Matching}

At the stellar surface,
the matching conditions (\ref{eq:matchingPi}) and 
(\ref{eq:matchingPiU}) for $\Pi$ and its derivatives
determine the boundary conditions for $\bar{\Pi}$, $P$, and $Q$ at 
$x=x_{\rm s}$ in region I, 
and for $\Phi$, $\tilde{Z}$ and $\tilde{W}$ at $\tilde{u}=\tilde{v}$ 
in region II.
In matching solutions of regions I and II, 
we have to be careful since 
the outermost several mass shells form a very disperse envelope
during the collapse. This implies that 
the grid resolution around the outermost envelope 
is not very good. If the matching 
between regions I and II is done 
at the location of the outermost mass shell, 
a large numerical error is produced. 
To suppress the numerical error, we match the solutions 
at the location of a mass shell which is not outermost one but 
is located slightly inside the outermost one. We find that 
this works quite well. We have also checked that 
the numerical results do not depend on the location of 
the matching if more than $\sim$ 95\% of the total mass 
is contained inside the 
mass shell for the matching.

At the junction null surface, the matching is not necessary, 
since the double-null coordinates are adopted in both regions. 
We only need to store the numerical data set on the junction 
null surface in region II, 
and use it as a part of initial data for the integration of 
region III.

\subsection{Initial data}

In the present formulation, 
the null cone composed of $u=u_{0}, \tilde{u}=\tilde{u}_{0}$, 
and $\bar{u}=\bar{u}_{0}$ should be taken as an initial null surface.
Since it is the characteristic surface 
of the wave equation, we 
only need to specify $\Pi_{\rm init}(R)$ and $\beta_{\rm init}(R)$
there. 

We provide the initial data set in the following manner: 
In region I, we first give the functions $\Pi_{\rm init}(R)$
and $\beta_{\rm init}(R)$. Then, $\bar{\Pi}(u_{0},x)$ and 
$j(x)$ are specified from 
Eqs.~(\ref{eq:j}) and (\ref{eq:barPi}). 
From the initial data $\bar{\Pi}(u_{0},x)$,
we can determine $Q(u_{0},x)$ on the initial slice 
by differentiation through Eq.~(\ref{eq:Q}).
Then, from $\bar{\Pi}(u_{0},x)$, $Q(u_{0},x)$, and $j(x)$, 
we can obtain $P(u_{0},x)$ by integrating
Eq.~(\ref{eq:evolveP}) with the central value 
given by the boundary condition (\ref{eq:centerPQ}) at the center.  

In region II
we first provide $\Phi(\tilde{u}_{0},\tilde{v}) 
=\Pi (\tilde{u}_{0},\tilde{v})R^{3}(\tilde{u}_{0},\tilde{v})$. 
Then, $\tilde{W}(\tilde{u}_{0},\tilde{v})$ 
is obtained by differentiation through 
Eq.~(\ref{eq:PhitildeV}), and $\tilde{Z}$ by integrating
Eq.~(\ref{eq:tildeZV}) with the initial 
value given by the matching condition (\ref{eq:matchingPiU}) 
at the surface. 

In region III the method for 
construction of the initial data sets of $\Phi$, $W$, and $Z$
is the same as that in region II, except for that 
the initial values needed for integration of 
Eq.~(\ref{eq:ZV}) are given at the point 
on the junction null surface. 

When the background spacetime is initially momentarily static, 
it is natural to choose momentarily static initial data sets. 
Thus, the initial data set is given in the following 
manner. First we specify $\beta_{\rm init}(R)$, and 
compute $j(x)$ from Eq.~(\ref{eq:j}).
Then, we determine $\bar{\Pi}(u_{0},x)$ and 
$Q(u_{0},x)$ by integrating Eqs.~(\ref{eq:evolveP})
and (\ref{eq:Q}) using the condition $P(u_{0},x)=P_{,x}(u_{0},x)=0$
with the initial value $\bar{\Pi}(u_{0},0)$ and 
$Q(u_{0},0)=0$. Here, we have to tune $\bar{\Pi}(u_{0},0)$ 
so that it matches the static exterior solution at
the stellar surface. The matching is written as 
\begin{equation}
e^{-\lambda/2}Q+\left[(2l+1)+\frac{2M}{R}\frac{F'_{l}}{F_{l}}
\right]\frac{\Gamma+U}{R}\bar{\Pi}-16\pi\bar{j}R=0,
\end{equation}
where $F(l-1,l+3,2l+2;2M/R)$ is abbreviated as $F_{l}$. 

After this procedure, the multipole moment $q$ is determined 
by matching the interior solution with the 
exterior static solution~(\ref{eq:exteriorstatic}). 
Then, using the moment $q$, the initial data set for the exterior 
solution is provided by Eq.~(\ref{eq:exteriorstatic}).

\subsection{Event horizon}

The effective potential term of the wave equation for 
$\Pi$ is regular on the event horizon. However,
it contains a term of the form ``zero divided by zero'' 
(for $R \rightarrow 2M$, $A \rightarrow \infty$, and $V \rightarrow 0$).
This implies that it is not possible to numerically integrate 
the wave equation for $\Pi$ 
on the null surfaces which are very close to the event horizon. 
To integrate the wave equations until the 
null surface reaches the event horizon, 
we use an extrapolation for the value of $\Pi$ 
on the junction null surface in the
neighborhood of the event horizon as
\begin{equation}
\Pi=\Pi^{N_{\rm max}}+\frac{\Pi^{\rm EH}-\Pi^{N_{\rm max}}}
{R^{\rm EH}-R^{N_{\rm max}}} 
(R-R^{N_{\rm max}}),
\end{equation}
where the value $\Pi^{\rm EH}$ on the event horizon is estimated as
\begin{equation}
\Pi^{\rm EH}=\Pi^{N_{\rm max}}+\frac{\Pi^{N_{\rm max}}-
\Pi^{N_{\rm max}-1}}{R^{N_{\rm max}}-R^{N_{\rm max}-1}} 
(R^{\rm EH}-R^{N_{\rm max}}),
\end{equation}
$\Pi^{N_{\rm max}}$ and $R^{N_{\rm max}}$ are the values for $\Pi$ and $R$
on the junction null surface at the $N_{\rm max}$th (final) 
time step in region II and $R^{\rm EH}=2M$. 
To compute gravitational waves in region III 
emitted in the neighborhood of
the event horizon, 
this extrapolation is necessary and quite helpful. 

\section{Code tests}

\subsection{Spherically symmetric hydrodynamic code}

As the first test of the spherically symmetric 
general relativistic hydrodynamic code, 
we performed a simulation for the homogeneous dust collapse. 
The numerical solution can be compared with 
the Oppenheimer-Snyder solution~\cite{os1939}. We chose 
the total grid number as 500 in this test. 
As the initial condition, we set a momentarily static 
dust ball of uniform density and of radius $4M$.
Precisely speaking, the dust surface is maximally expanded
on the initial null slice.
In this case, the event horizon is located 
at $\tau_{\rm s}=7.243M$, where 
$\tau_{\rm s}$ denotes the proper time of an observer comoving with
the stellar surface. 

\begin{figure}[htbp]
\includegraphics[scale=0.6]{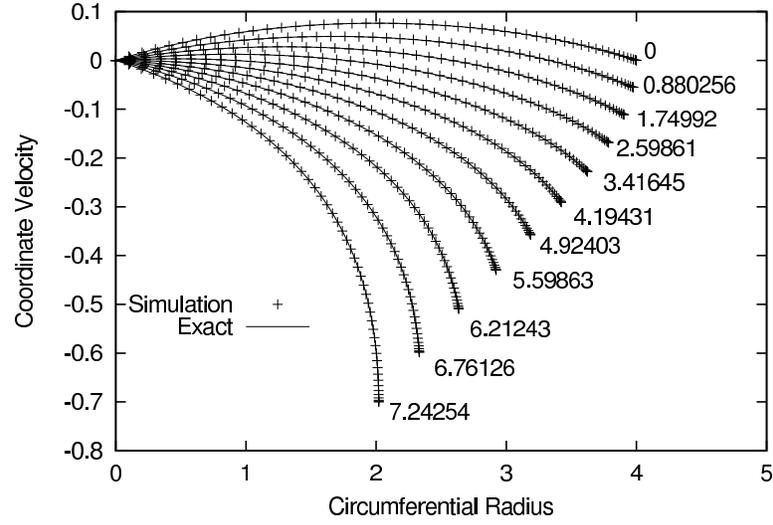}
\caption{\label{fg:oppenheimer_snyder} 
Snapshots of the velocity profile in the numerical simulation of 
homogeneous dust collapse with the initial radius $R=4M$.
Crosses and solid curves denote the numerical and exact solutions. 
The initial data set is put on the initial outgoing null cone 
on which the dust surface is momentarily static. 
The vertical and horizontal axes are the coordinate velocity 
$U=e^{-\psi}R_{,u}$ and the circumferential 
radius $R$, respectively.
The attached labels denote 
the proper time $\tau_{\rm s}$ of a comoving
observer at the stellar surface. All the quantities are shown 
in units of $M=1$.}
\end{figure}

In Fig.~\ref{fg:oppenheimer_snyder} we show the time evolution of 
the coordinate velocity profiles as a function of
circumferential radius at selected time steps. The crosses denote 
the numerical results and the solid curves the exact solutions.
The figure shows that the numerical results agree with 
the exact solutions within $0.01\%$ error. 
We also note that the computation can be 
continued until the null hypersurface reaches a 
surface very close to the event horizon. 

To demonstrate that the code works well even in the presence of 
pressure, we carried out a long-term evolution of 
spherically symmetric stars in equilibrium. 
Here, we adopt the polytropic equations of state as 
\begin{eqnarray}
\epsilon&=&n(1+e), \\
p&=&Kn^{\Gamma_{\rm a}}, \\
n e&=&\frac{p}{\Gamma_{\rm a}-1},
\end{eqnarray}
where $e$ is the specific internal energy, and 
$\Gamma_{\rm a}$ an adiabatic constant, for which 
we choose as 2 and $\approx 4/3$. 
With $\Gamma_{\rm a}=2$,
moderately stiff equations of state for neutron stars are 
qualitatively approximated. 
With $\Gamma_{\rm a} \approx 4/3$, the equation of state for 
supermassive stars of mass $\agt 10^{6}M_{\odot}$,
in which the radiation pressure dominates over the gas pressure, 
is well approximated. 
According to~\cite{st1983}, the equation of state for supermassive
stars may be approximated by the polytropic equation of state 
with $\Gamma_{\rm a}$ as 
\begin{equation}
\Gamma_{\rm a}=
\frac{4}{3}+0.00142\left(\frac{M}{10^{6}M_{\odot}}\right)^{-1/2}.
\end{equation}
Thus, we adopt $\Gamma_{\rm a}=(4/3)+0.00142$ to model a supermassive
star of mass $10^6M_{\odot}$. The equilibrium
configurations are obtained by solving the 
Tolman-Oppenheimer-Volkoff equation. 

\begin{figure}[htbp]
\begin{center}
\begin{tabular}{cc}
(a)\includegraphics[scale=0.45]{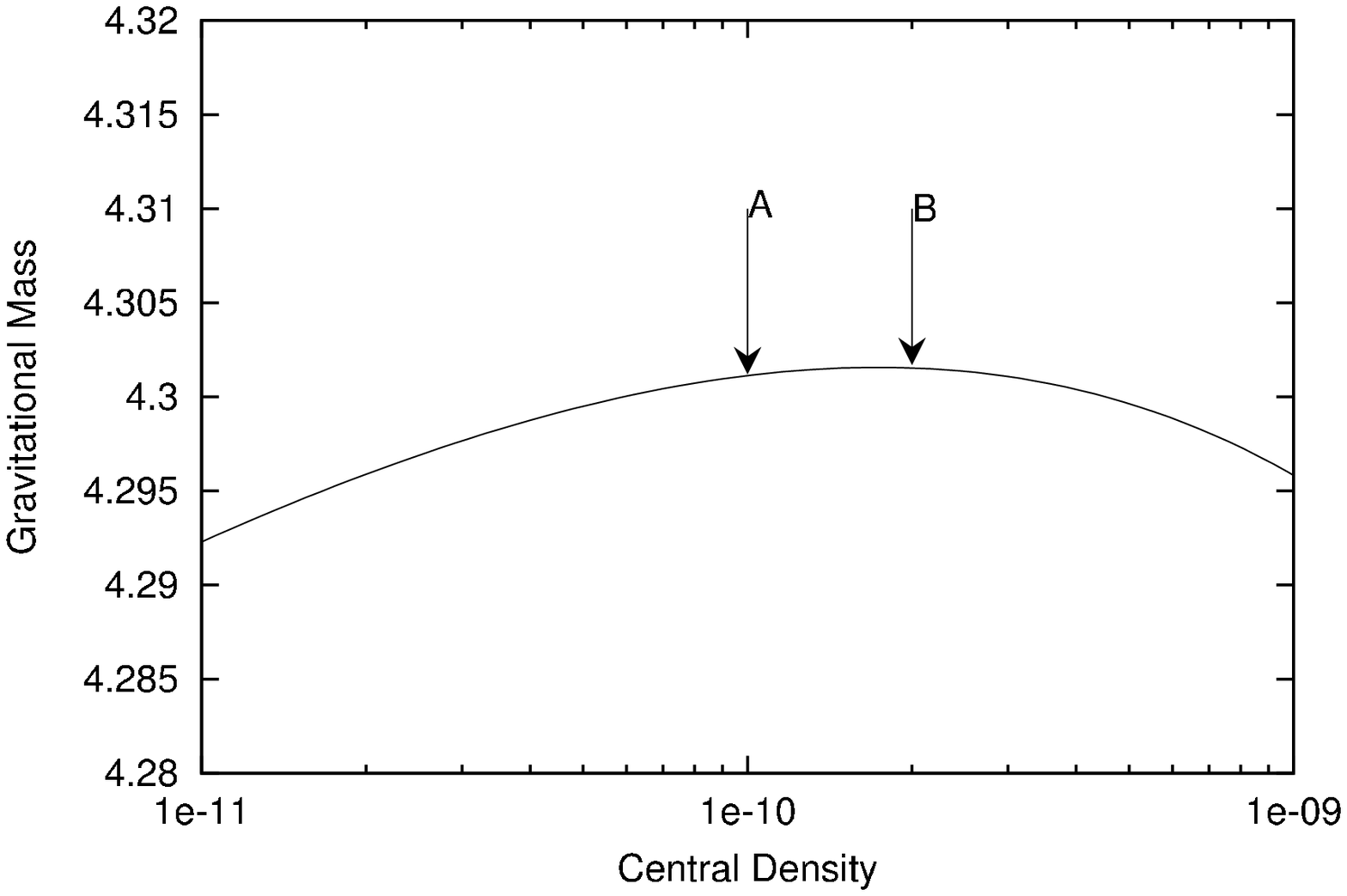} &
(b)\includegraphics[scale=0.45]{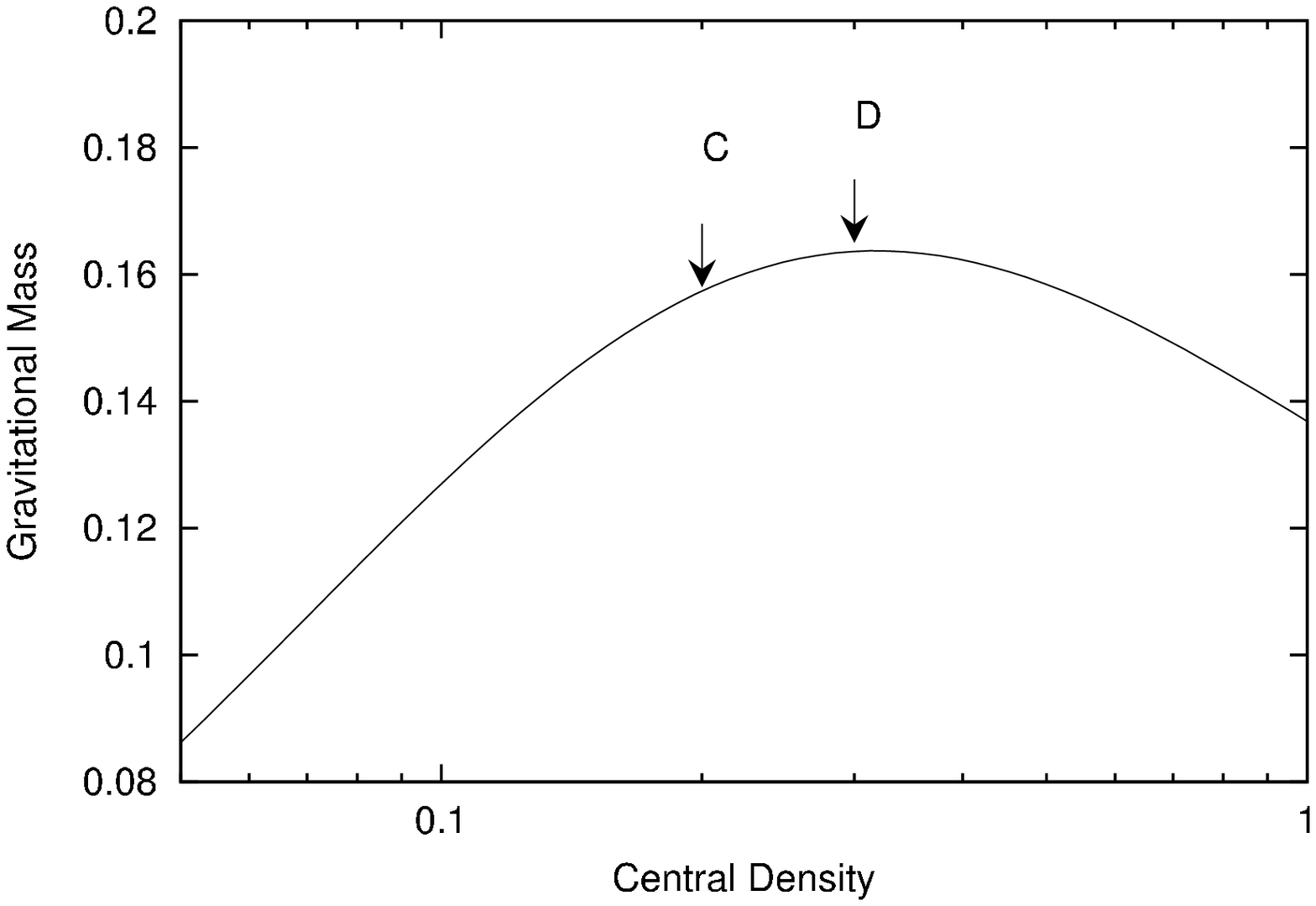} \\
\end{tabular}
\end{center}
\caption{\label{fg:sequence} 
Gravitational mass as a function of central density of 
spherical equilibrium stars 
with polytropic equations of state $P=K n^{\Gamma_{\rm a}}$, where 
(a) $\Gamma_{\rm a}=(4/3)+0.00142$ and (b) $\Gamma_{\rm a}=2$. 
Models A and C are stable while 
models B and D are marginally stable against gravitational collapse. 
The numerical values are shown in units of $c=G=K=1$.}
\end{figure}

In Fig.~\ref{fg:sequence} 
the gravitational mass as a function of the central density for the 
equilibrium configurations with $\Gamma_{\rm a}=(4/3)+0.00142$ 
and $2$ are shown. For simulations in this paper, 
we adopted four static configurations of nearly 
maximum mass along the equilibrium sequences 
(see A--D in Fig.~\ref{fg:sequence}) 
as initial data sets. Models A and C are stable 
against gravitational collapse, while B and D are
marginally stable. 
We summarize these models in Table~\ref{tb:model}.

\begin{table}[htbp]
  \caption{\label{tb:model} Models for equilibrium solutions. The unit 
of $c=G=M=1$ is adopted.}
\begin{center}
    \begin{tabular}{c|ccc} 
\hline 
\hline 
      Model  & $\Gamma_{\rm a}$ & Central energy density & Radius \\ \hline  
      A & $(4/3)+0.00142$ & $1.852\times 10^{-9}$  & $1.900\times 10^{3}$\\
      B & $(4/3)+0.00142$ &  $3.707\times 10^{-9}$ & $1.508\times 10^{3}$\\
      C & 2 & $5.944\times 10^{-3}$ & 5.501 \\
      D & 2 & $1.044\times 10^{-2}$ & 4.745 \\
\hline 
\hline 
    \end{tabular}
  \end{center}
\end{table}

\begin{figure}[htbp]
\begin{center}
\begin{tabular}{cc}
(a)\includegraphics[scale=0.45]{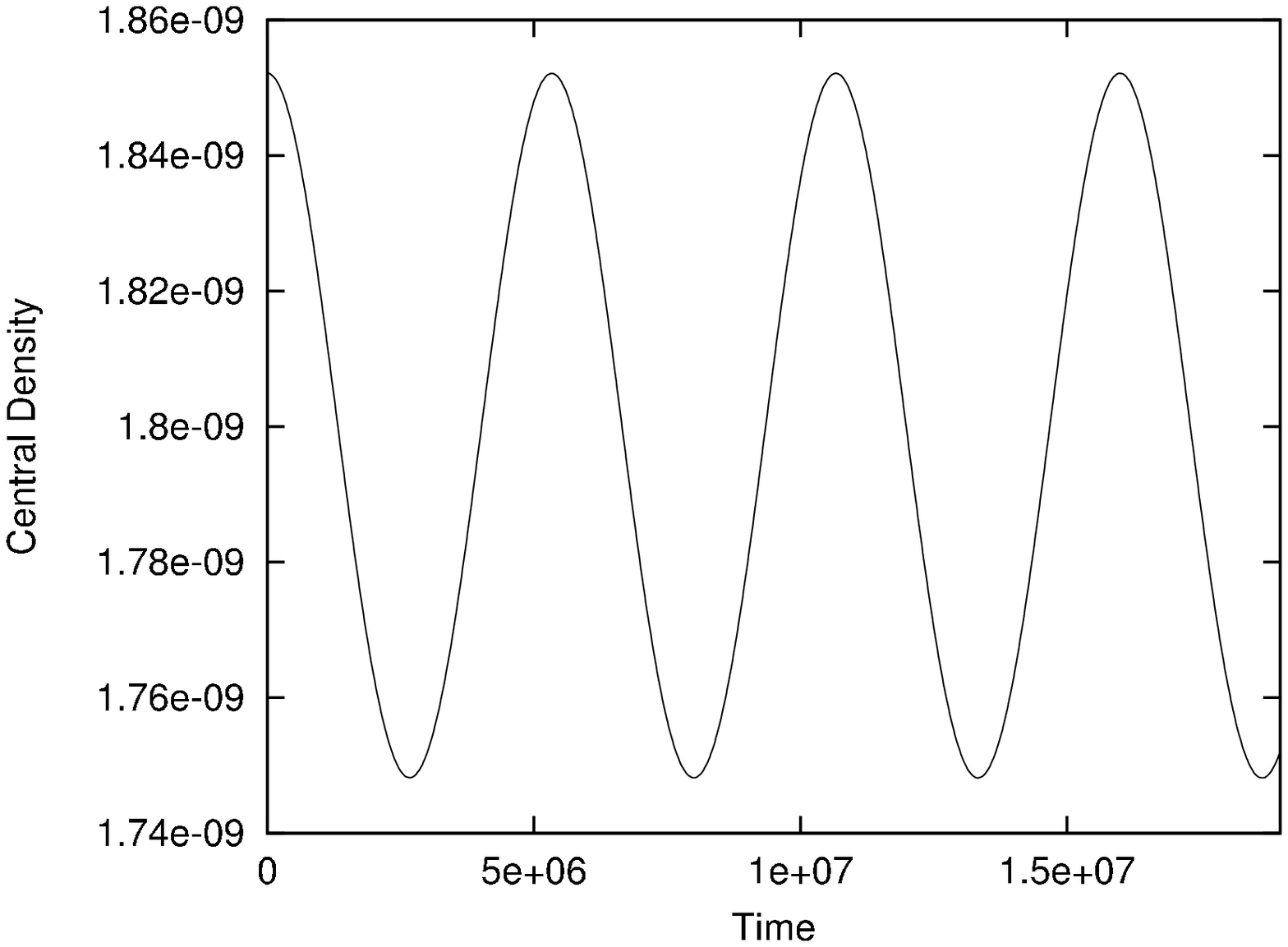} &
(b)\includegraphics[scale=0.45]{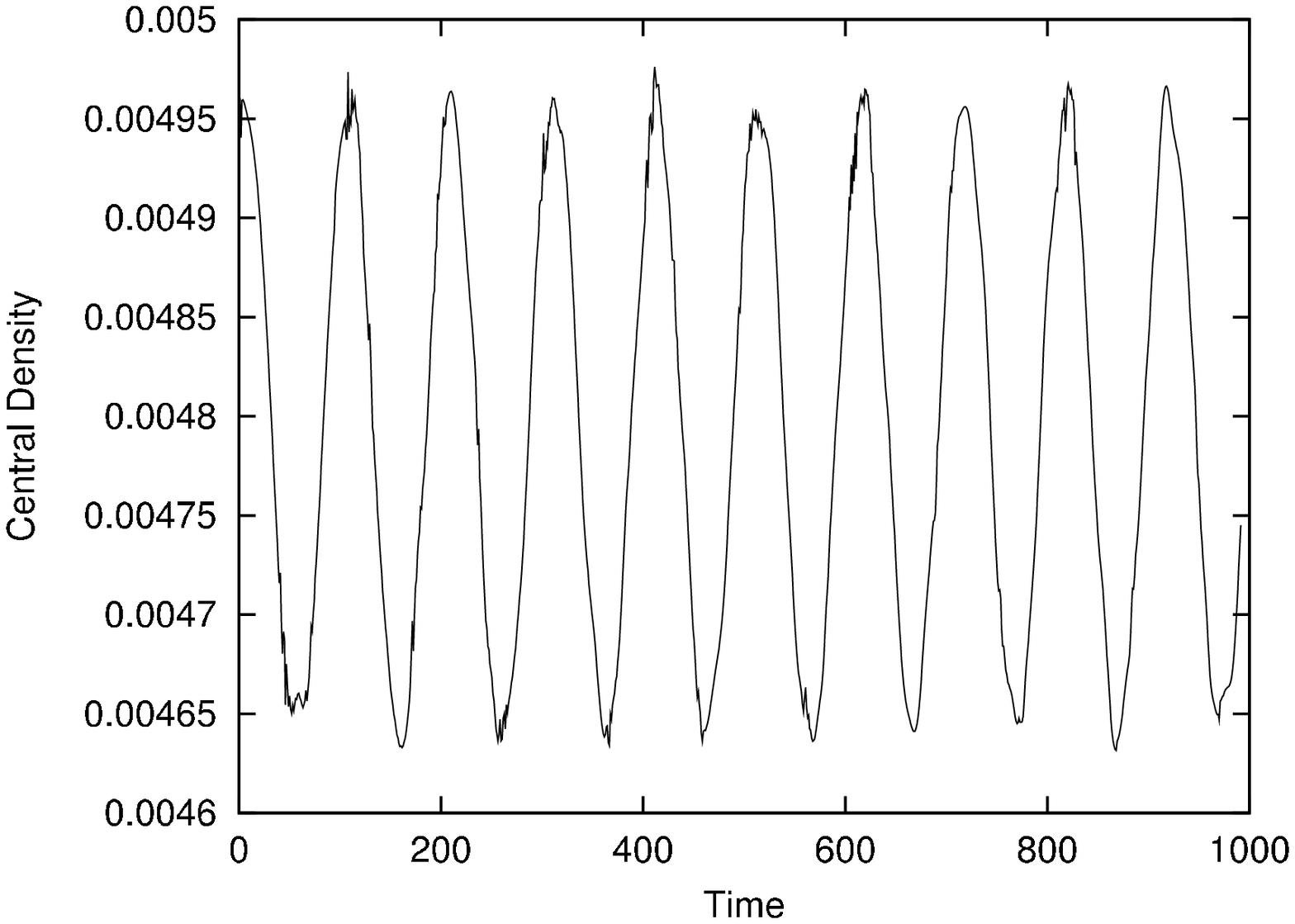} \\
\end{tabular}
\end{center}
\caption{\label{fg:oscillation} 
Evolutions of the central density 
(a) for model A with $\Gamma_{\rm a}=(4/3)+0.00142$ 
and (b) for model C with $\Gamma_{\rm a}=2$.
The horizontal axis is the time $\bar{u}$ 
for an observer at infinity. 
The internal energy $e$ is initially increased 
uniformly by 1\% both for (a) 
and for (b) from the equilibrium configuration.
All the quantities are shown in units of $M=1$.}  
\end{figure}

For long-term test simulations, we adopt models A and C. 
To induce a small oscillation, we 
initially increased the specific internal energy $e$ uniformly 
by 1\% both for model A and for model C.
In Fig.~\ref{fg:oscillation} we show the time evolution of 
the central density. 
It is found that the stars oscillate in a periodic manner. 
By performing the Fourier analysis, 
we measured the period of this stellar oscillation, and
found that the oscillation angular frequency is 
\begin{equation}
\omega \approx \bar \omega {M^{1/2} \over R^{3/2}},
\end{equation}
where $\bar \omega=9.59 \times 10^{-2}$ 
for model A and $0.788$ for model C, where
the initial stellar radii are given by $1.900\times 10^{3}M$ and 
$5.501M$ for models A and C, respectively (see Table I).

According to a post-Newtonian theory, 
the angular frequency of the radial oscillation 
for supermassive stars can be estimated as~\cite{st1983}
\begin{equation}
\omega^{2}=\frac{3|W|}{I}\left(\Gamma_{\rm a}-\Gamma_{\rm crit}\right),
\end{equation}
where $W$, $I$, and $\Gamma_{\rm crit}$ 
are the gravitational energy, moment of
inertia, and critical polytropic index, respectively.
Here, $\Gamma_{\rm crit}$ is estimated in the 
post-Newtonian approximation as~\cite{Chandra}
\begin{equation}
\Gamma_{\rm crit}=\frac{4}{3}+1.125\biggl(\frac{2M}{R}\biggr).
\end{equation}
From this analysis, $\bar \omega$ should be $9.62\times 10^{-2}$ 
for model A. 
Thus, the numerical result agrees with the analytic one 
within 0.5~\% error.

The angular frequency of the radial oscillation for neutron stars can 
be also calculated using the semianalytic formula derived 
by Chandrasekhar~\cite{Chandra}.
From this approximate formula, we can predict the 
oscillation frequency should
be $\approx 0.9 \rho_c^{1/2}$~\cite{gr3d} and 
hence $\omega \approx 0.063 M^{-1}$. 
On the other hand, the numerical result indicates that 
$\omega \approx 0.0616 M^{-1}$, 
which is again in good agreement with 
the analytic result within a few percent error. 

Models B and D are marginally stable against 
gravitational collapse. Thus, if the internal energy is 
reduced, they start collapsing. 
We have checked that they indeed collapse to black holes. 
The detailed results of the gravitational collapse are described in 
Sec.~\ref{sec:demonstration}.

\subsection{Perturbation code}

\begin{figure}[htbp]
\begin{center}
\begin{tabular}{cc}
(a)\includegraphics[scale=0.45]{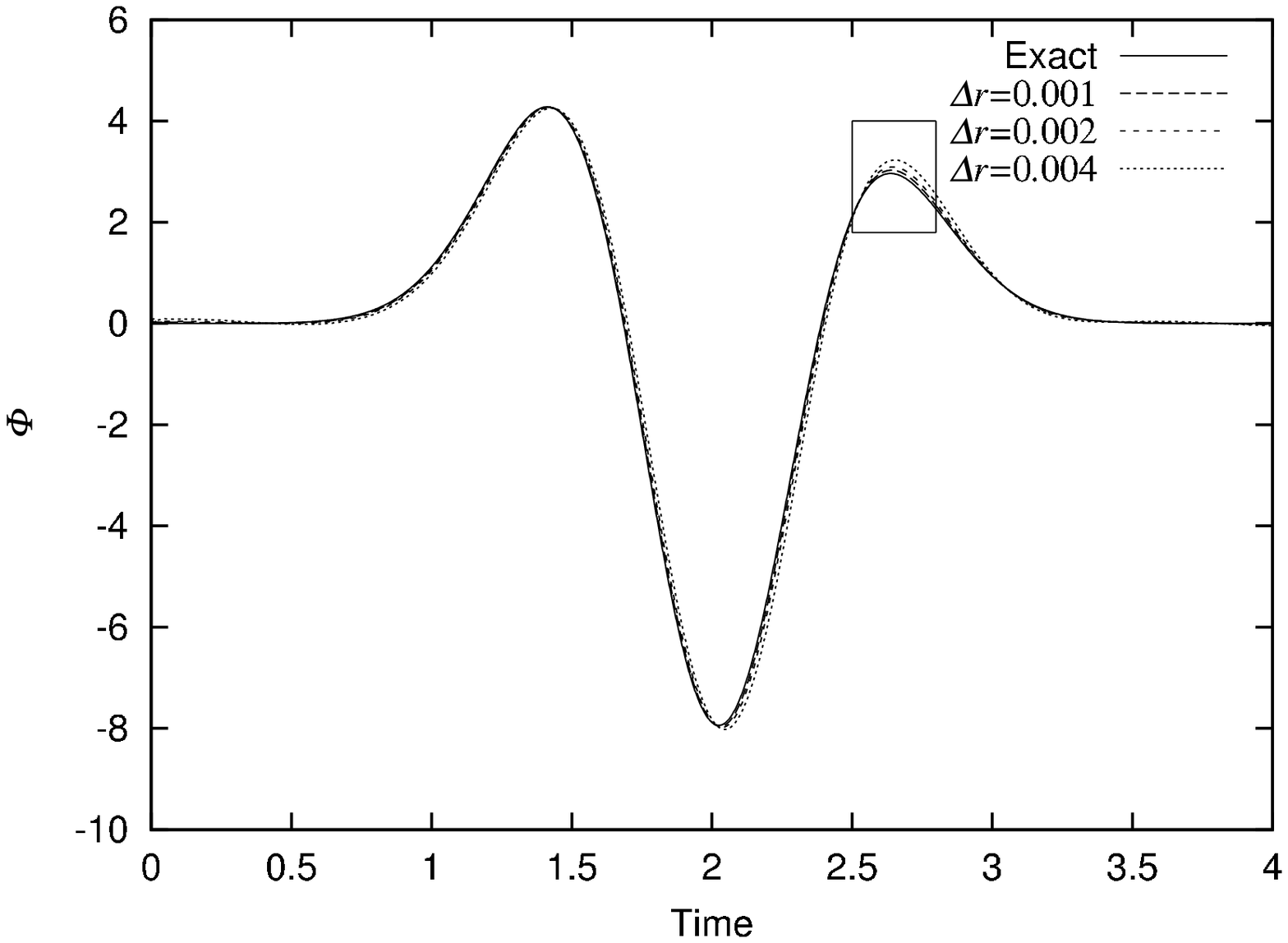} &
(b)\includegraphics[scale=0.45]{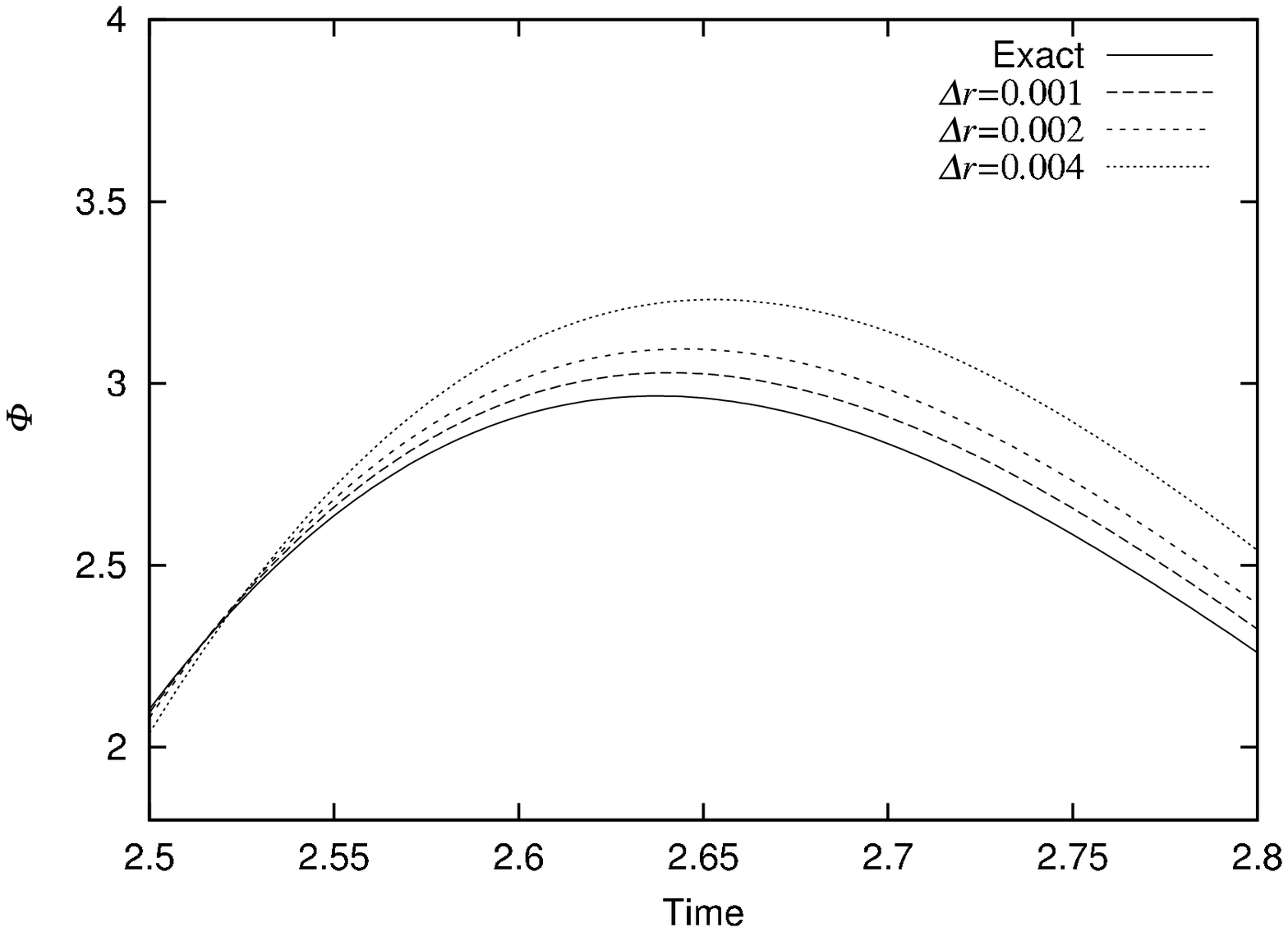} \\
\end{tabular}
\end{center}
\caption{\label{fg:flat} 
Linear gravitational waveforms for $l=2$ in the Minkowski spacetime 
extracted at $r=5$.
The initial data set is provided on the initial null surface $t=r$.
Three levels of the grid resolution 
as $\Delta r=\Delta u=\Delta v=0.001$, 0.002, and 0.004 are adopted. 
The solid line denotes the exact solution.
(b) is the magnification of the region encompassed by the square in (a).
}
\end{figure}

We have carried out a wide variety of test simulations 
for our perturbation code. First, we checked that 
gravitational waves accurately propagate on the flat 
Minkowski background. For $l=2$, the exact solutions for 
linear gravitational waves in the Minkowski spacetime 
are obtained as~\cite{Teukolsky}
\begin{equation}
\bar{\Pi}=3\frac{I(t-r)-I(t+r)}{r^{5}}+3\frac{I'(t-r)+I'(t+r)}{r^{4}}
+\frac{I''(t-r)-I''(t+r)}{r^{3}},
\end{equation}
where $I$ is an arbitrary function, and $I'$ its derivative. 
Here, we choose 
\begin{equation}
I(y)=\exp[-4(y-2)^2]. 
\end{equation}
Since the background spacetime is flat,
the matching surface between regions I and II is artificial. 
As the matching surface, we chose the surface of $r=1$ 
in this test. In Fig.~\ref{fg:flat}, we show numerical results 
with the exact solution for the waveforms observed at $r=5$. 
To demonstrate the convergence of the numerical solutions,
we performed simulations with three levels of the grid resolution 
as $\Delta r=\Delta u=\Delta v=0.001$, 0.002, and 0.004. 
Figure~\ref{fg:flat} shows that numerical results converge at first order 
to the exact solution. We note that 
the number of grid points per wavelength is $\sim$ 
1000 for the best-resolved case. 

Next, we computed the propagation of gravitational waves for $l=2$ 
in the collapse of a homogeneous dust ball. 
The same type of computation was already carried
out~\cite{cpm1978,sm1987}, 
so that we can calibrate our code by comparing the present results with
previous ones. 

In~\cite{cpm1978},
the momentarily static perturbations 
are provided at the initial hypersurface. 
Thus, we also choose the momentarily static initial data 
on the null hypersurface (i.e., the dust surface is assumed to 
reach the maximum expansion at initial slice).
However, the word of caution is appropriate here. 
The previous computation was carried out using 1+1 coordinate 
system (not single-null coordinate system). On the other hand, 
we choose the single-null coordinate system. 
Namely, the coordinate system and also
the time slicing are different between the two. 
Thus, even if we set the same function of $\bar{\beta}$ 
in different coordinate systems, this does not imply that 
we give the identical initial condition. 
Moreover, the former formulation has ambiguity in 
determining the second-order 
time derivative of gravitational perturbation,
while the latter one does not. 
Hence, the momentarily static initial data sets for these two
formulations are different from each other.
It implies that it is difficult to precisely compare 
the results obtained in these two formulations. 

To calibrate the effect of different initial conditions on the results, 
we provided two perturbation profiles fixing $q$. 
In the first one, we gave $\bar{\beta}_{\rm init}(R)=\mbox{const}$ 
on the initial hypersurface, and in the second one, 
$\bar{\beta}_{\rm init}(R)\propto \exp[-(R/R_c)^{2}]$, 
where $R_{c}$ is chosen to be one-third of 
the initial surface radius.

\begin{figure}[htbp]
\includegraphics[scale=0.6]{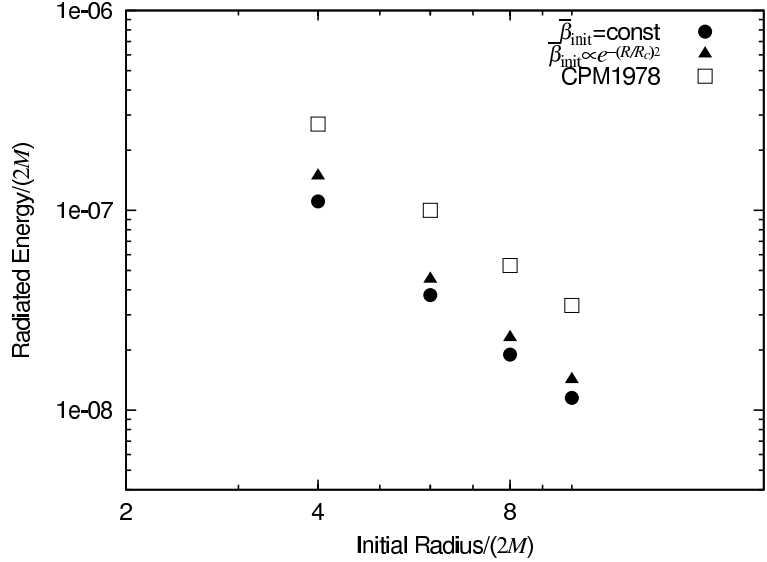}
\caption{\label{fg:radius_energy_dust_collapse} 
Total radiated energy of gravitational waves 
for $l=2$ from homogeneous dust collapse for different 
initial (maximally expanded) radii. The initial moment $q$ is 
normalized so that $q=2M$.
We set the momentarily static initial data 
on the initial null slice.
The results for two kinds of initial distribution for
matter perturbation $\bar{\beta}_{\rm init}$ are plotted.
For the case of $\bar{\beta}_{\rm init}\propto \exp[-(R/R_{c})^{2}]$, 
the radius $R_{c}$
is chosen to be one-third of the initial radius of the dust ball.
The present results are compared with those of Cunningham,
Price, and Moncrief~\cite{cpm1978} (CPM1978), in which 
the initial data sets were set on the 
spacelike surface of maximum expansion.}
\end{figure}

In Fig.~\ref{fg:radius_energy_dust_collapse} 
the total radiated energy of gravitational waves are 
summarized. In these numerical computations, 
gravitational waves are extracted at $R=40M$. 
For comparison, 
we also plot the results of~\cite{cpm1978}. We note that 
in this figure, $q$ is normalized so that $q=2M$. 
It is found that numerical results of two different profiles of 
$\beta$ differ by a factor of $\sim 1.4$, and 
that the results of~\cite{cpm1978} are greater than our results 
by a factor of $\sim 3$. 
However, the total radiated energy systematically decreases as 
the initial dust radius increases 
for all the cases in the same manner. 
Thus, we conclude that our numerical results
agree with the previous ones besides a possible systematic error 
associated with the difference of the coordinate conditions. 

\begin{figure}[htbp]
\begin{center}
\begin{tabular}{cc}
(a)\includegraphics[scale=0.45]{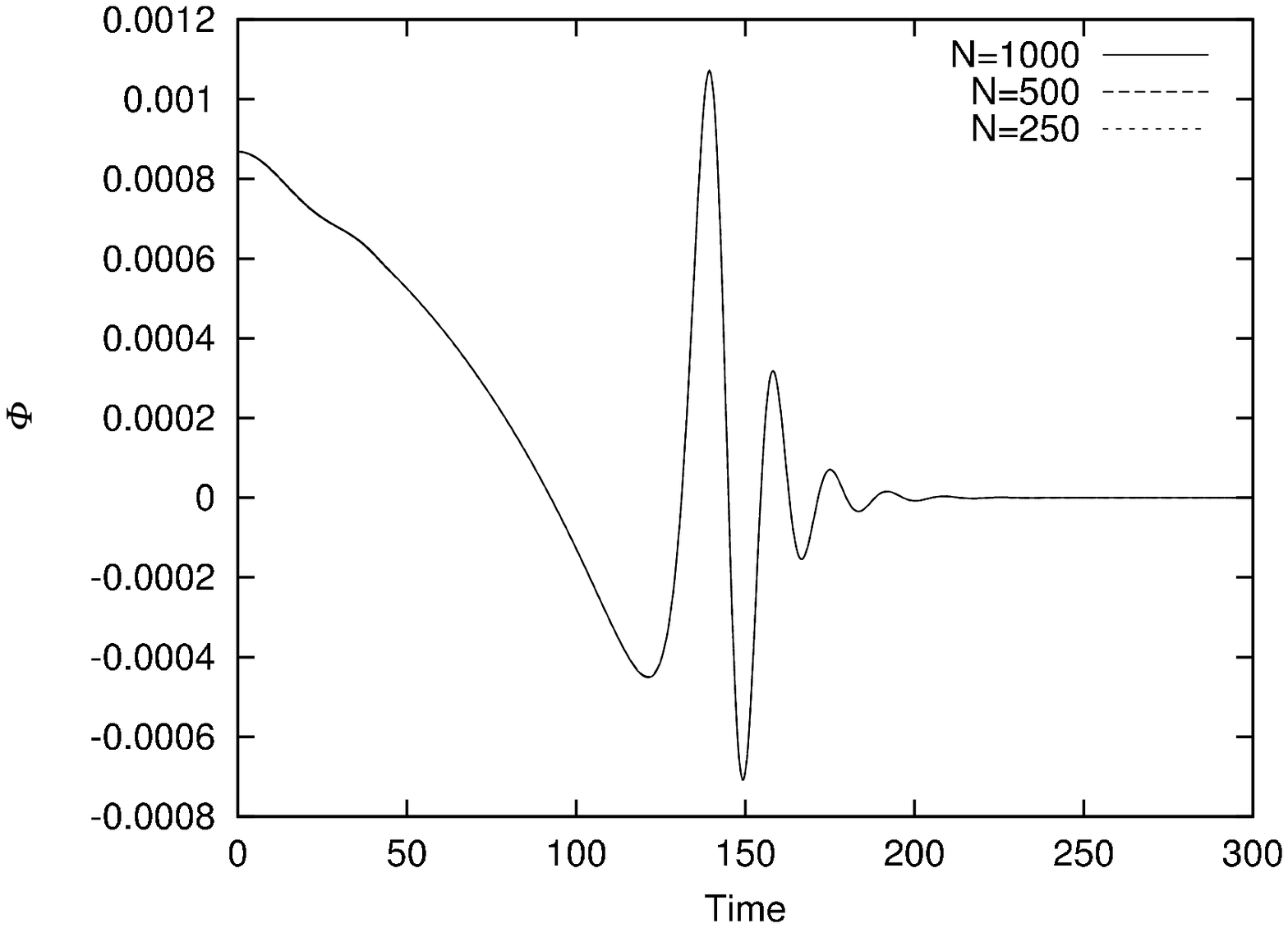} &
(b)\includegraphics[scale=0.45]{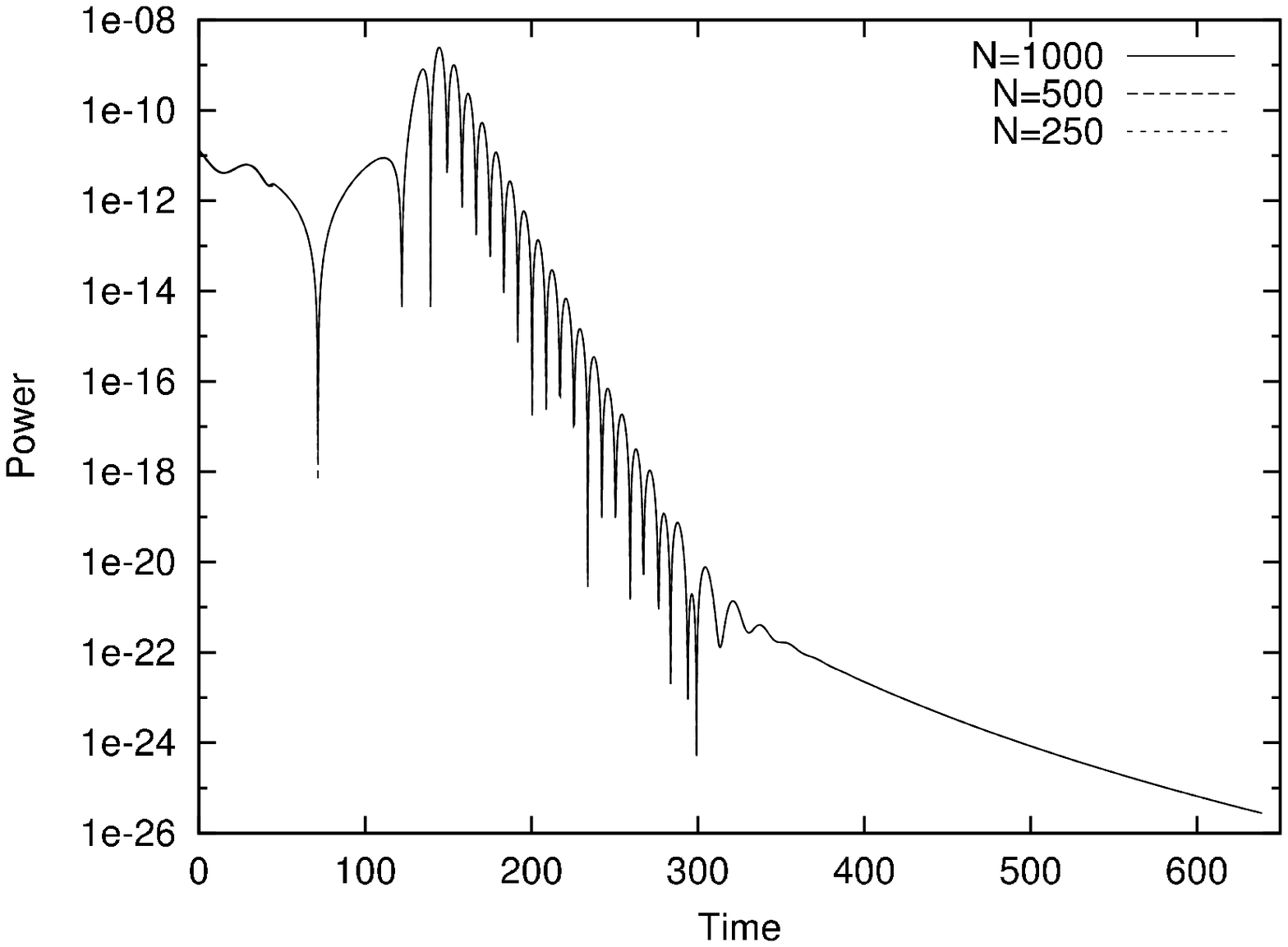} \\
(c)\includegraphics[scale=0.45]{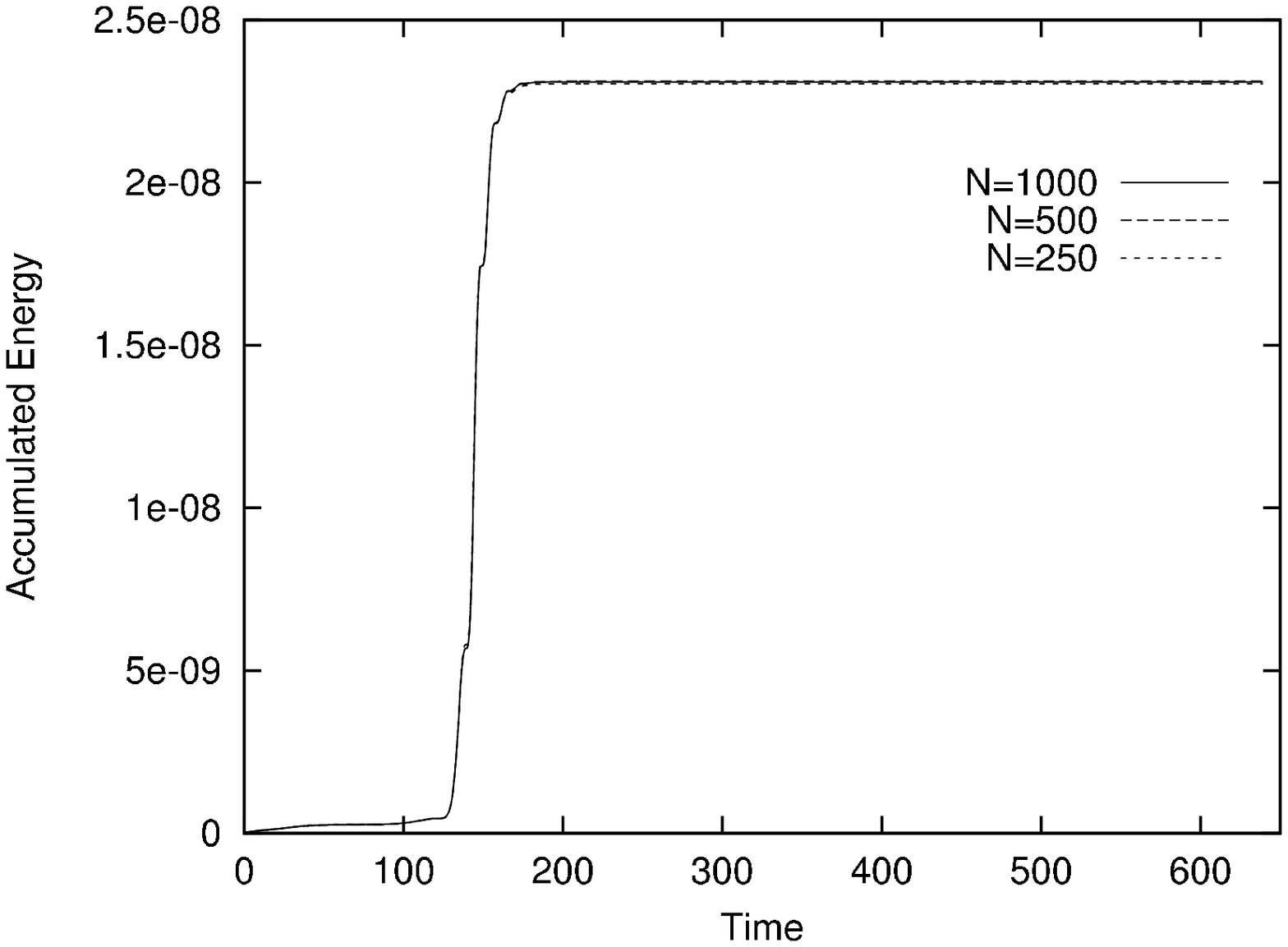} & \\
\end{tabular}
\end{center}
\caption{\label{fg:gw_L2R20_dust} 
(a) Waveform, (b) luminosity, and (c) accumulated energy
of gravitational waves for $l=2$ radiated from 
homogeneous dust collapse with the initial radius $R=20M$. 
These are estimated at $R=40M$ and
normalized so that $q=2M$.
The horizontal axis is the observer time $\bar{u}$. 
We set the momentarily static initial data 
on the initial null surface with the matter perturbation
$\bar{\beta}_{\rm init}=\mbox{const}$. 
The solid, long-dashed, and dashed lines denote
the results for $N=1000$, $500$, 
and $250$, respectively,
where $N$ is the number of spatial grid points 
inside the dust ball.
The matching is done at the outermost mass shell.
All the quantities are 
shown in unit of $M=1$.
The long-dashed and dashed lines are almost indistinguishable
because they lie on top of the solid line.}
\end{figure}

The waveform, luminosity, and integrated total energy of 
gravitational waves from a collapsing 
dust ball with the initial radius $R=20M$ for $l=2$ are 
plotted in Fig.~\ref{fg:gw_L2R20_dust}. 
Gravitational waves are extracted at $R=40M$. 
As found in~\cite{cpm1978}, the quasinormal ringing 
oscillation and subsequent 
power-law tail characterize the gravitational waveforms. 
The complex frequency of the fundamental quasinormal mode
was calculated to be
$2M\omega =0.74734+0.17792i$ by solving the eigenvalue 
equations~\cite{cd1975}.
On the other hand, our numerical results show that the 
complex frequency of the damped oscillation is given by 
$2M\omega =0.752+0.179i$. 
Thus, the numerical results agree with 
the theoretical value within 1~\% error 
both for the frequency and for the damping rate. 
It is found that the waveform is characterized 
by the power-law tail for $\bar{u}\agt 350M$.
The power-law index numerically computed is 
$\approx 7.0$ and also agrees with that analytically derived 
in~\cite{price1972} as $-(2l+3)=-7$. 

We also computed gravitational waves from a static star.
As the stellar models, we adopted models A and C.
Since the odd-parity matter perturbation is time independent 
in this case, gravitational waves propagate freely.
This implies that with the momentarily static initial perturbation,
no gravitational radiation should be emitted. We checked that 
this is indeed the case except for the tiny amount 
emitted soon after the simulation started, which is 
due to numerical errors 
associated with the finite differencing and relaxation of the
numerical system.

\section{Stellar collapse to black holes and gravitational waves} 
\label{sec:demonstration}

We have computed stellar collapse of 
models B and D to black holes. Since they are 
marginally stable against gravitational collapse, 
we extracted the internal energy $e$ by 1~\% initially 
to induce the collapse.
Here the energy extraction is done on the initial null cone. 
The difference between the energy extractions on the 
initial null cone and on the spacelike hypersurface is very small for 
nonrelativistic stars but may be significant
for highly relativistic stars.
During the collapse, we adopt the $\Gamma$-law equation of state as 
\begin{equation}
p=(\Gamma_{\rm a}-1)n e.
\end{equation}
The simulations were carried out using both the May-White 
and the Hernandez-Misner schemes. 
We note that with the latter scheme, 
we can follow the evolution only outside an event horizon. 
This implies that we cannot find apparent horizon and event 
horizon in the following simulation. 
We stop the calculation of the stellar collapse 
and estimate the value of the perturbation field 
on the event horizon using the extrapolation 
when $R_{\rm s}/2M=1.01$ is satisfied
for the surface radius $R_{\rm s}$.
We have confirmed that the obtained waveform is not so sensitive
to the choice of the criterion.  
In the simulation of the 
collapse of models B and D, 
we have observed no evident shock wave until 
the black hole forms.   

\subsection{Collapse of a supermassive star}

Supermassive stars are quasistable objects of mass 
$5\times 10^{5}M_{\odot}\alt M \alt 10^{10}M_{\odot}$
and possible direct progenitors of supermassive black holes~\cite{ZN,st1983}. 
They quasistationarily contract due to radiative cooling 
to the onset of radial instability~\cite{ZN,st1983}, resulting in 
formation of supermassive black holes. 
The quasistatic evolution of rotating supermassive stars 
was recently investigated by Baumgarte and Shapiro~\cite{bs1999}. 
Taking into account that 
supermassive stars are likely to be rigidly rotating and 
that the adiabatic index is $\approx 4/3$, they clarified 
that the ratio of the rotational energy to the 
gravitational energy is at most $\sim 0.009$ at
the onset of the gravitational collapse. They also found that 
oblateness of supermassive stars around the central region 
is very small (see Fig. 1 in~\cite{bs1999}), justifying that 
the Roche model~\cite{st1983} 
for rotating supermassive stars is adequate. 
Shibata and Shapiro~\cite{ss2002} numerically computed 
the collapse of a marginally stable rotating supermassive star
to a Kerr black hole 
in the two-dimensional fully general relativistic simulation. 
They indicated that more than 90\% of 
the stellar mass collapses directly into the black hole
in the dynamical time scale as in the spherical collapse. 
These results suggest that if we pay attention only to the 
inner region, the collapse proceeds in a nearly spherical manner. 
Motivated by this fact, we apply the present perturbation 
analysis to compute gravitational waves emitted during the collapse 
of supermassive stars. 

\subsubsection{Spherical collapse}

We adopted model B with $\Gamma_{\rm a}=(4/3)+0.00142$ 
to model a supermassive star of mass $10^6M_{\odot}$. 
In the numerical simulation, 
we typically take 1000 grid points to cover the supermassive star. 
For the collapse of model B, with the May-White scheme, 
apparent horizon was located near the center 
in a late time of collapse, but 
soon after the formation, numerical accuracy deteriorates and 
as a result computation crashed before the event horizon 
swallowed all the fluid elements. 
On the other hand, the computation can be continued until 
the null hypersurface reaches the event horizon 
with the Hernandez-Misner scheme; i.e.,
the whole region outside the event horizon is computed
numerically. 
This clearly indicates the robustness of the Hernandez-Misner scheme. 
The matching is done on the mass shell 
within which 99.7\% of the total mass is enclosed. 
In the following, we deal with this matching surface as the 
stellar surface.

\begin{figure}[htbp]
\begin{center}
\begin{tabular}{cc}
(a)\includegraphics[scale=0.45]{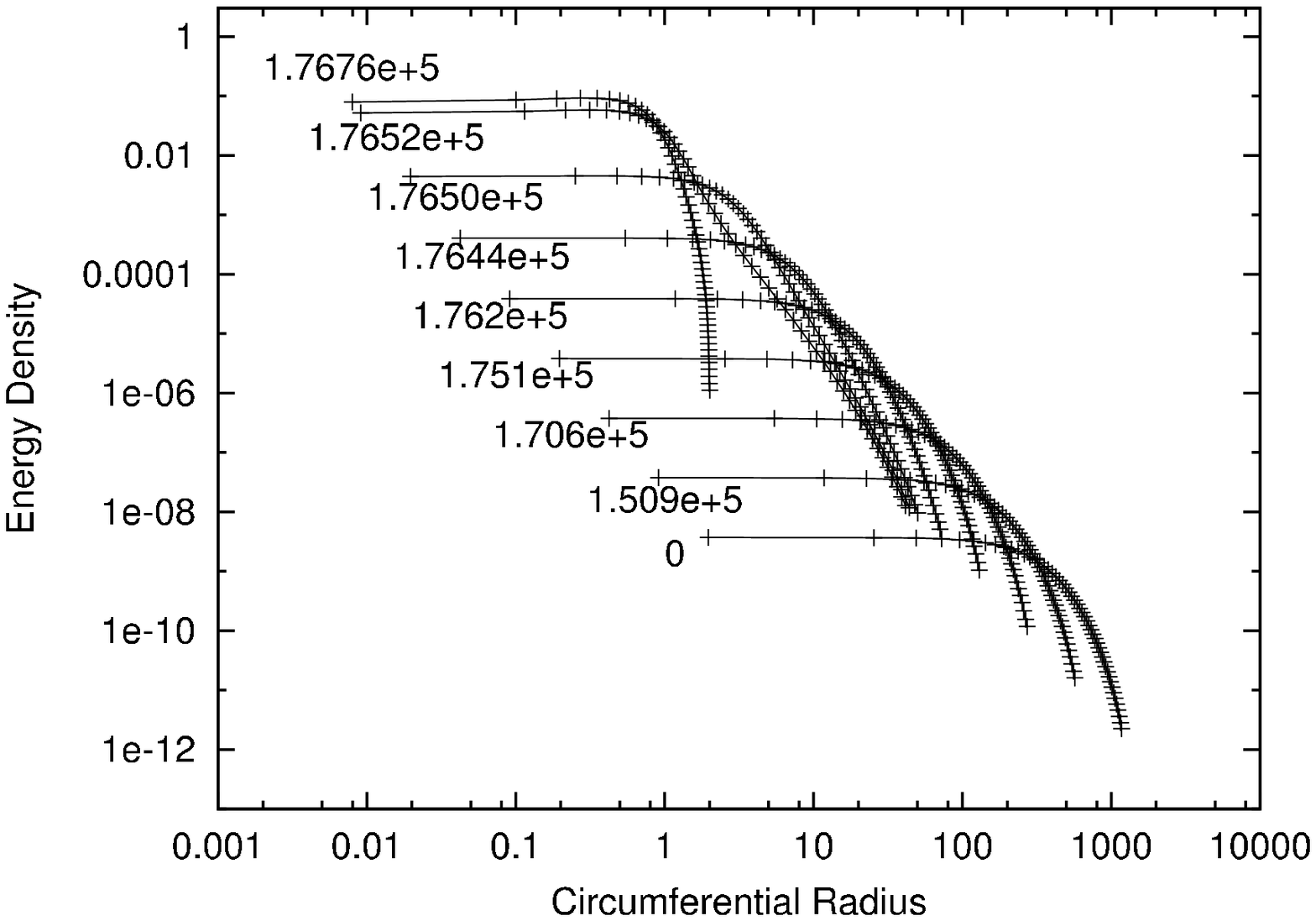} &
(b)\includegraphics[scale=0.45]{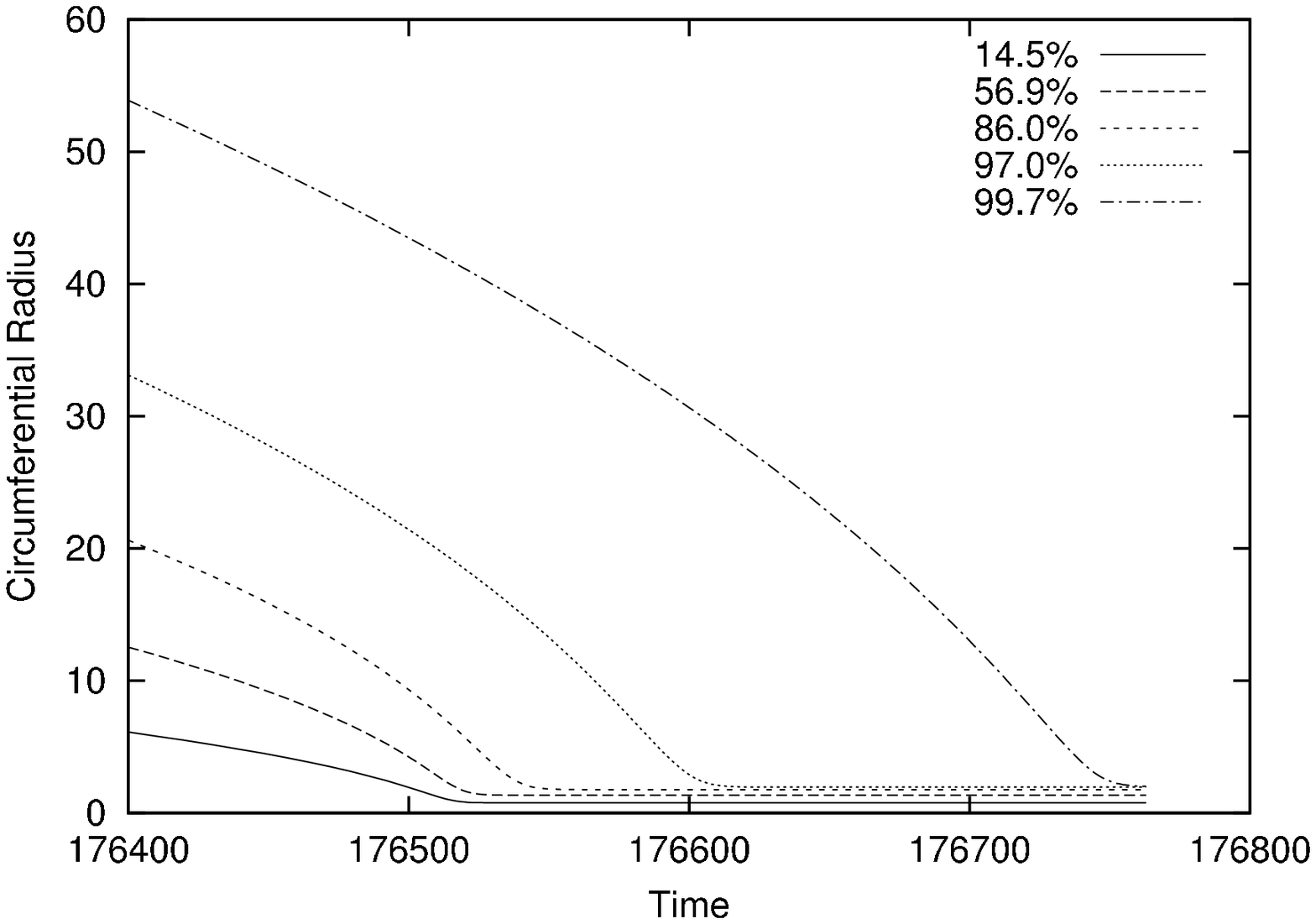} \\
(c)\includegraphics[scale=0.45]{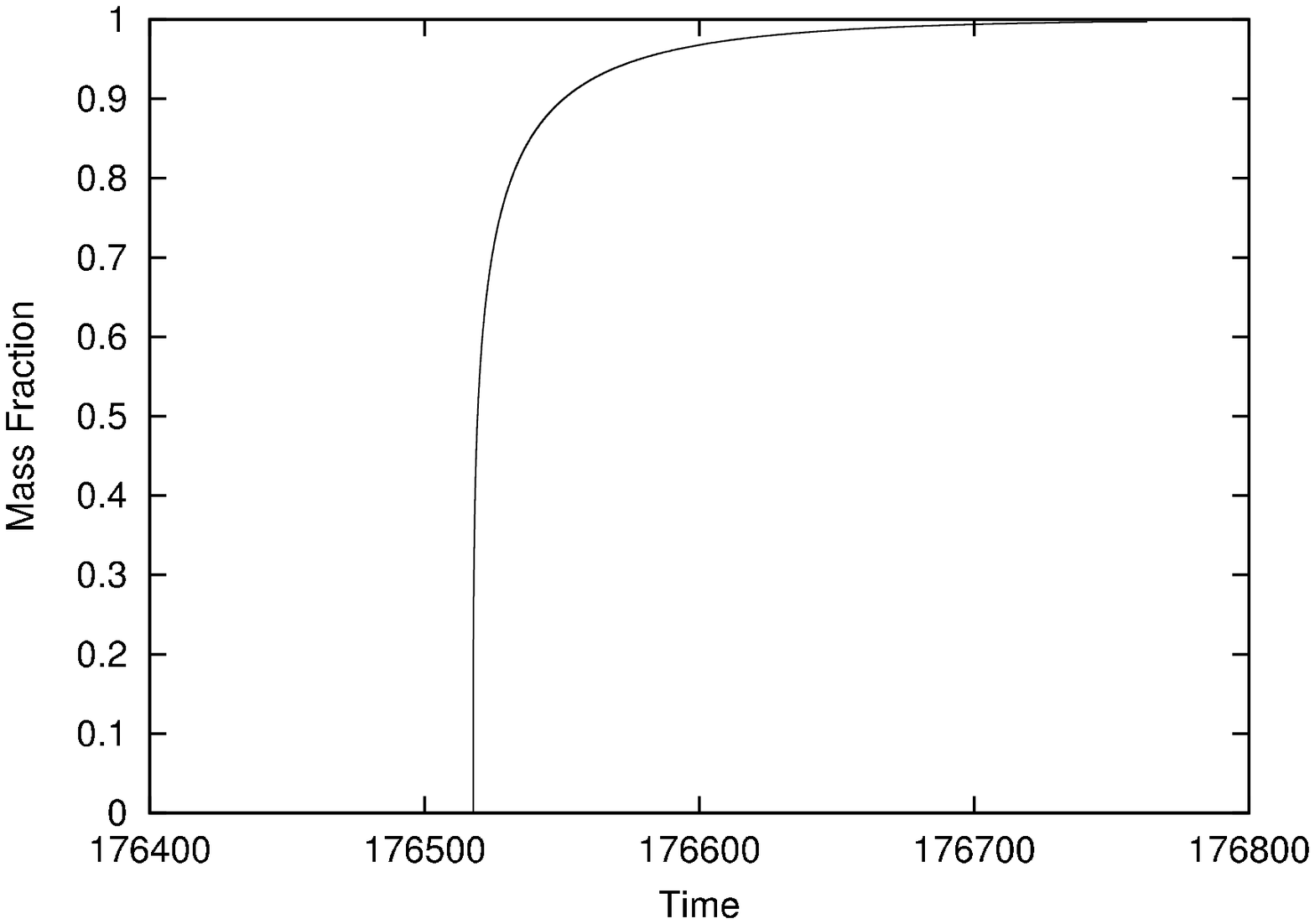} & \\
\end{tabular}
\end{center}
\caption{\label{fg:collapse_B} 
(a) Snapshots of the density profile, 
(b) the trajectories of mass shells, and 
(c) the mass fraction contained in a high-redshift region are plotted
for the gravitational collapse of model B 
with $\Gamma_{\rm a}=(4/3)+0.00142$. 
We extract 1~\% of the internal energy from the 
equilibrium configuration to induce the collapse.
In (a) the label for each curve denotes the observer time $\bar{u}$.
In (b) the vertical and horizontal axes denote the circumferential radius
and observer time, respectively. The labels denote 
the mass fractions enclosed by mass shells.
In (c) the high-redshift region is defined as the one 
in which the lapse function $\alpha$ is less than 0.1.
We deal with the mass shell which encloses 99.7\% of the total mass
as the stellar surface.
All the quantities are shown in units of $M=1$.}  
\end{figure}

In Fig.~\ref{fg:collapse_B}(a) we display the 
snapshots of the density profile at selected time steps. 
In this simulation, 
the spacetime settles down to a static one 
at $\bar{u} \simeq 176760M$. 
The calculation has been stopped
at $\bar{u} \simeq 176763M$ or $\tau_{\rm s}\simeq 175294M$.
At this moment, we have obtained the redshift 
$(1+z_{\rm s})\simeq 190.7$ at the surface.
Since the equation of state of supermassive stars is soft, 
the mass is highly concentrated around the center. 
In the late stage of the collapse, the increase of the 
central density is accelerated, and hence the 
collapse proceeds in a runaway manner. (This makes the
simulation without null formulation technically difficult.)
Figure~\ref{fg:collapse_B}(b) shows the trajectories of 
mass shells for $\bar{u}\agt 176400M$. 
Each mass shell asymptotically approaches
a constant value greater than $2m$ 
because the lapse function decreases to zero.
This figure shows that the central region collapses earlier, while
the outer envelope accretes slowly after the evolution of the
central region is almost frozen.
Figure~\ref{fg:collapse_B}(c) shows the total mass 
contained in the high-redshift region in which 
the lapse function is less than 0.1 [or equivalently  
$(1+z)$ is greater than 10 where $z$ denotes the gravitational redshift].
This region first appears at the center at $\bar{u}\simeq 176510 M$.
We find that 80\% of the stellar mass is 
swallowed into this high-redshift region
within the time interval $\Delta\bar{u}\sim 15M$. 
After the inner region collapses, 
surrounding atmosphere falls into this high-redshift 
region spending a much longer time $\simeq 200M$. 

\subsubsection{Gravitational radiation}

For computation of the nonspherical perturbations,
a static initial condition for $\bar{\beta}$ should be given. 
However, the realistic perturbation profile of the initial data set
is not clear for the odd-parity perturbation.
The purpose of this paper is to study the gravitational waveforms
during the formation of a black hole qualitatively. 
Thus, to investigate the dependence of
the gravitational waveforms on the initial perturbation profile, 
we gave three kinds of the initial data sets as 
(1) with $\bar{\beta}_{\rm init}=$const,
(2) with $\bar{\beta}_{\rm init}=\exp[-(R/R_{\rm c})^{2}]$, and 
(3) with $\bar{\beta}_{\rm init}=\exp\{-[(R-R_{\rm s})/R_{\rm c}]^{2}\}$,
where the scale length of the inhomogeneity, $R_{\rm c}$,
for the matter perturbation is chosen to be $R_{\rm c}=R_{\rm s}/3$. 
For (1), the perturbation is uniformly distributed.  
For (2), the amplitude of the perturbation in the  
inner region is larger than that in the outer region. For (3), 
the amplitude of the perturbation is the largest 
near the stellar surface. 

To check the convergence, computations for case (1)
were carried out with 
$N=1000$, $500$ and $250$, where $N$ is the number of
grid points in the stellar interior.
The wavelength of gravitational waves is roughly comparable with 
the stellar size for the early stage and with the stellar
size or the size of the formed black hole for the late
stage. Since we adopt the comoving coordinate, the number of 
grid points per wavelength is kept almost constant and of the order
1000 for 1000-zone calculations.

\begin{figure}[htbp]
\begin{center}
\begin{tabular}{cc}
(a)\includegraphics[scale=0.45]{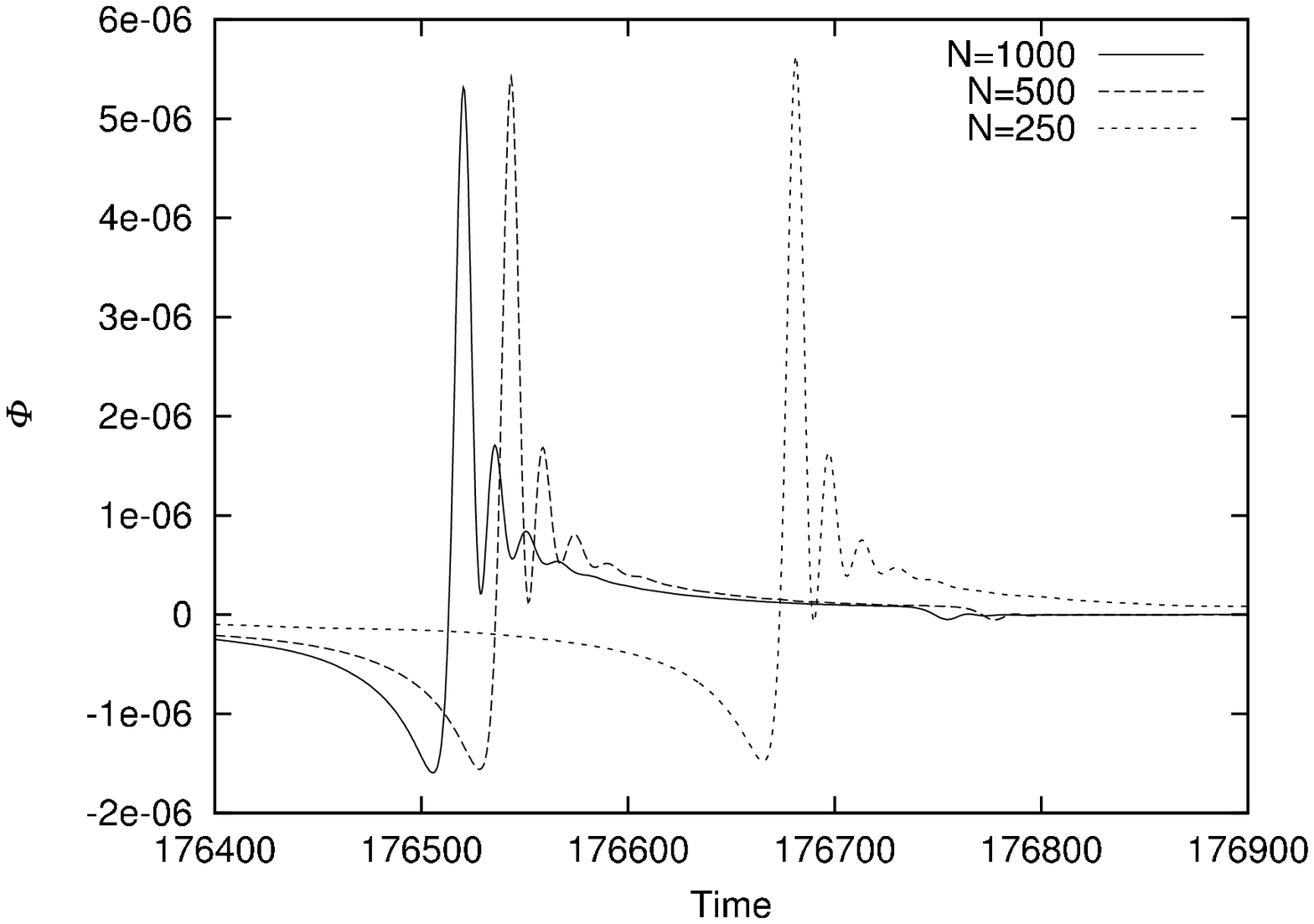} &
(b)\includegraphics[scale=0.45]{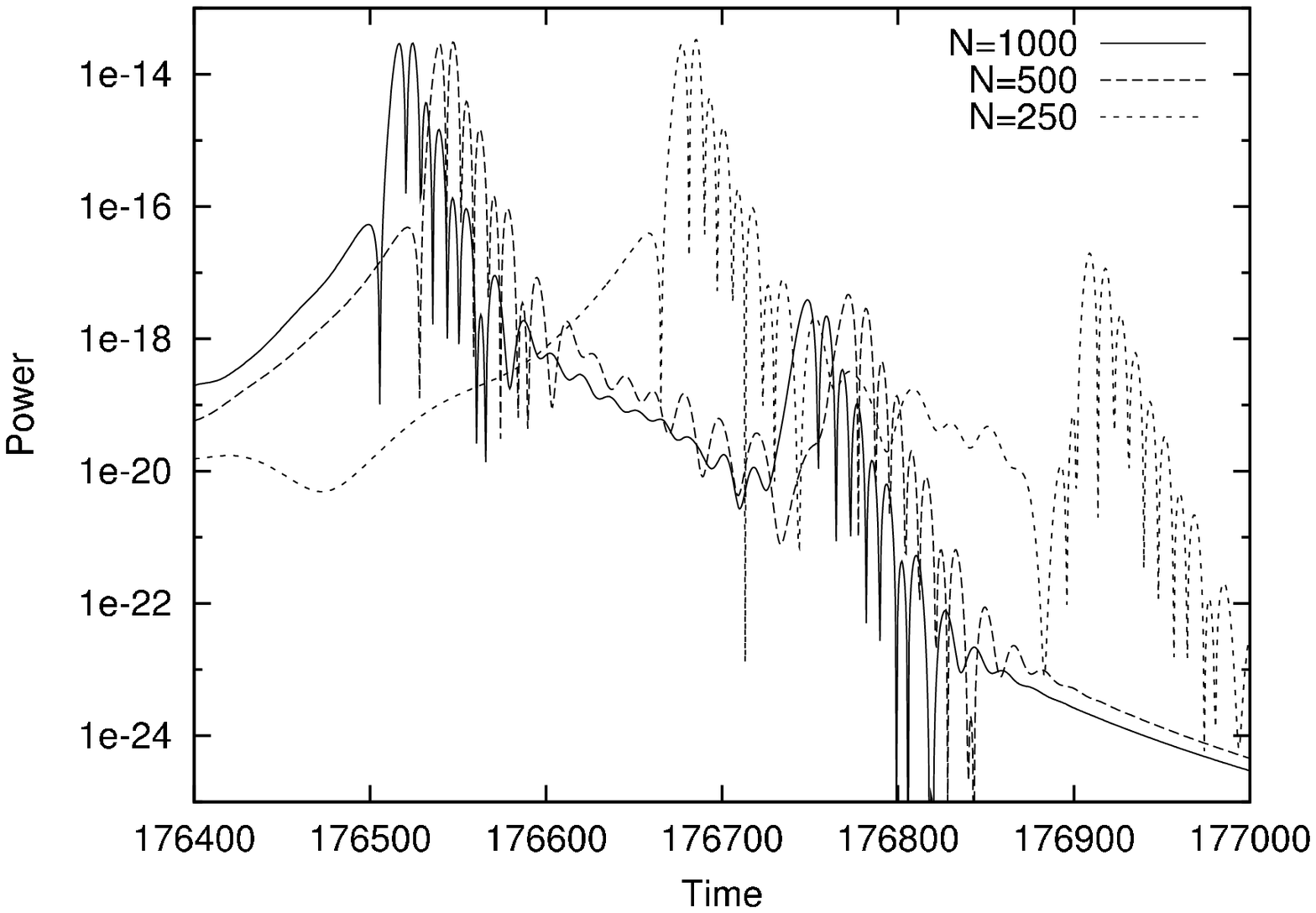} \\
(c)\includegraphics[scale=0.45]{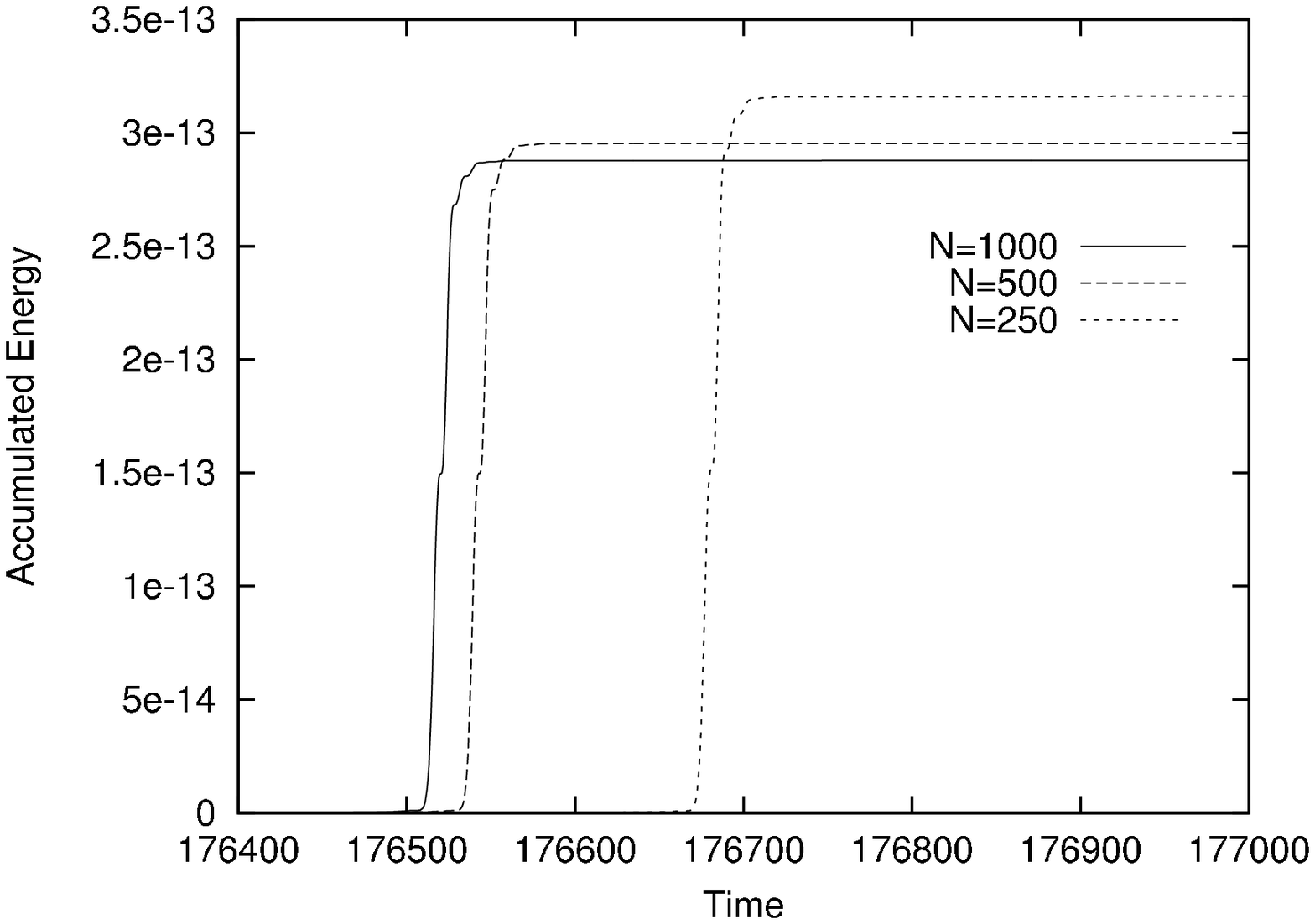} & \\
\end{tabular}
\end{center}
\caption{\label{fg:gw_L2B_n} 
(a) Waveform, (b) luminosity and (c)
accumulated energy of gravitational waves for $l=2$ radiated from 
the collapse of model B with $\Gamma_{\rm a}=(4/3)+0.00142$. 
Gravitational waves are extracted at $R=2000M$.
The initial matter perturbation is chosen so that 
$\bar{\beta}_{\rm init}=\mbox{const}$.
The momentarily static initial perturbation is given. 
The amplitude of the perturbation is normalized so that
$q=2M$. 
The solid, long-dashed, and dashed lines denote
the results for $N=1000$, $500$, and $250$, respectively,
where $N$ is the number of spatial grid points inside the star.
The matching is done on the mass shell which encloses 
99.7\% of the total mass. All the quantities are 
shown in units of $M=1$.}
\end{figure}

The waveform, luminosity, and integrated total energy 
of gravitational waves for $l=2$ are plotted in Fig.~\ref{fg:gw_L2B_n}. 
Since the oscillation period of gravitational waves 
is much shorter than the total time of integration, 
we show them only around the time of major emission. 
Figure~\ref{fg:gw_L2B_n} demonstrates that the convergence is 
achieved well.
For example, it is found that 
the totally radiated energy shows the first-order convergence.
Although the numerical error in the time of the major emission
might look very large, it is only due to the extremely long 
time for the collapse compared to the duration of the major 
emission. 
It is seen in Fig.~\ref{fg:gw_L2B_n}(a) 
that the waveform is made up of the precursory wave, 
quasinormal ringing, long-time-scale decay,
quasinormal ringing again, and power-law tail.
Figures~\ref{fg:gw_L2B_n}(b) and (c) indicate that 
the primary and secondary contributions to the totally 
radiated energy come from the first and second bursts of 
quasinormal ringing, respectively.
The reason of the existence of this second burst 
will be explained later. 

\begin{figure}[htbp]

\begin{center}
    \begin{tabular}{cc}
(a)   \includegraphics[scale=0.4]{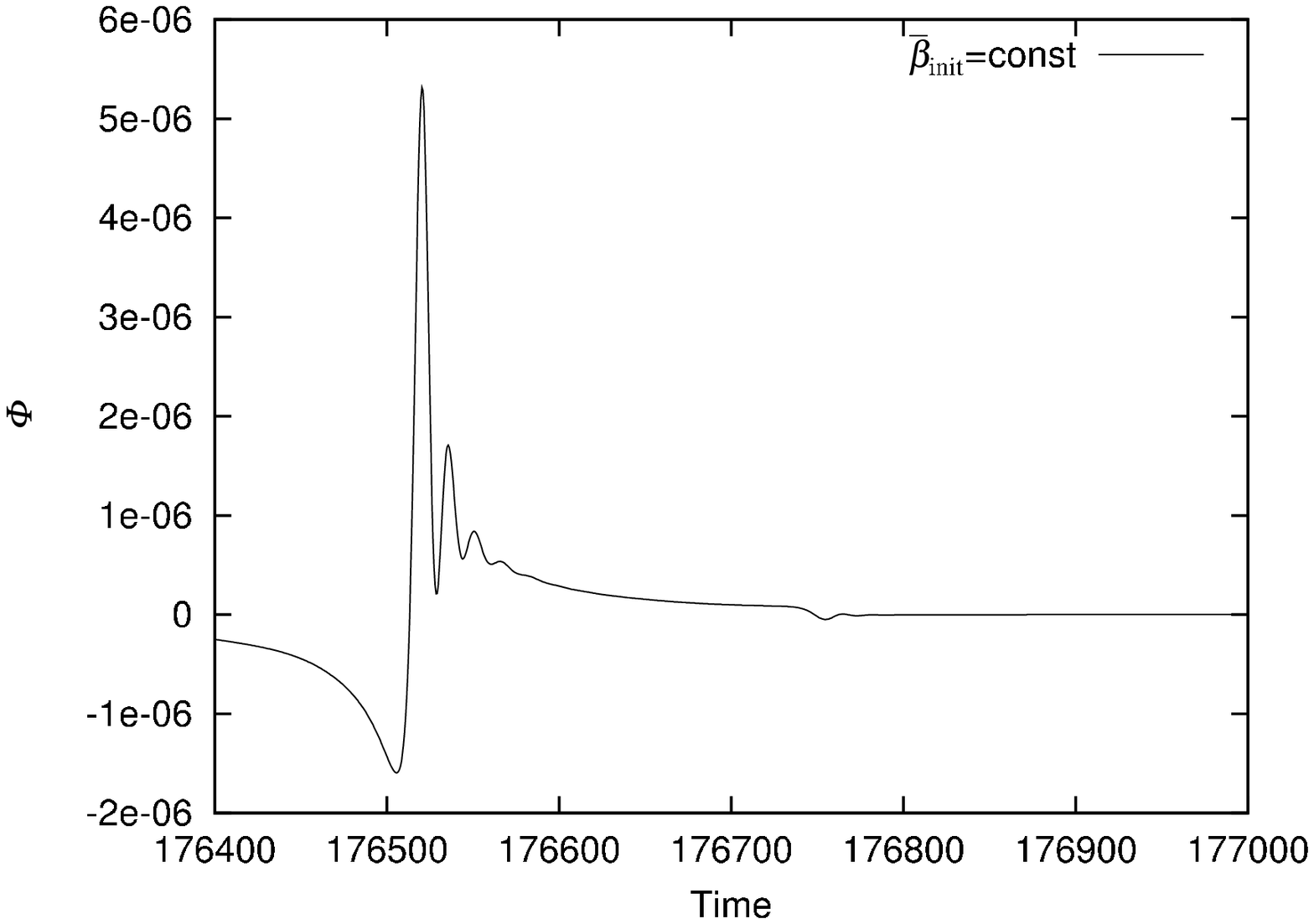} &
(b)   \includegraphics[scale=0.4]{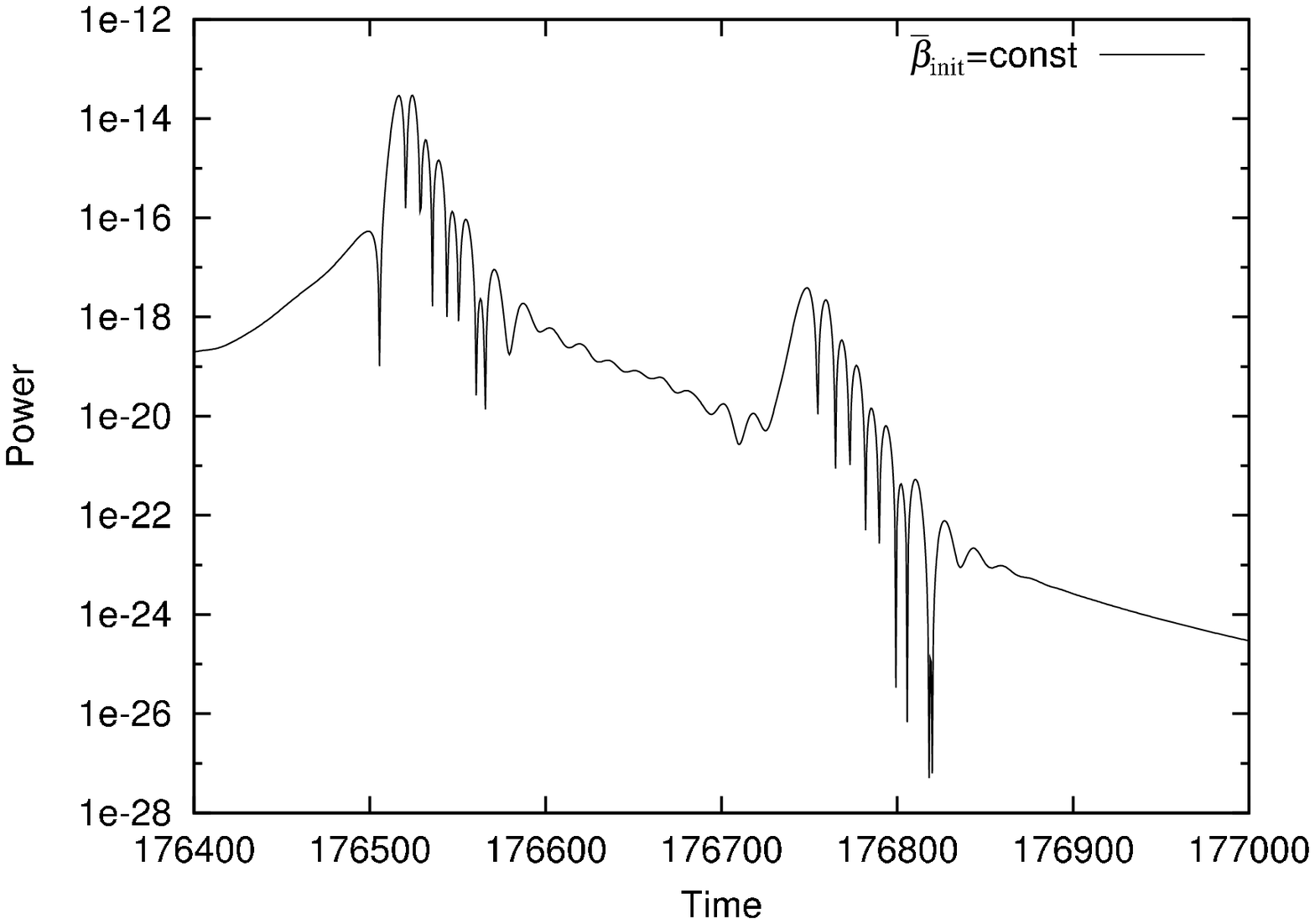} \\
(c)      \includegraphics[scale=0.4]{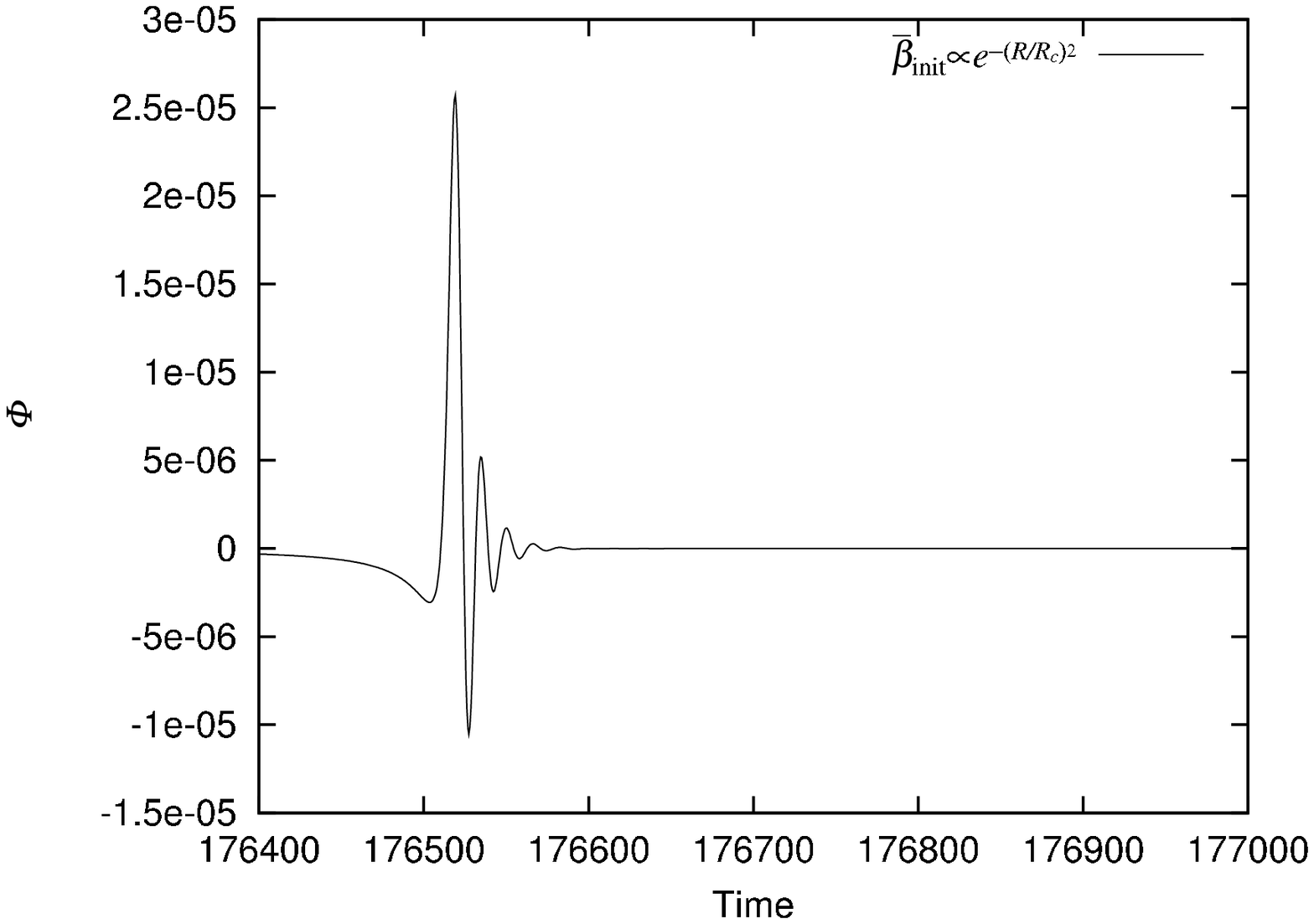} &
(d)      \includegraphics[scale=0.4]{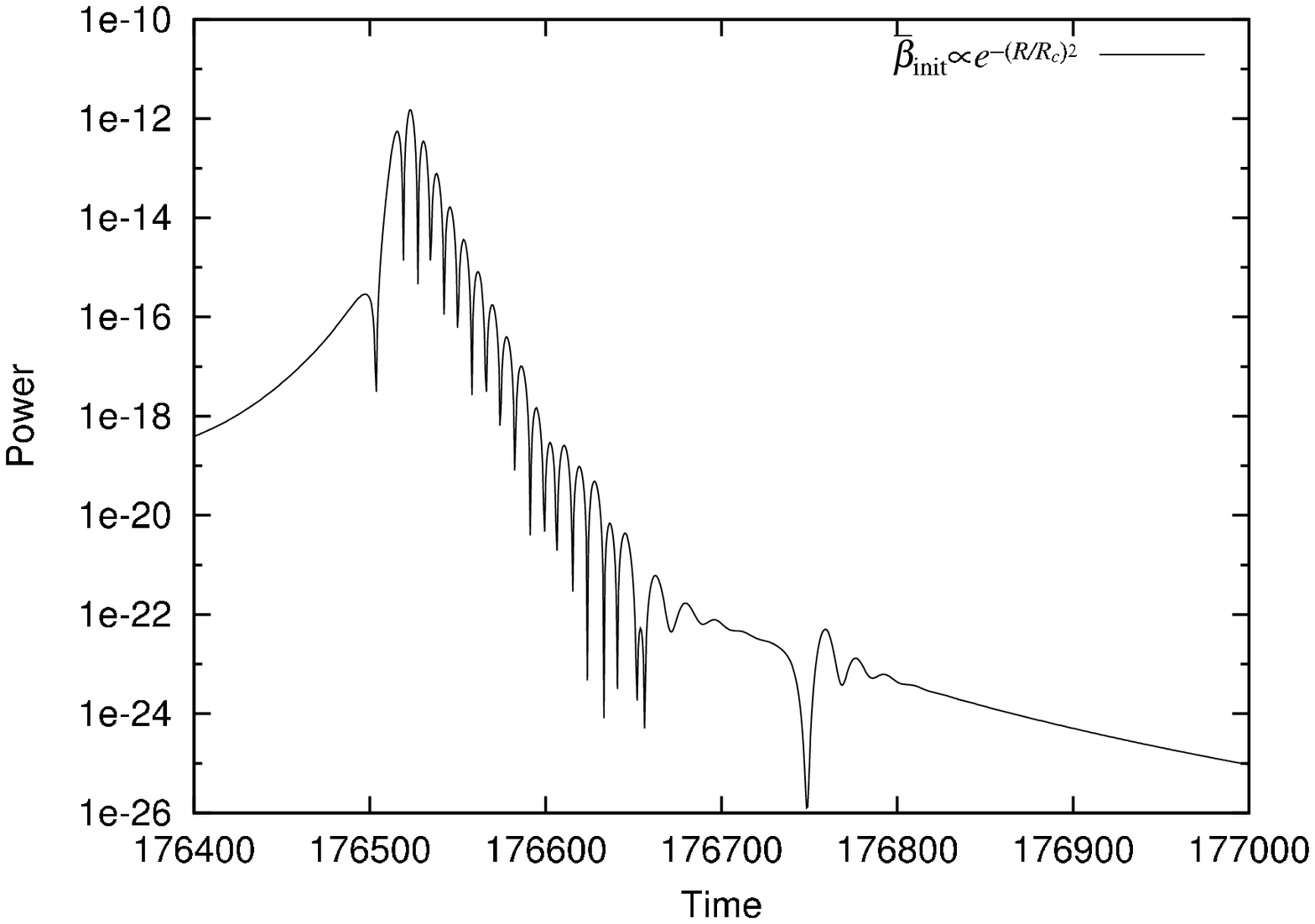} \\
(e)      \includegraphics[scale=0.4]{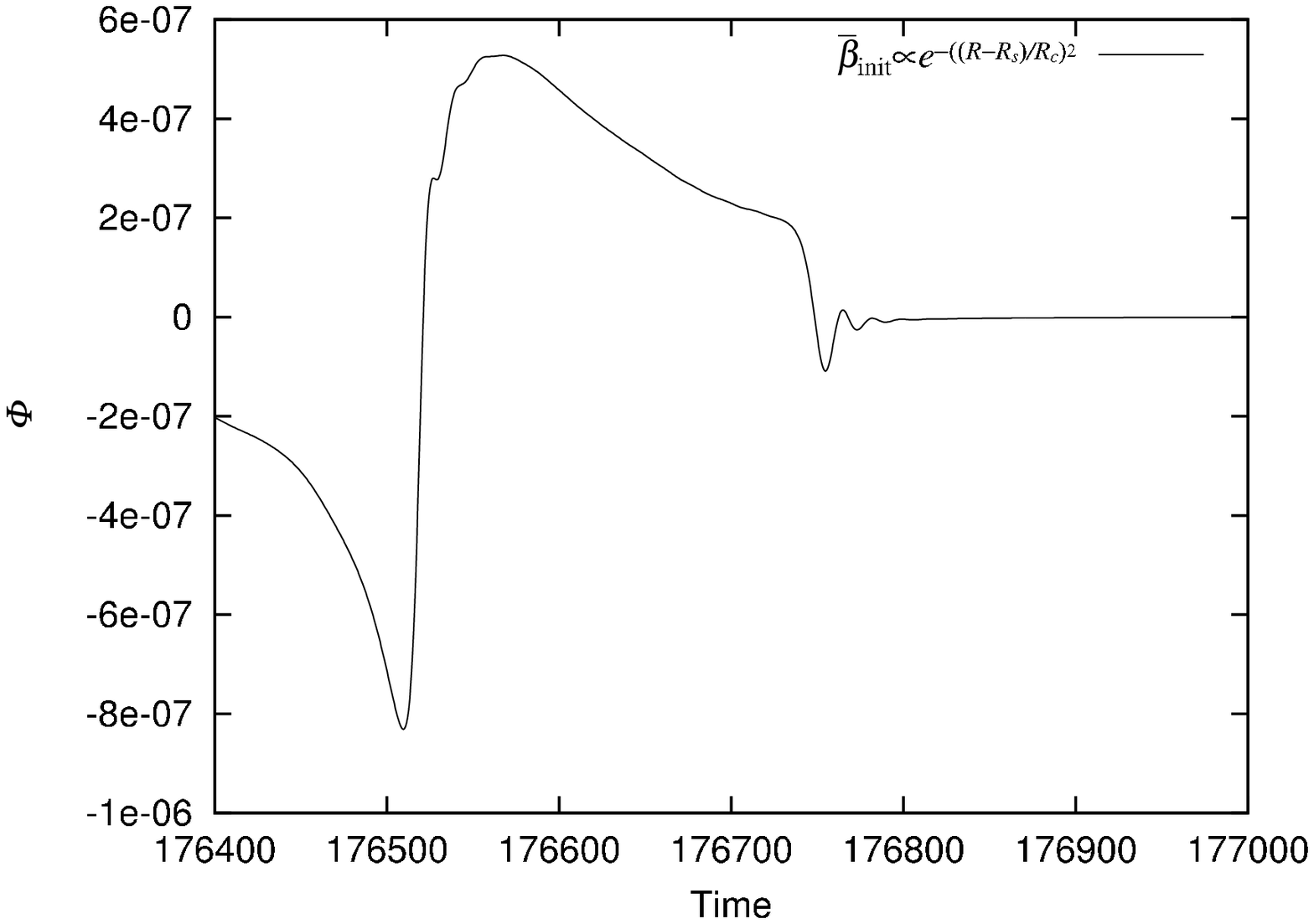} &
(f)      \includegraphics[scale=0.4]{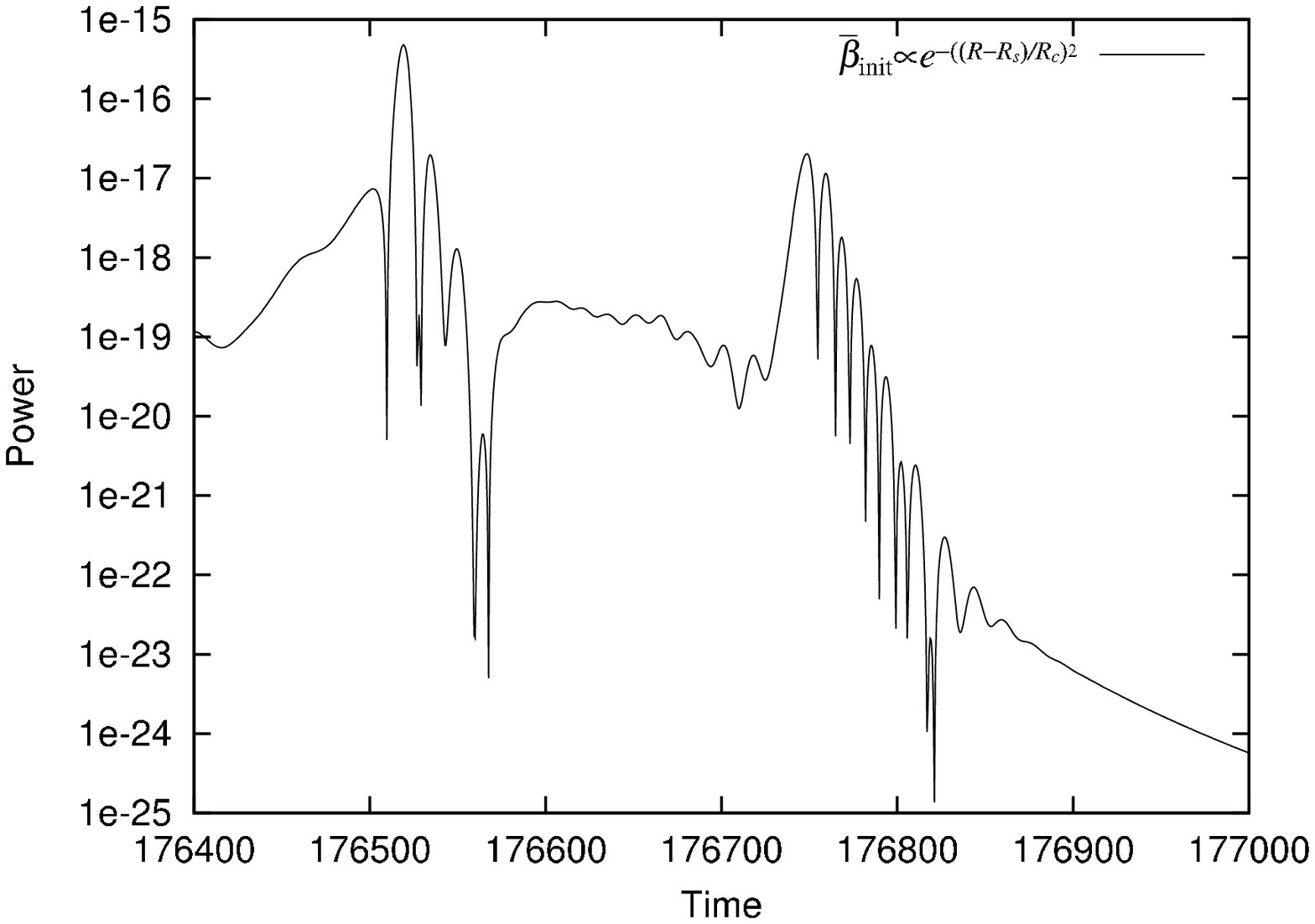} \\
(g) \includegraphics[scale=0.4]{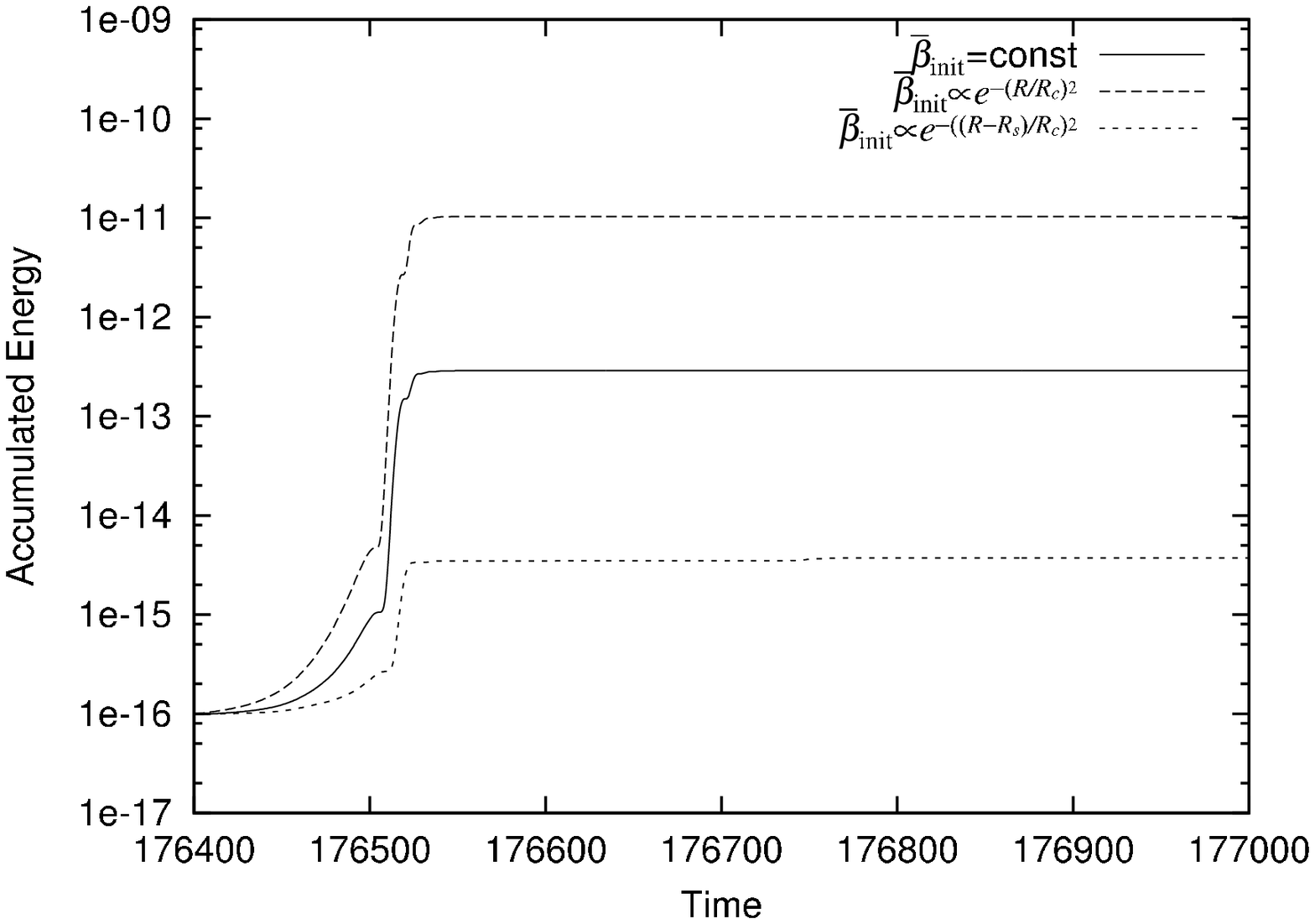} & 
(h) \includegraphics[scale=0.4]{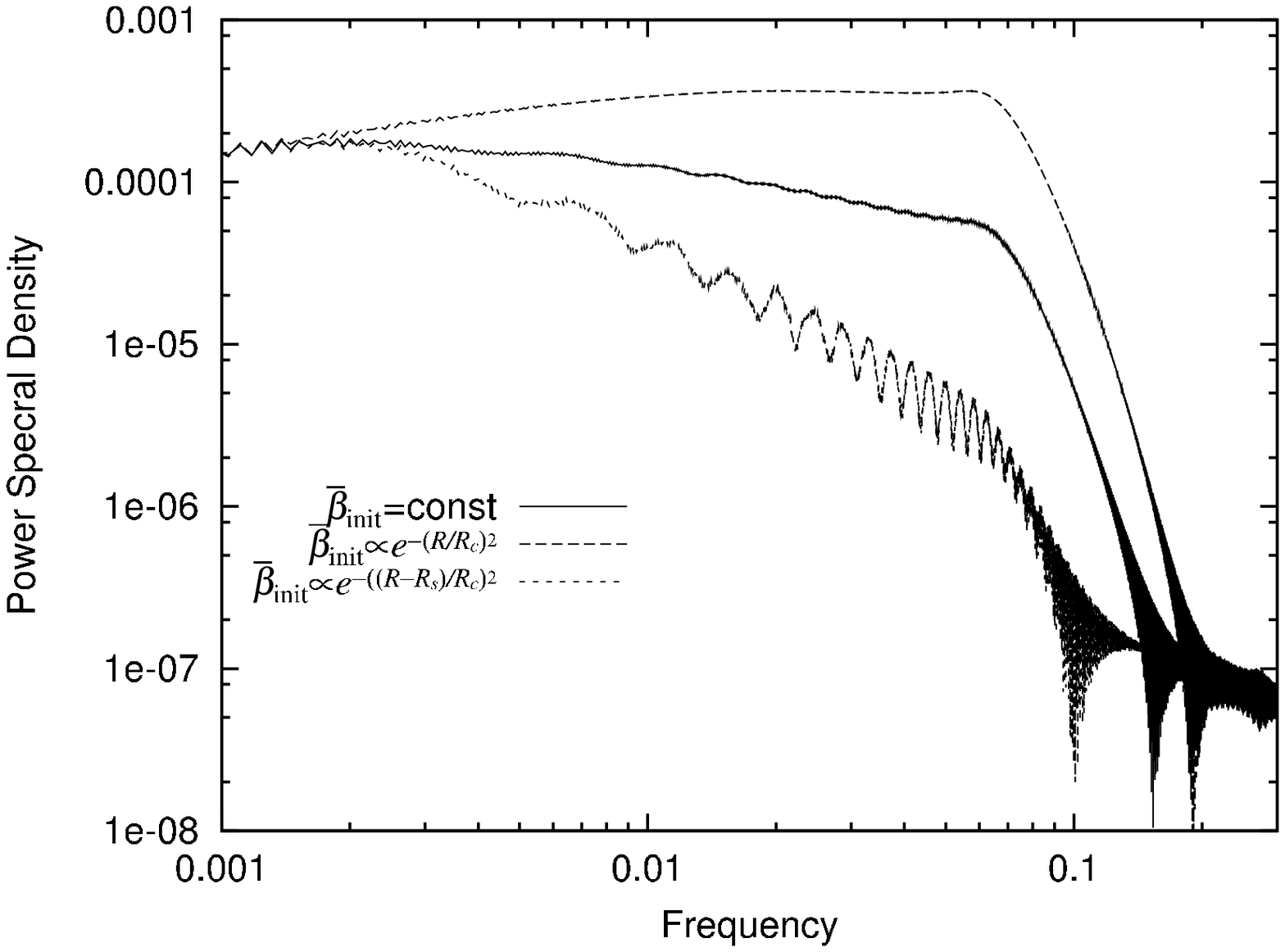} \\
    \end{tabular}
  \end{center}
\caption{\label{fg:gw_L2B_n_ih} 
(a), (c), and (e) Waveform, (b), (d), and (f) luminosity,
(g) accumulated energy, and 
(h) the one-sided power spectral density 
of gravitational waves for $l=2$ radiated from 
the collapse of model B with $\Gamma_{\rm a}=(4/3)+0.00142$. 
Gravitational waves are extracted at $R=2000M$.
The momentarily static initial perturbation is given. 
The amplitude of the perturbation is normalized so that
$q=2M$. 
In (g) and (h), the solid, long-dashed, and dashed lines denote
the results for 
$\bar{\beta}_{\rm init}=\mbox{const}$, $\exp[-(R/R_{\rm c})^{2}]$, and 
$\exp\{-[(R-R_{\rm s})/R_{\rm c}]^{2}\}$, respectively,
where $\bar{\beta}_{\rm init}$ is the initial distribution of the renormalized
matter perturbation and $R_{\rm c}$ is chosen to be $R_{\rm c}=R_{\rm s}/3$.
For waveforms and luminosities, we display the results for 
each case in each figure for clarity.
Note that the vertical axis of (g) is 
in logarithmic scale.
The matching is done on the mass shell which encloses 
99.7\% of the total mass. 
All the quantities are shown in units of $M=1$.}
\end{figure}

We show the waveform, luminosity, 
and integrated total energy 
of quadrupole ($l=2$) gravitational waves 
for different initial perturbations (1)--(3) in Fig.~\ref{fg:gw_L2B_n_ih}.
In all three cases, the amplitude of the 
perturbation is normalized so that $q=2M$. 
Here, we display the results with $N=1000$. 

As seen in Fig.~\ref{fg:gw_L2B_n_ih}, the gravitational waveform 
depends strongly on the perturbation profile initially given. 
For case (2), as seen in Figs.~\ref{fg:gw_L2B_n_ih}(c) and (d),
gravitational waves of high amplitude 
are emitted around $\bar{u}\simeq 176510M$, 
approximately at the same time as 
the formation of the high-redshift region. 
The waveform is characterized mainly by 
a quasinormal mode of the formed black hole. 
The duration of this major emission is roughly $\sim 50 M$, 
i.e., approximately equal to the period and/or
the damping time of the quasinormal mode. 
Indeed, the waveform for $\bar{u}\agt 176520M$
is well fitted by a damped oscillation
of the complex frequency $2M\omega \approx 0.74 +0.19i$, 
which agrees with the theoretically predicted value
$2M\omega = 0.74734 +0.17792i$~\cite{cd1975}
within a few percent error.
It is possible in principle in the present scheme 
to observe the modulation
of gravitational waveforms due to mass accretion as 
Papadopoulos and Font~\cite{pf2001} demonstrated for the 
spherically symmetric Klein-Gordon field on the accreting
black hole background.
However, in the present numerical results, it is difficult to distinguish 
the effects of the mass increase from the gravitational waveforms, 
possibly because the mass accretion rate is rather high and 
the duration of the accretion is not so long.

For case (1), as seen in Figs.~\ref{fg:gw_L2B_n_ih}(a) and (b),
gravitational waves look as the linear combination of a 
quasinormal mode of a black hole and a long-time-scale
and nonoscillative component
(note that the amplitude of gravitational waves does not
settle down to zero for a long-time duration $\simeq 200M$ after
$\bar{u}\simeq 176510M$). 
The long-time-scale component is produced due both to 
the perturbation profile initially given 
and to the nature of the collapse of the background spherical star: 
In the collapse of supermassive stars, 
the central region collapses first and subsequently, 
the outer envelope gradually falls into the black hole spending a
long time duration $\simeq 200M$. 
Since $\bar{\beta}_{\rm init}=$const, the outer envelope 
retains a considerable fraction of the perturbation for case (1). 
As a result, the quasinormal modes are likely to be
continuously excited for the duration $\simeq 200M$
during which the matter falls into a black hole.
However, the quasinormal modes continuously excited 
should cancel each other due to the phase cancellation effect
\cite{NS81}.
This suppresses the amplitude of gravitational waves and
produces a long-time-scale component, which results 
in the suppression of radiated energy as seen in 
Fig.~\ref{fg:gw_L2B_n_ih}(g).

For case (3), in which the matter perturbation is retained 
mainly near the stellar surface, 
as seen in Figs.~\ref{fg:gw_L2B_n_ih}(e) and (f), the effect
of the phase cancellation is more outstanding. 
In this case, the amplitude of gravitational waves
is highly suppressed and
the amplitude of the quasinormal mode ringing is much smaller
than that of the long-time-scale component.
We note that the time scale of the long-time-scale component 
is in approximate agreement with the time duration in 
which the accretion of the surrounding envelope continues. 

Figure~\ref{fg:gw_L2B_n_ih}(h) 
shows the one-sided power spectral density 
for the obtained gravitational waveforms for all three cases.
It is clear that the phase cancellation effects 
significantly suppress the high-frequency component
for cases (1) and (3)
while the low-frequency component 
depends not on the initial distribution of matter perturbation
but on the initial moment of perturbation. 

\begin{figure}[htbp]

\begin{center}
    \begin{tabular}{cc}
(a)   \includegraphics[scale=0.45]{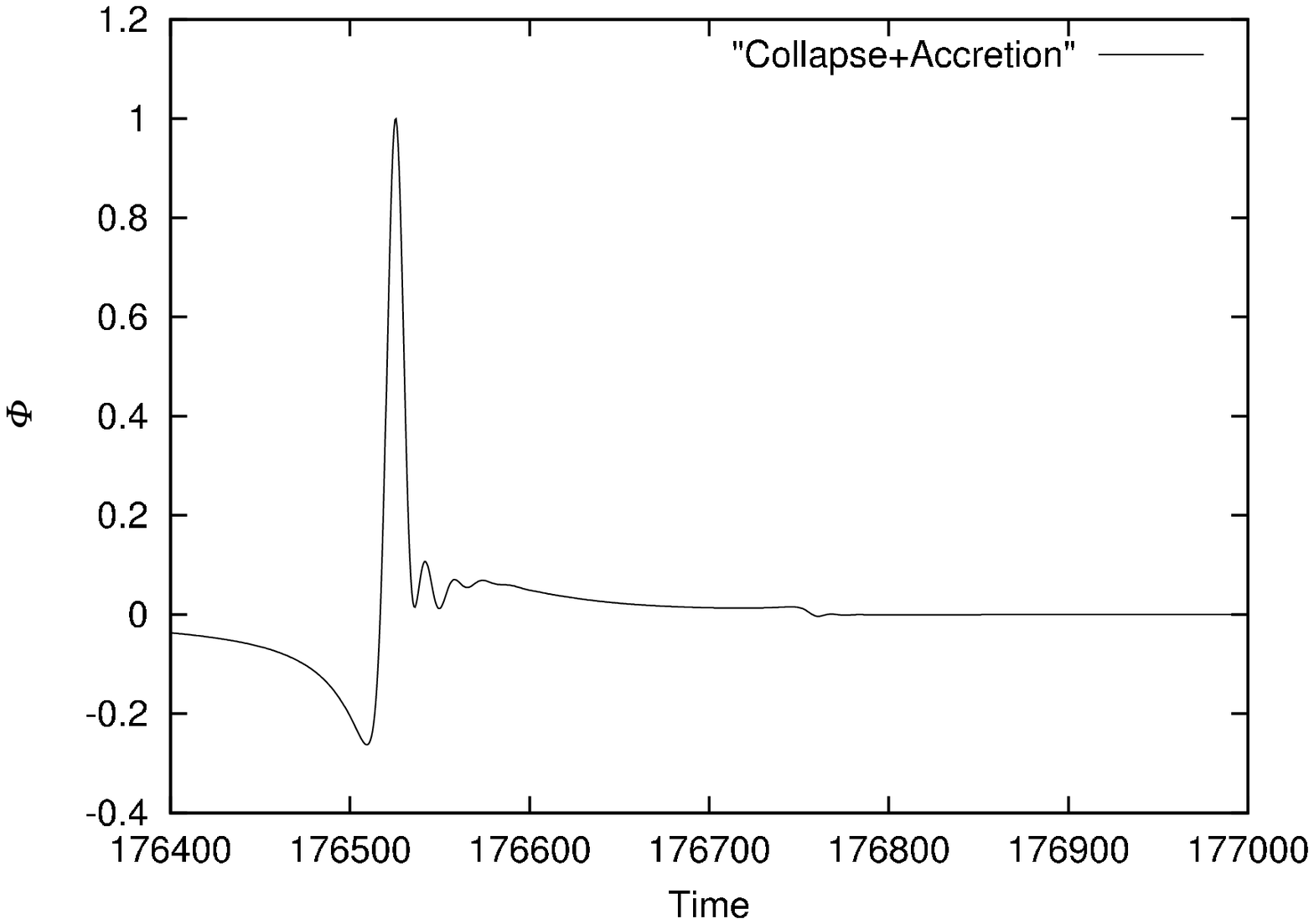} &
(b)   \includegraphics[scale=0.45]{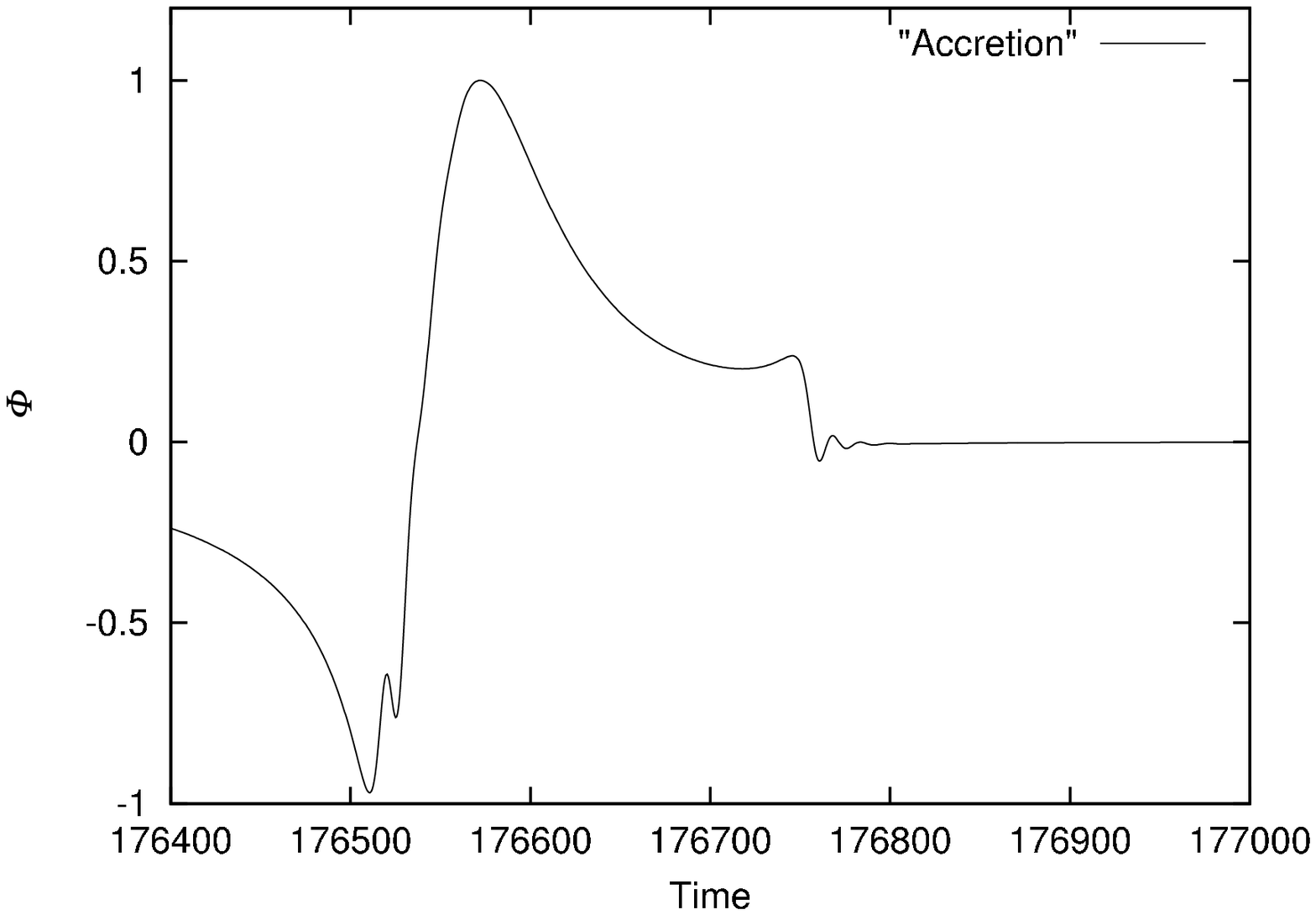} \\
(c)   \includegraphics[scale=0.45]{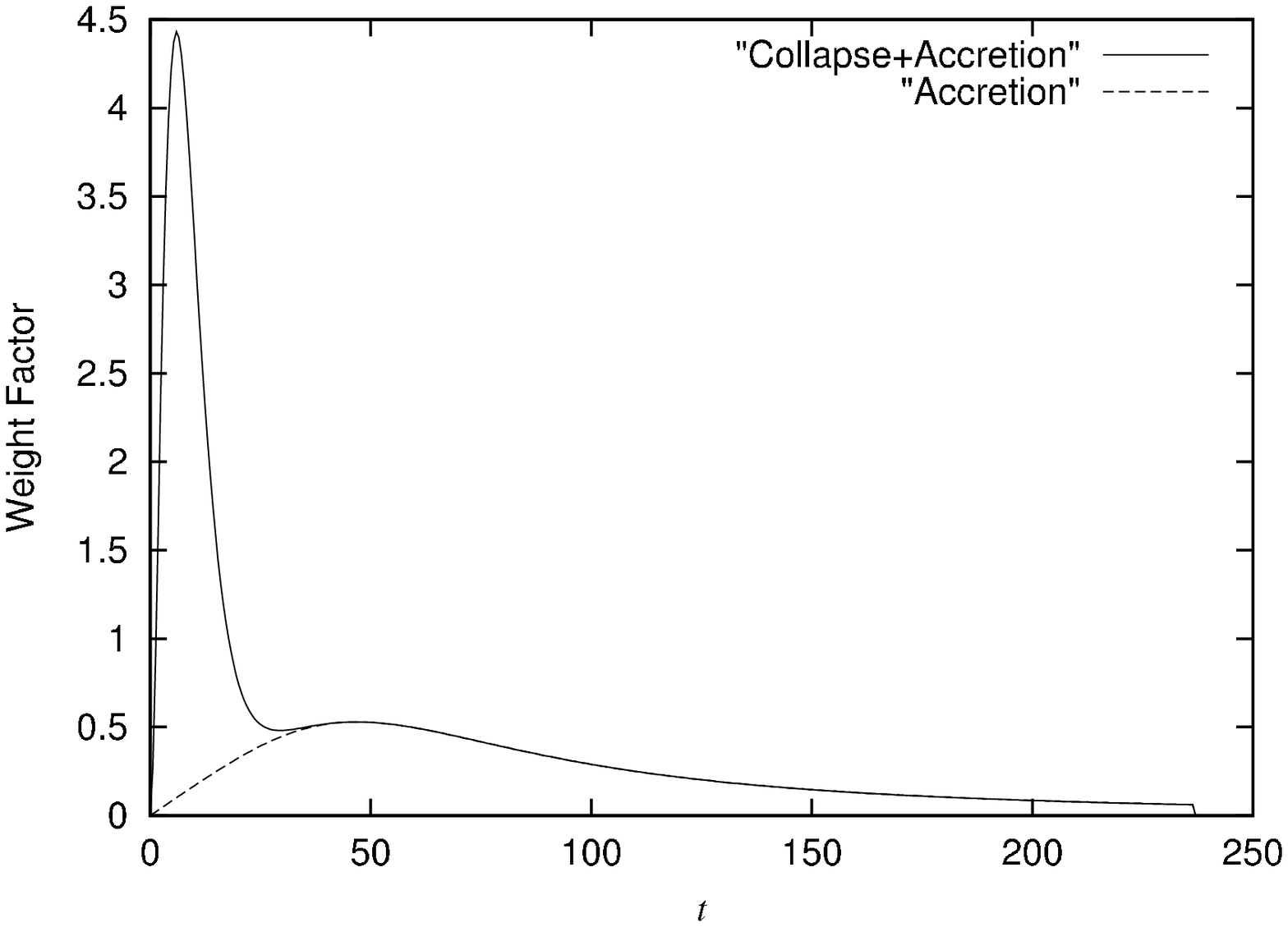} & \\
    \end{tabular}
  \end{center}
\caption{\label{fg:superimpose}
Waveforms generated by superimposing the waveform obtained
for case (2). The shape of weight factors are displayed in (c).
(a) and (b) show the waveforms for the weight factor labelled
``Collapse+Accretion'' and for that 
labelled ``Accretion.'' The waveforms are normalized so that 
the maximum amplitude is unity. See text for details.}
\end{figure}

To explain more clearly why the long-time-scale and 
nonoscillative component
appears, we artificially superimpose the
waveform $\Phi(t)$ obtained for case (2) displayed in 
Fig.~\ref{fg:gw_L2B_n_ih}(c),
which is characterized by quasinormal ringing, 
with the weight factor $w(t)$ as
\begin{equation}
\int_{-\infty}^{\infty}dt'w(t')\Phi(t-t'), 
\end{equation}
where $w(t)$ is chosen by trial as
\begin{equation}
w(t)=\left\{
\begin{array}{cc}
8\left(\frac{t}{t_{\rm c}}\right)\exp\left(
-\frac{t}{t_{\rm c}}\right)+\left(\frac{t}{t_{\rm a}/4}\right)\left[
\left(\frac{t}{t_{\rm a}/4}\right)^3+1\right]^{-1}
 & \mbox{for } 0\le t\le t_{\rm a}, \\
0                                   & \mbox{for } t<0, t_{\rm a}<t, 
\end{array}
\right.
\end{equation}
and we set $t_{\rm c}=2.95M$ and $t_{\rm a}=236M$.
The $t_{\rm c}$ and $t_{\rm a}$ correspond to 
the time scales of the collapse of the inner region
and of the accretion of the outer envelope, respectively.
See Fig.~\ref{fg:superimpose}(c) for the shape of this function.
In the above functional form, the first and second terms imitate
the effects of inner collapse and subsequent 
accretion of the outer envelope, 
respectively.
In Fig.~\ref{fg:superimpose}(a), 
we display the result for the superimposed
waveform. 
This result is qualitatively the same as that obtained 
for case (1) displayed in Fig.~\ref{fg:gw_L2B_n_ih}(a).
We note that the long-time-scale and nonoscillative 
component is produced by the long-term superposition 
of quasinormal ringing, including a precursory ``burst'' wave. 

Next we choose another weight factor $w(t)$ defined as
\begin{equation}
w(t)=\left\{
\begin{array}{cc}
\left(\frac{t}{t_{\rm a}/4}\right)\left[
\left(\frac{t}{t_{\rm a}/4}\right)^{3}+1
\right]^{-1} & \mbox{for } 0\le t\le t_{\rm a}, \\
0                                   & \mbox{for } t<0, t_{\rm a}<t, 
\end{array}
\right.
\end{equation}
and we again set $t_{\rm a}=236M$.
See Fig.~\ref{fg:superimpose}(c) for the shape of this function.
This functional form imitates the effect of accretion of the 
outer envelope alone.
In Fig.~\ref{fg:superimpose}(b), we display the result for the superimposed
waveform. This result is qualitatively the same as that obtained 
for case (3) displayed in Fig.~\ref{fg:gw_L2B_n_ih}(e).
It should be again emphasized that 
the superposition not of the
pure damped oscillation but of the full waveform for case (2),
including the precursory burst wave, 
produces the long-time-scale component. 
The above consideration confirms our speculation.

We note that the second 
quasinormal ringing seen around $\bar{u}\simeq 176750M$ 
in Figs.~\ref{fg:gw_L2B_n_ih}(a) and (b) for case (1) 
and in Figs.~\ref{fg:gw_L2B_n_ih}(e) and (f) for case (3)
are excited by the stellar surface which falls into the
black hole finally.
Since the phase cancellation is not effective at this stage,
the quasi-normal ringing dominates the waveform
for the latest phase.
After this ringing, a power-law tail is seen, and
the power-law index agrees with the theoretically predicted
value $-7$. 

Finally we note the following point.
In this simulation, all the matter is swallowed into
a black hole eventually because the zeroth-order solution
is spherically symmetric.
That is the reason why the accretion ends at some finite moment.
However, in a realistic nonspherical problem, the matter 
around the stellar surface does not fall into the black hole,
and would form surrounding disks. This implies that
the gravitational waveforms calculated for the latest phase
$\bar{u}\agt 176750M$
will be modified in a realistic simulation.

\subsection{Collapse of a neutron star}

We performed a simulation for the collapse of 
a neutron star to a black hole adopting model D. 
As in the collapse of a supermassive star, 
1000 grid points are typically taken to cover the neutron star interior. 
For the collapse of model D,
both the May-White and Hernandez-Misner schemes can
evolve the collapse to the formation of the event horizon 
with reasonable accuracy.
In the following, the results in the Hernandez-Misner scheme
will be described.
In this simulation, the matching is done on the mass shell 
within which 96.1\% of the total mass is enclosed.
In the following,
we deal with this matching surface as the stellar surface.

\begin{figure}[htbp]
\begin{center}
\begin{tabular}{cc}
(a)\includegraphics[scale=0.45]{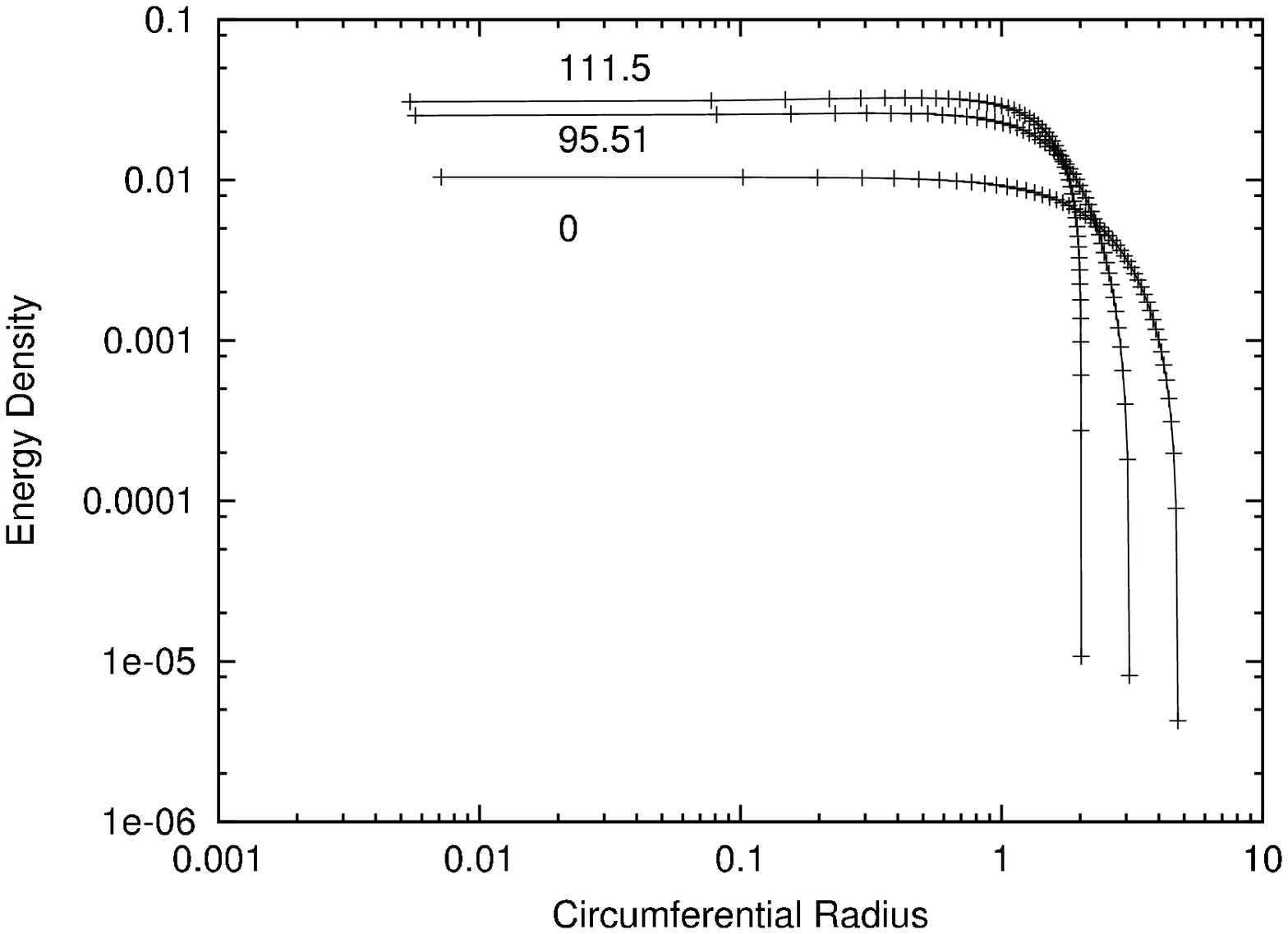} &
(b)\includegraphics[scale=0.45]{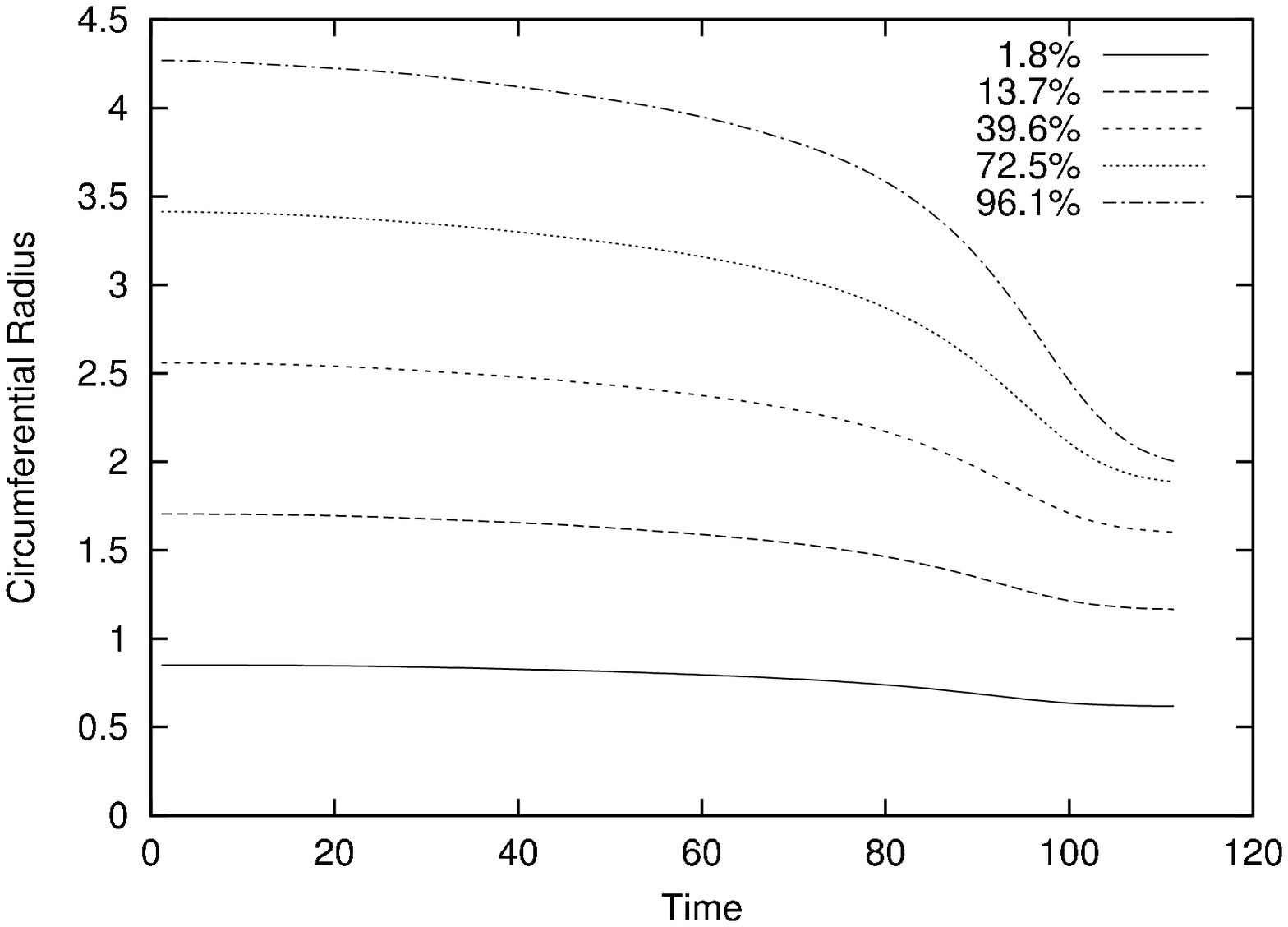} \\
(c)\includegraphics[scale=0.45]{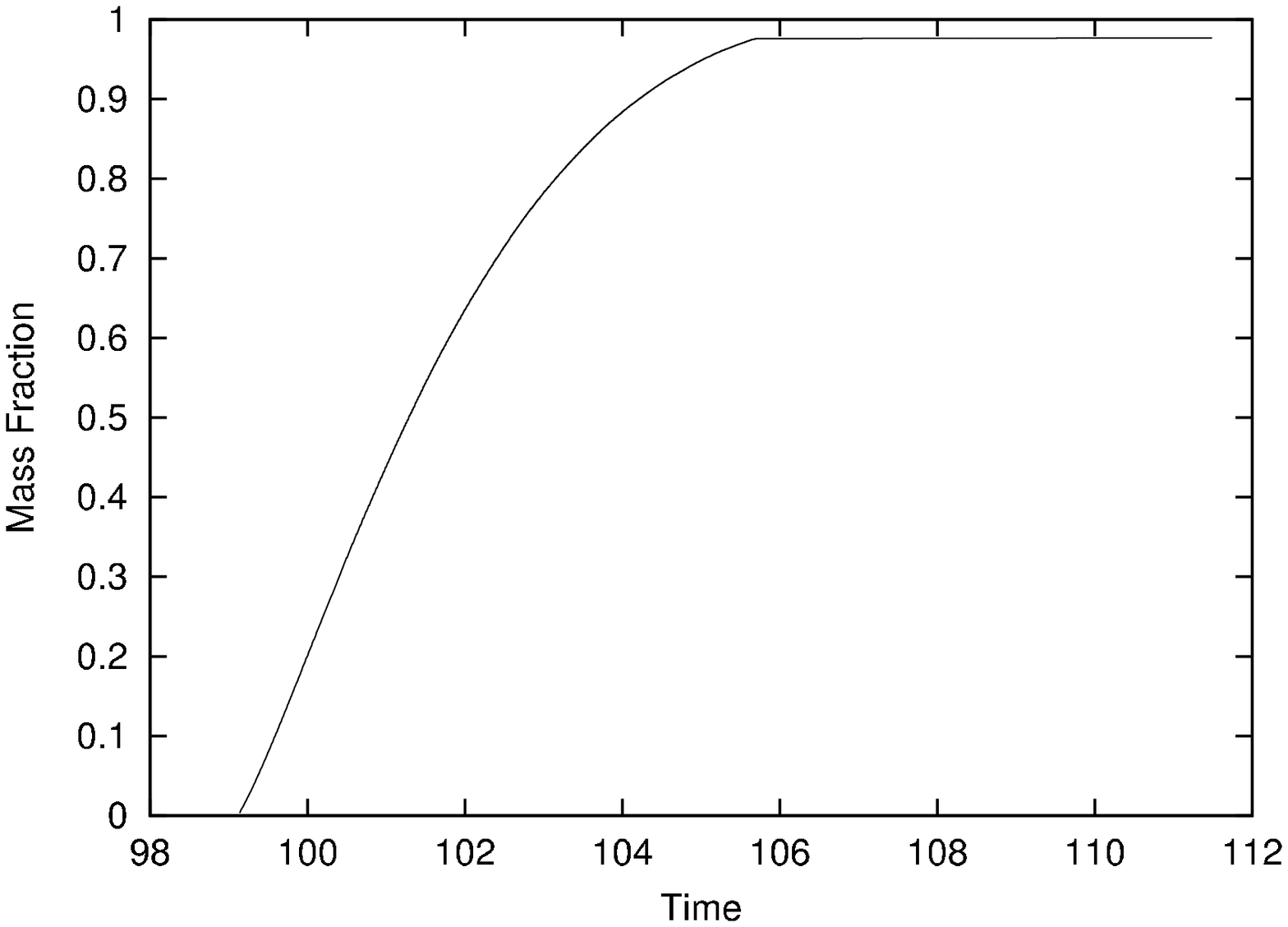} & \\
\end{tabular}
\end{center}
\caption{\label{fg:collapse_D} 
Same as Fig.~\ref{fg:collapse_B} but for the collapse of 
model D with $\Gamma_{\rm a}=2$.
We deal with the mass shell which encloses 96.1\% of the total mass
as the stellar surface.
}
\end{figure}

In Fig.~\ref{fg:collapse_D} we show the snapshots of 
the density profile at selected time steps, trajectory of 
the mass shells as a function of time, and 
the mass fraction of the fluid elements in the region of 
strong gravitational field in which the magnitude of 
the lapse function is less than 0.1. We note that 
in this simulation, the event horizon is just about to form at 
$\bar{u}\simeq 111.5M$. 
The calculation has been stopped at $\bar{u}\simeq 111.5M$
or $\tau_{\rm s}\simeq 68.04M$.
At this moment, we have obtained the redshift 
$(1+z_{\rm s})\simeq 22.89$ at the surface.
In contrast with the collapse of the supermassive stars, 
the central density increases only by a factor of $\approx 3.0$ 
throughout the collapse.
Thus, the evolution of the central density does not show 
the runaway behavior. Also, the collapse proceeds very coherently: 
Figure~\ref{fg:collapse_D}(c) indicates 
that the high-redshift region first appears at the center 
at $\bar{u}\simeq 99.1M$ and 
that almost all the fluid elements collapse to the high-redshift 
region in a short time interval $\Delta \bar{u}\sim 6M$.

To study the evolution of the perturbation, 
we gave the momentarily static initial data sets 
with matter perturbations (1)--(3) as
in the collapse of a supermassive star. The 
amplitude of the perturbation is normalized so that $q=2M$. 

\begin{figure}[htbp]
\begin{center}
\begin{tabular}{cc}
(a)\includegraphics[scale=0.45]{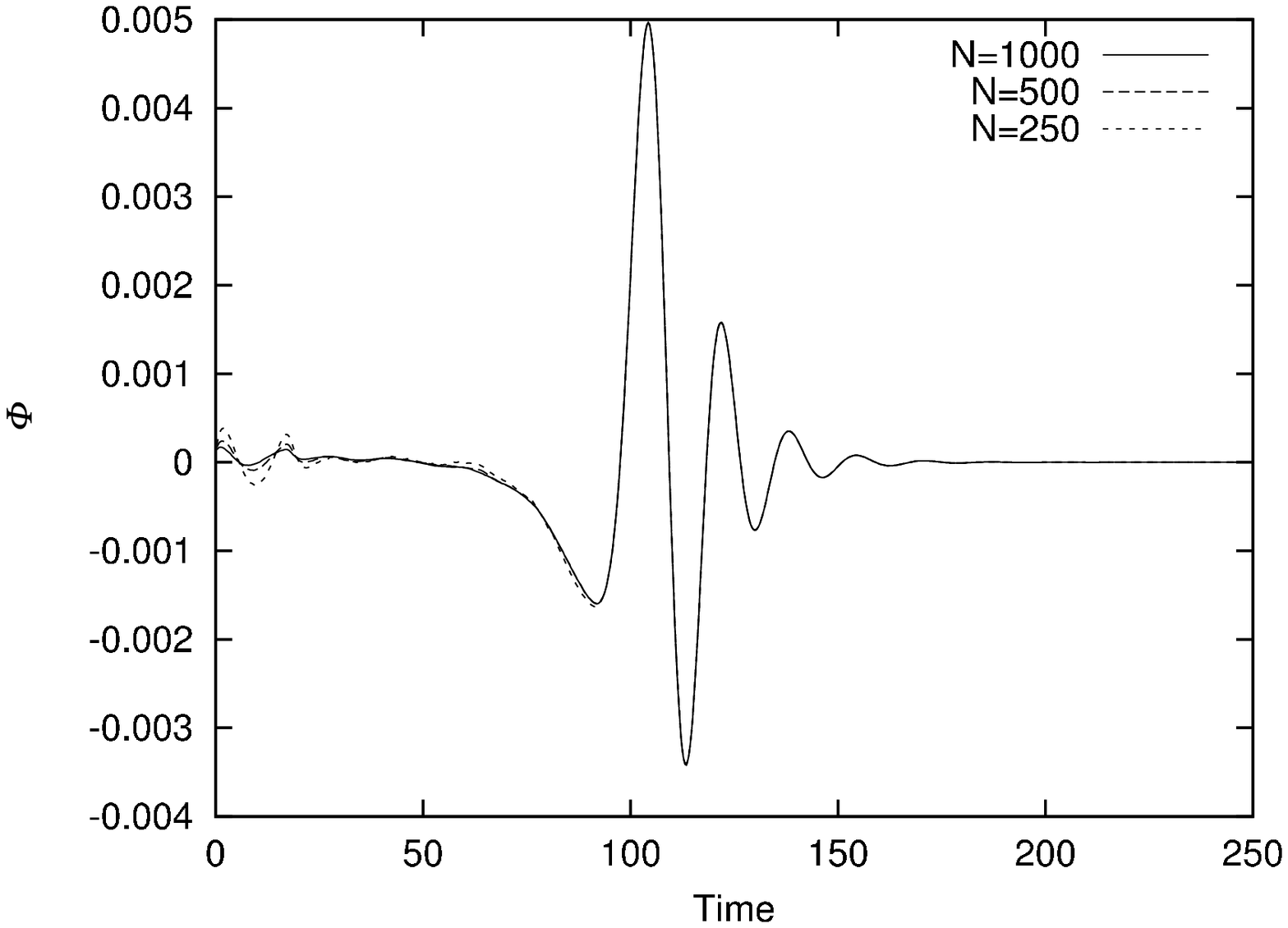} &
(b)\includegraphics[scale=0.45]{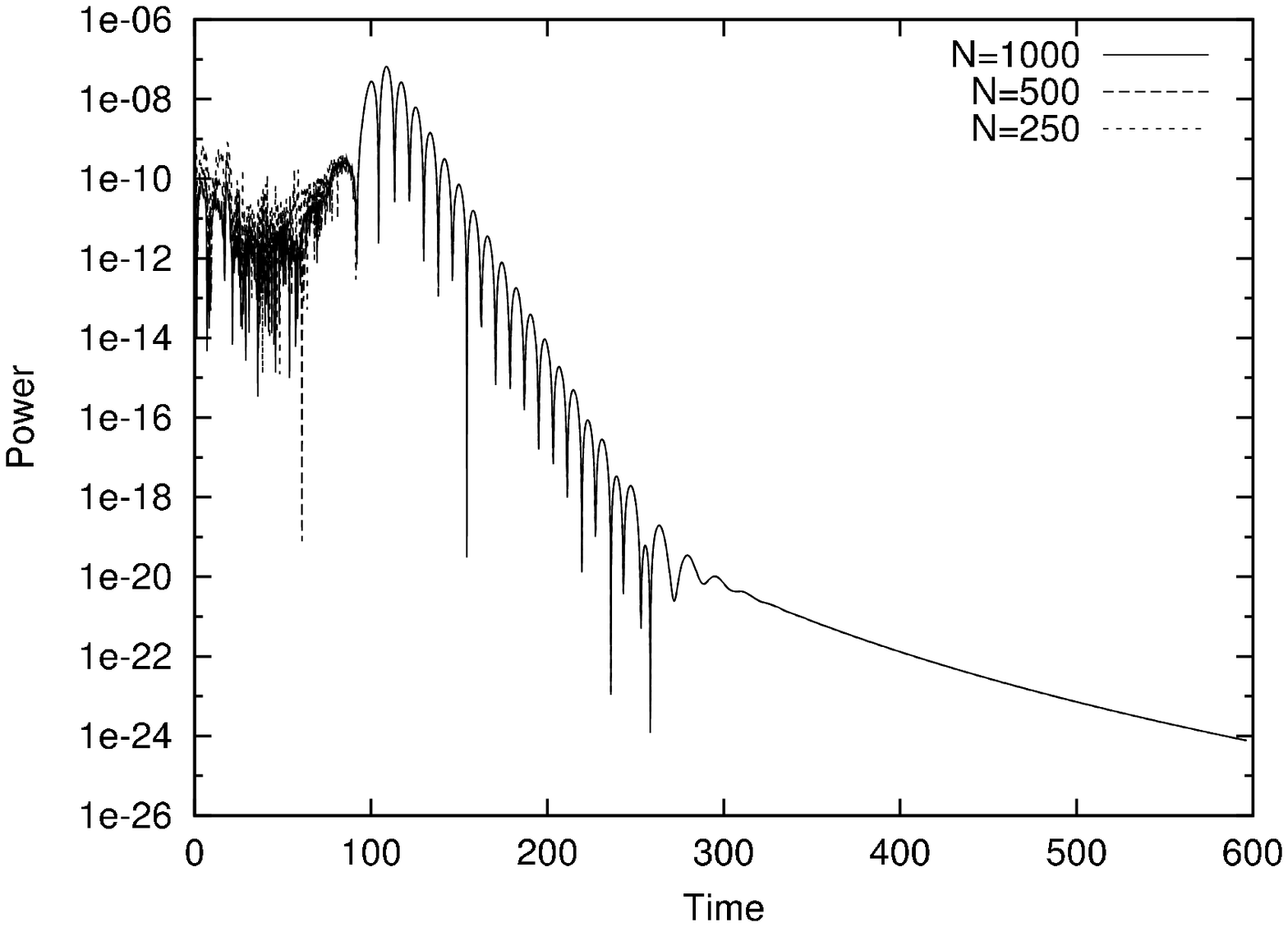} \\
(c)\includegraphics[scale=0.45]{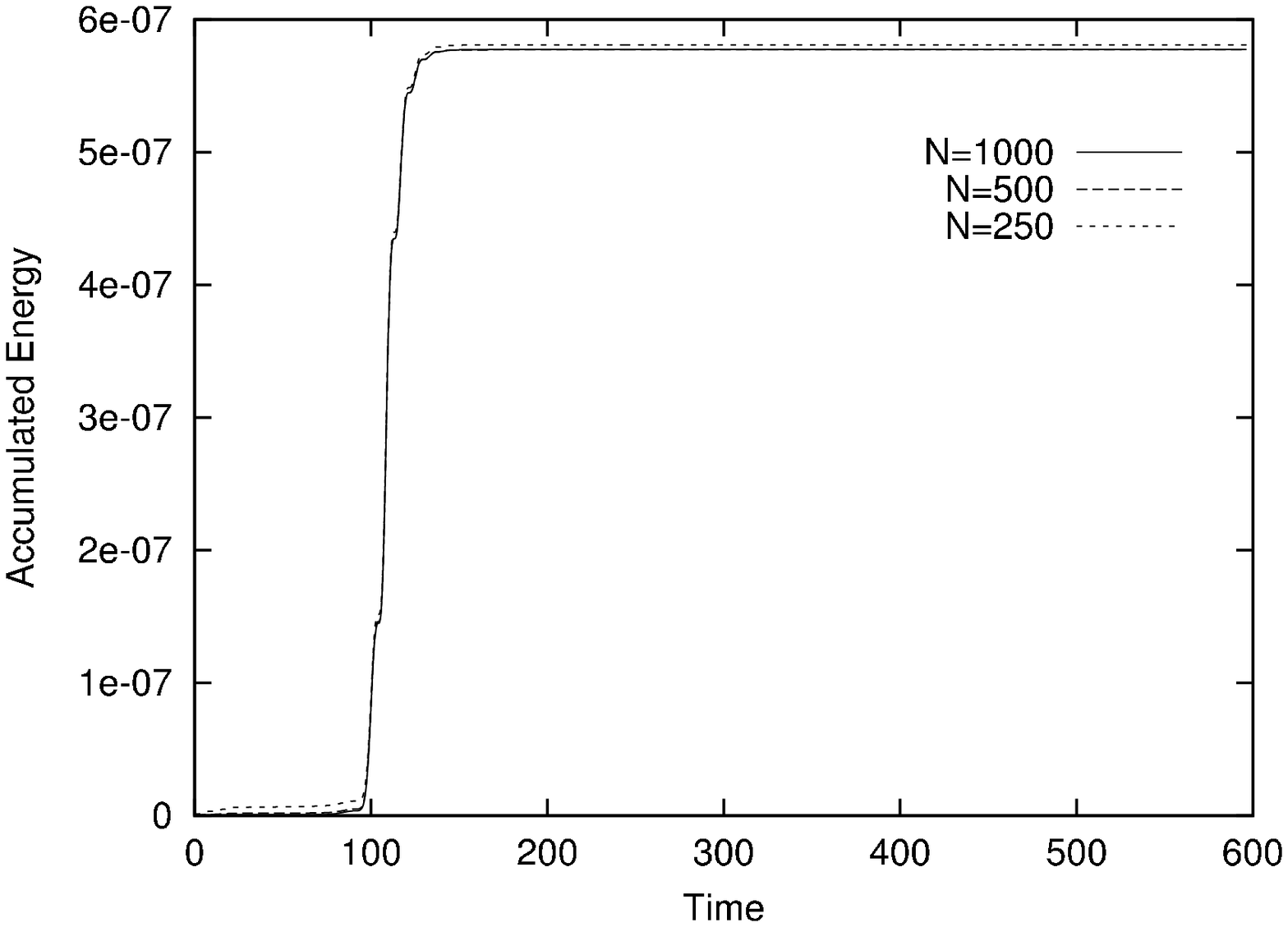} & \\
\end{tabular}
\end{center}
\caption{\label{fg:gw_L2D_n}
Same as Fig.~\ref{fg:gw_L2B_n} but for  
the collapse of model D with $\Gamma_{\rm a}=2$.
Gravitational waves are extracted at $R=100M$.
The matching is done on the mass shell which encloses 
96.1\% of the total mass. All the quantities are shown 
in units of $M=1$.}
\end{figure}

In Fig.~\ref{fg:gw_L2D_n}, we show 
the waveform, luminosity and accumulated energy of
gravitational waves for case (1). 
Computations were carried out with $N=1000$, $500$, and $250$,
and Fig.~\ref{fg:gw_L2D_n} demonstrates that 
convergence is achieved (note that results with $N=1000$ and 500
agree well with each other).
A small modulation is found just after the beginning of computation.
This is because the initial data sets which are
numerically constructed 
are not precisely static solutions of the 
finite differencing equation for the evolution of perturbations.
The above is confirmed by the fact that this modulation
becomes smaller as $N$ increases.
After this spurious precursor disappears, a black hole 
quasinormal mode is excited and a power-law tail 
follows~\footnote{The similar results were obtained for linearized 
gravitational waves during the spherically symmetric homogeneous dust 
collapse~\cite{cpm1978}
and for scalar waves during the 
neutron star collapse in the fully nonlinear spherically symmetric 
Einstein-fluid-Klein-Gordon
system~\cite{siebel2002}.}. 
No other outstanding feature is found in the gravitational waveform 
in contrast with the case of the supermassive star collapse. 
This is likely due to that 
the collapse proceeds very coherently for $\Gamma_{\rm a}=2$. 
The first and major gravitational emission 
of the quasinormal oscillation occurs 
approximately just after
the appearance of the high-redshift region.
From the numerical result, 
the quasinormal frequency and 
the index of the power-law tail are calculated as 
$2M\omega = 0.752+0.176 i$ and $-7.1$.
These values agree well with the theoretical values
$2M\omega=0.74734+0.17792i$ and $-(2l+3)=-7$~\cite{cd1975,price1972}
both within a few percent error.

\begin{figure}[htbp]
\begin{center}
\begin{tabular}{cc}
(a)\includegraphics[scale=0.45]{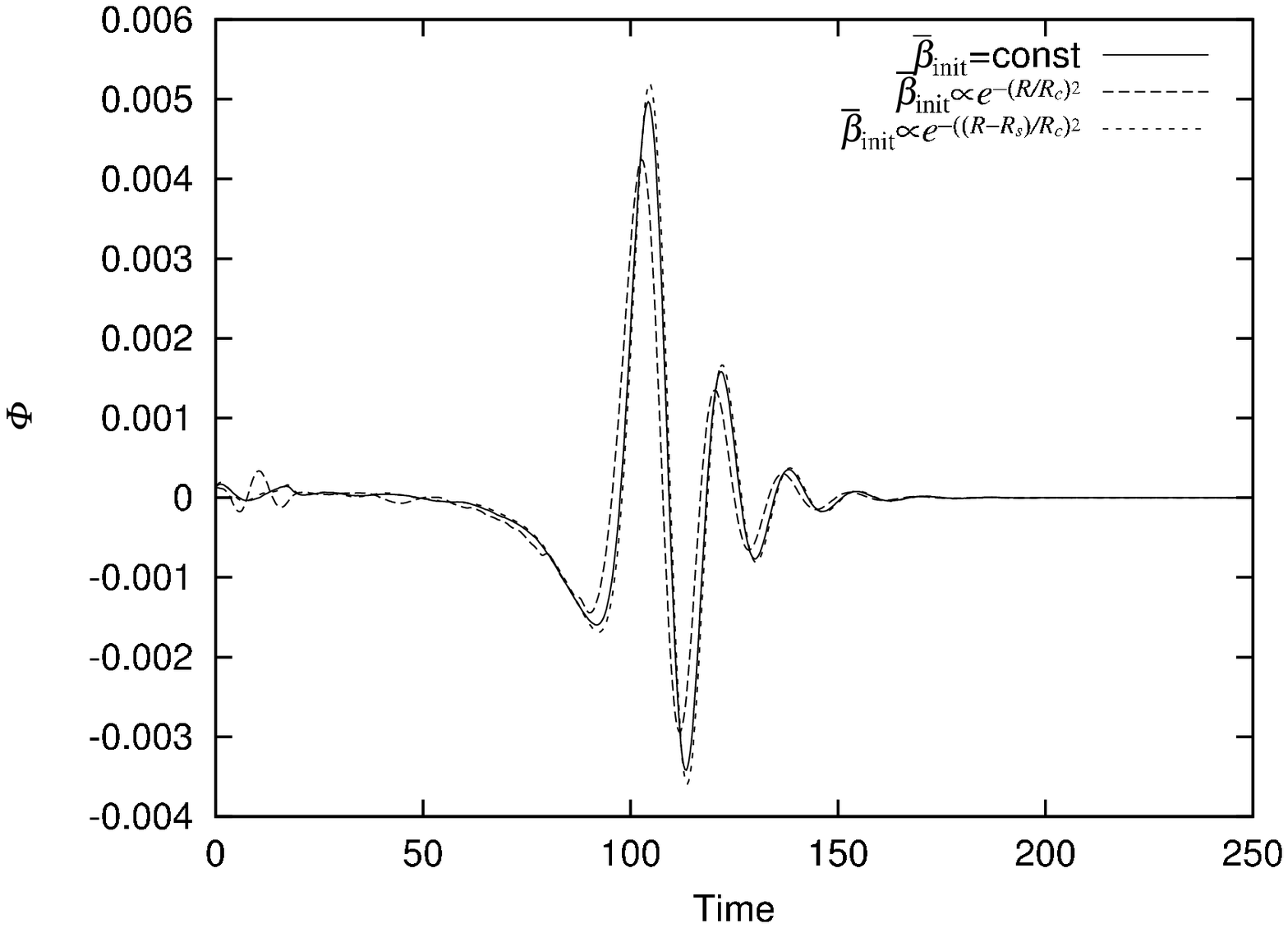} &
(b)\includegraphics[scale=0.45]{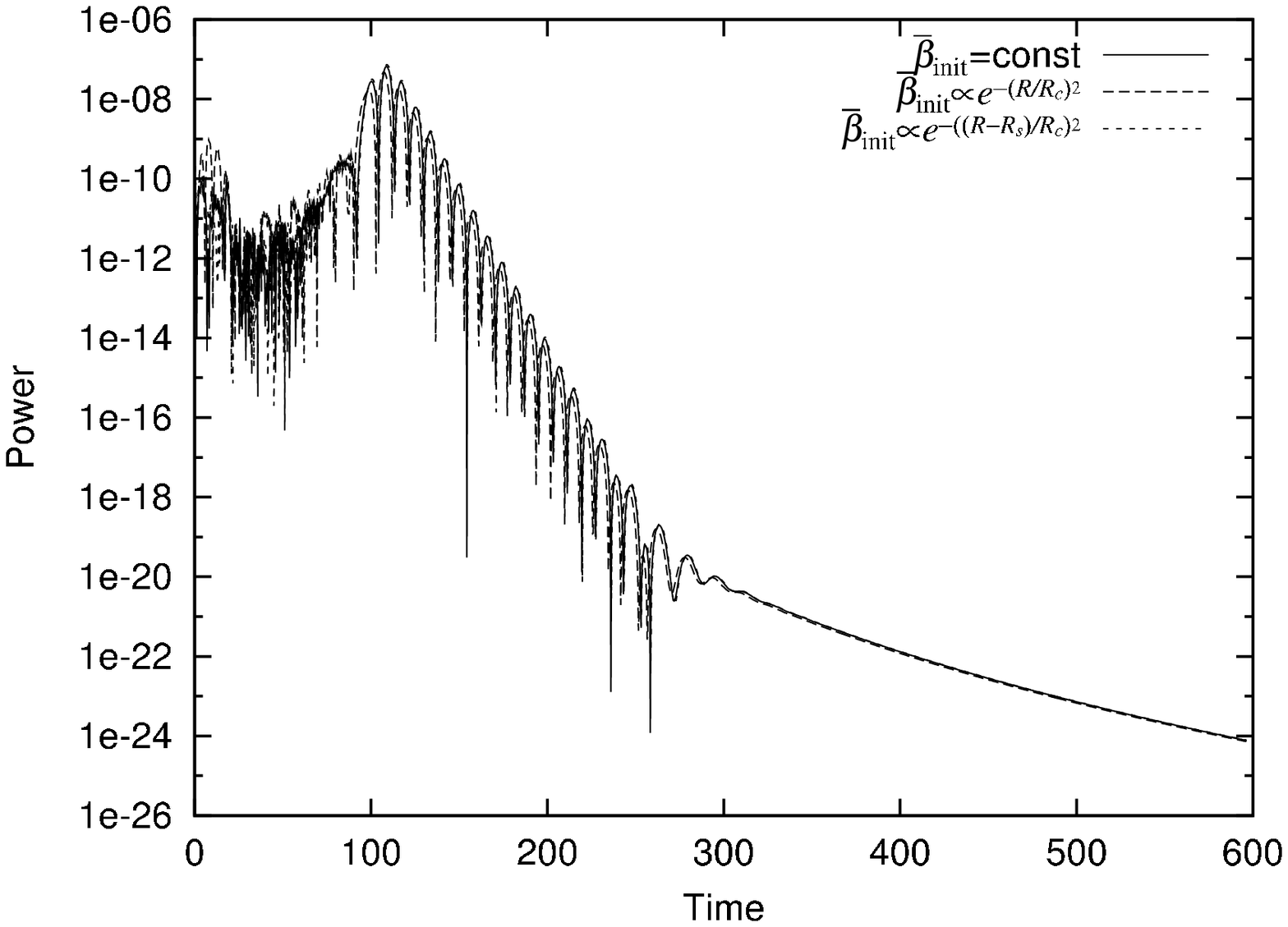} \\
(c)\includegraphics[scale=0.45]{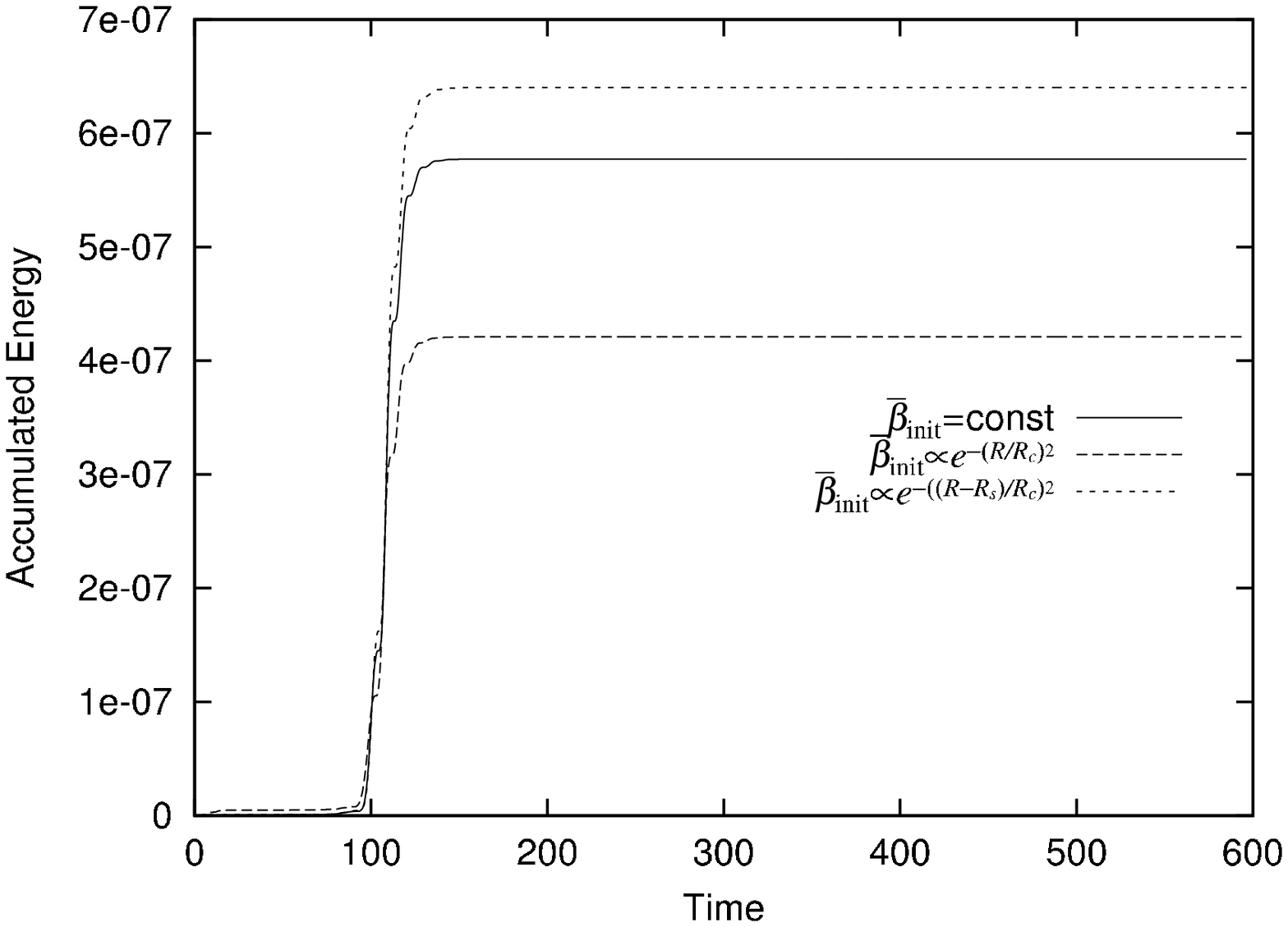} & 
(d)\includegraphics[scale=0.45]{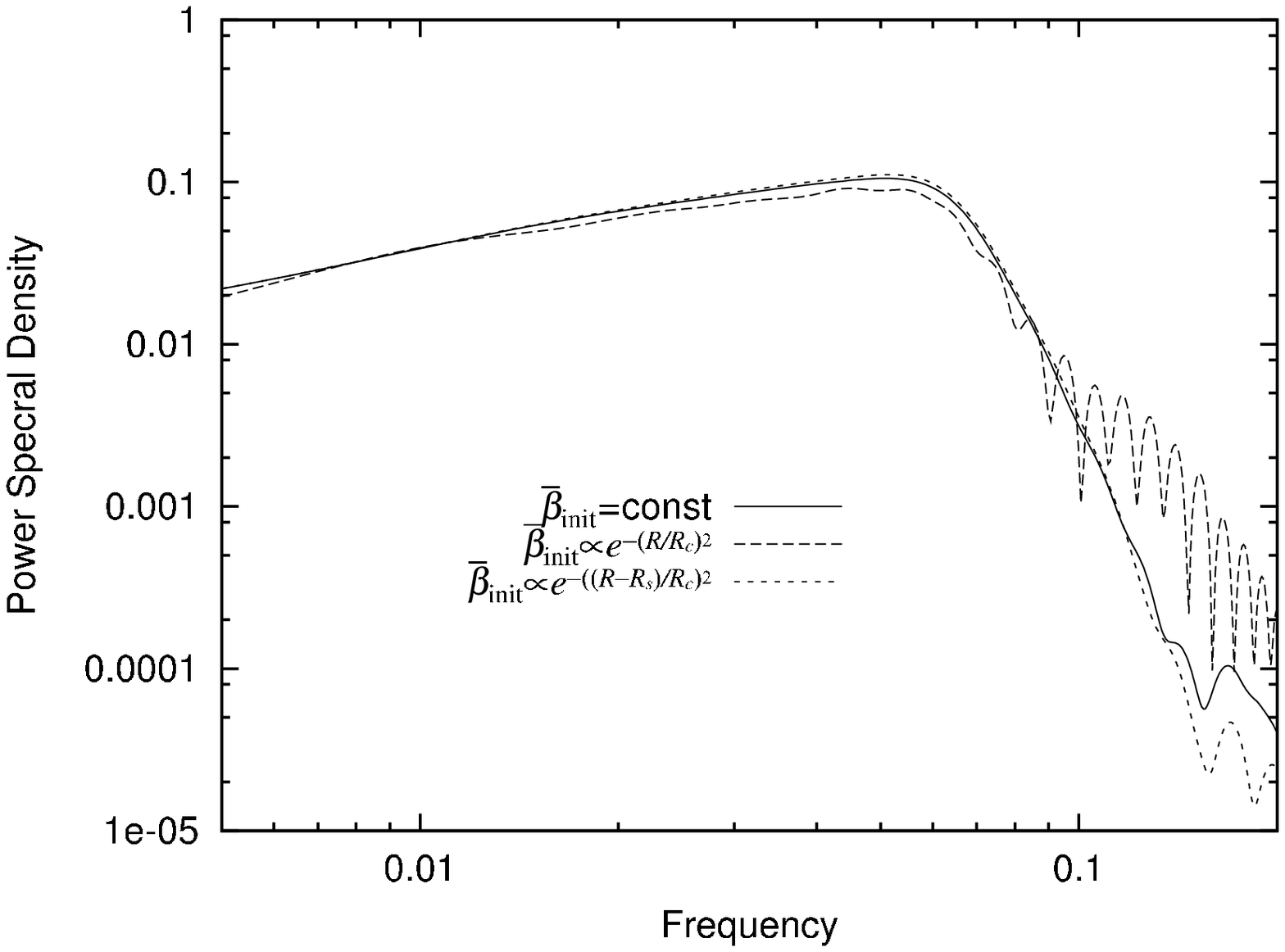} \\
\end{tabular}
\end{center}
\caption{\label{fg:gw_L2D_n_ih}
(a) Waveform, (b) luminosity,
(c) accumulated energy, and (d) one-sided 
power spectral density 
of gravitational waves for $l=2$ radiated from 
the collapse of model D with $\Gamma_{\rm a}=2$.
Gravitational waves are extracted at $R=100M$.
The momentarily static initial perturbation is given. 
The amplitude of the perturbation is normalized so that
$q=2M$. 
The solid, long-dashed and dashed lines denote
the results for 
$\bar{\beta}_{\rm init}=\mbox{const}$, $\exp[-(R/R_{\rm c})^{2}]$, and 
$\exp\{-[(R-R_{\rm s})/R_{\rm c}]^{2}\}$, respectively,
where $\bar{\beta}_{\rm init}$ is the initial 
distribution of the renormalized
matter perturbation and $R_{\rm c}$ is chosen to 
be $R_{\rm c}=R_{\rm s}/3$.
The matching is done on the mass shell which encloses 
96.1\% of the total mass. All the quantities are shown 
in units of $M=1$.}
\end{figure}

Figure~\ref{fg:gw_L2D_n_ih} shows gravitational waves for different
initial matter perturbations (1)--(3). 
The waveform depends only very weakly on the initial 
perturbation distribution.
The reason is that the collapse proceeds in a very coherent manner
and hence the difference of the perturbation distribution does not
generate outstanding differences.
There is no phase cancellation effects on the neutron star collapse
irrespective of the shape of the perturbation as opposed
to the supermassive star case as is seen in the shape of spectrum.
A small time delay of the oscillational 
phase is seen for cases (1) and (3) compared with case (2).
This is simply because the matter perturbation enters the black hole
for case (2) earlier than for cases (1) and (3).

\section{summary}

We have reported a new implementation in linearized Einstein theory.
In this code, the Hernandez-Misner scheme is adopted to compute 
a spherically symmetric zeroth-order solution. As a result, we
can compute stellar collapse to a static black hole until
the null hypersurface reaches the event horizon, and
the whole region outside the event horizon is numerically generated. 
We emphasize that the collapse of a supermassive star
to a black hole proceeds in a runaway manner, i.e., the central density
grows rapidly although the surrounding atmosphere does not collapse
very rapidly. It is not technically easy to compute such collapse
with a 1+1 numerical scheme. The Hernandez-Misner coordinate
system is essential to enable thorough computation of 
the black hole formation. 

We have also proposed a new numerical method to compute gravitational waves
in perturbation theory. We divide the 
computational domain into three regions. For computation of 
the perturbations in the exterior region, we adopt double-null
coordinates which agree with the characteristic
curves of gravitational waves. This choice 
enables the computation of gravitational waves 
emitted during the entire history of black hole formation. 

To study the qualitative nature of 
gravitational waves from stellar collapse, 
we performed simulations for the collapse of a supermassive star 
and a neutron star to black holes. 
In the gravitational collapse of the neutron star, 
gravitational waveforms are characterized by a 
black hole quasinormal mode, as 
demonstrated in the fully general relativistic 
simulations~\cite{SP}. On the other hand, for gravitational collapse of 
the supermassive star, the waveform depends strongly on the 
perturbation profile that we give initially. 
For a centrally concentrated matter perturbation, 
the waveforms are characterized mainly by a 
black hole quasinormal mode, as in the collapse 
of neutron stars. 
However, when the matter perturbation is distributed uniformly, 
the waveform is determined by a linear combination of 
the black hole quasinormal mode and 
a long-time-scale component which results from
the superposition of the quasinormal ringing component. 
Moreover, when the matter perturbation is located around the surface, 
the long-time-scale component dominates the waveform. 
This is likely due to the less-coherent nature of 
the collapse of supermassive stars: 
the central part collapses earlier and subsequently 
the outer envelope accretes on to the central black hole. 

We have determined the choices of initial 
distribution of perturbation not assuming physically 
realistic situations because of the lack of our knowledge
on physically realistic odd-parity perturbation.
Therefore, at present, we do not
claim that the gravitational waves obtained here are
realistic in particular for the collapse of supermassive
stars. However, our present results strongly suggest
that the gravitational radiation is not so sensitive 
to the initial condition for neutron star collapse 
but is highly sensitive for supermassive star collapse.

In the formation of intermediate-mass black holes
(mass $\agt 200M_{\odot}$) formed from quite massive stars,
which is destabilized 
by electron-positron pair creation~\cite{Fryer01},
collapse would proceed in the same manner
as for supermassive stars. 
Thus, gravitational waveforms from the formation
of intermediate-mass black holes
also depend strongly on the state of the precollapse star. 

We have focused on perturbations of odd parity. 
The dominant modes of gravitational waves are likely to be 
of even parity in most cases, and thus the study of the 
even-parity perturbations seems to be more important.
In a rotating stellar collapse,
quadrupole deformation that rotating stars retain before collapse
will be the source of gravitational waves of even parity. 
The study of such effects on gravitational waves emitted
in a black hole formation is now in progress. 

\acknowledgments

We are grateful to K.~Nakao and C.~Gundlach for helpful discussion.
We are also grateful to J.M.~Overduin for carefully reading the 
manuscript.
This work was partly supported by the 
Grant-in-Aid for Scientific Research (Nos. 05540, 11217, 13740143, and 
14047207) from the Japanese Ministry of
Education, Culture, Sports, Science and Technology. 

\appendix

\newpage

\clearpage

\clearpage

\end{document}